\documentclass[twocolumn]{aastex631}
\usepackage{amsmath}
\usepackage{graphicx}
\usepackage{soul}

\newcommand{\kms}{\mbox{${\rm km\,s^{-1}}$}}

\newcommand{\fescLyC}{\mbox{$f_{\rm LyC}$}}
\newcommand{\fescLya}{\mbox{$f_{\rm Ly\alpha}$}}
\newcommand{\fescLyCL}{\mbox{$\left<f_{\rm LyC}\right>$}}
\newcommand{\fescLyaL}{\mbox{$\left<f_{\rm Ly\alpha}\right>$}}
\newcommand{\fion}{\mbox{$f_{\rm dust}^{\rm ion}$}}
\newcommand{\fescLyCa}{\mbox{$f_{\rm LyC}'$}}

\newcommand{\msun}{\mbox{$M_{\odot}$}}

\newcommand{\Lya}{\mbox{Ly$\alpha$}}
\newcommand{\cmq}{\mbox{${\rm cm^{-3}}$}}
\newcommand{\vsep}{\mbox{$v_{\rm sep}$}}
\newcommand{\nH}{\mbox{$n_{\rm H}$}}
\newcommand{\tSFE}{\mbox{SFE$_{\rm tot}$}}

\received{\today}

\shorttitle{LyC--LyA escape from GMCs}
\shortauthors{Kimm et al.}

\begin{document}

\title{A systematic study of the escape of LyC and \Lya\ photons from star-forming, magnetized turbulent clouds}

\author[0000-0002-3950-3997]{Taysun Kimm}
\affiliation{Department of Astronomy, Yonsei University, 50 Yonsei-ro, Seodaemun-gu, Seoul 03722, Republic of Korea}

\author[0000-0002-4554-4488]{Rebekka Bieri}
\affiliation{Max Planck Institute for Astrophysics, Karl-Schwarzschild-Stra{\ss}e 1, 85748, Garching, Germany}

\author[0000-0002-3150-2543]{Sam Geen}
\affiliation{Anton Pannekoek Institute for Astronomy, Universiteit van Amsterdam, Science Park 904, NL-1098 XH Amsterdam, the Netherlands}

\author[0000-0002-7534-8314]{Joakim Rosdahl}
\affiliation{Univ Lyon, Univ Lyon1, Ens de Lyon, CNRS, Centre de Recherche Astrophysique de Lyon UMR5574, F-69230 Saint-Genis-Laval, France}

\author[0000-0003-1609-7911]{J\'er\'emy Blaizot}
\affiliation{Univ Lyon, Univ Lyon1, Ens de Lyon, CNRS, Centre de Recherche Astrophysique de Lyon UMR5574, F-69230 Saint-Genis-Laval, France}

\author{L\'eo Michel-Dansac}
\affiliation{Univ Lyon, Univ Lyon1, Ens de Lyon, CNRS, Centre de Recherche Astrophysique de Lyon UMR5574, F-69230 Saint-Genis-Laval, France}

\author[0000-0002-9613-9044]{Thibault Garel}
\affiliation{Observatoire de Gen\`eve, Universit\'e de Gen\`eve, 51 Ch. des Maillettes, 1290 Versoix, Switzerland}

\email{tkimm@yonsei.ac.kr}

\begin{abstract}
Understanding the escape of Lyman continuum (LyC) and Lyman $\alpha$ ($\Lya$) photons from giant molecular clouds (GMCs) is crucial if we are to study the reionization of the Universe and to interpret spectra of observed galaxies at high redshift.  To this end, we perform high-resolution, radiation-magneto-hydrodynamic simulations of GMCs with  self-consistent star formation and stellar feedback. We find that a significant fraction (15--70\%) of ionizing radiation escapes from the simulated GMCs with different masses ($10^5$ and $10^6\,\msun$), as the clouds are dispersed within about $2$--$5\,{\rm Myr}$ from the onset of star formation. The fraction of LyC photons leaked is larger when the GMCs are less massive, metal-poor, less turbulent, and less dense.  The most efficient leakage of LyC radiation occurs  when the total star formation efficiency of a GMC is about $20\%$. The escape of $\Lya$ shows a trend similar to that of LyC photons, except that the fraction of \Lya\ photons escaping from the GMCs is larger ($\fescLya\approx f_{900}^{0.27}$) and that a GMC with strong turbulence shows larger $\fescLya$. The simulated GMCs show a characteristic velocity separation of $\Delta v\approx 120 \,\kms$ in the time-averaged emergent \Lya\ spectra, suggesting that \Lya\ could be useful to infer the kinematics of the interstellar and circumgalactic medium. We show that \Lya\ luminosities are a useful indicator of the LyC escape, provided the number of LyC photons can be deduced through stellar population modeling. Finally, we find that the correlations between the escape fractions of \Lya, ultraviolet photons at 1500\AA, and the Balmer $\alpha$ line are weak.
\end{abstract}

\keywords{Giant molecular clouds (653) --- Photoionization (2060) -- Reionization (1383) -- Lyman-apha galaxies (978)}


\section{Introduction} 

Strong Lyman continuum (LyC) radiation produced by young massive stars is known to create low-density channels in giant molecular clouds (GMCs) \citep[e.g.,][]{dale12,geen18,kimjg18,grudic21a}. After escaping from their place of origin, the photons not only regulate the dynamics of the interstellar medium (ISM), but also drive the reionization of the Universe by ionizing neutral hydrogen in the intergalactic space \citep[e.g.,][]{madau99,iliev06,mesinger07,gnedin14,kimm14,norman15,oshea15,ocvirk16,pawlik17,rosdahl18,finlator18,doussot19,hutter21}. Therefore, understanding the propagation of LyC photons within GMCs is an important first step to build a coherent picture of star formation, galaxy evolution, and reionization. 

Reionization theory requires a significant fraction ($\sim10\%$) of LyC photons to escape from their host dark matter haloes in order to explain the end of reionization as well as the electron optical depth \citep{robertson15,dayal18}. The escape fraction of 10\% may seem low and even trivial, but this is not the case. 
For a simple stellar population with a Kroupa initial mass function,  the majority of LyC radiation is produced by short-lived massive stars, and hence it is necessary to disrupt  star-forming clouds before they evolve off the main sequence ($\approx 5\,{\rm Myr}$) to ionize the intergalactic medium (IGM). Rapidly rotating stars \citep{topping15}, stars stripped of their hydrogen envelope or blue stragglers in binary systems can provide extra photons on a longer timescale \citep[][]{stanway16,gotberg20,secunda20}, but the disruption timescale of the GMC required is still  short ($\la 10\,{\rm Myr}$) to  have high escape fractions and explain the reionization history of the Universe \citep[e.g.,][]{yoo20}. 

Previous radiation-hydrodynamics (RHD) simulations of reionization achieved this primarily by including strong supernova (SN) feedback \citep[e.g.,][]{kimm14,trebitsch17}. Although photoionization heating due to LyC radiation is an efficient mechanism to over-pressurize the star-forming clouds \citep{matzner02,krumholz07b}, the Stromgren sphere is usually under-resolved in galactic-scale RHD simulations. 
Consequently, star formation histories become more extended, and early SNe destroy star-forming clouds before young massive stars that produce the majority of ionizing radiation explode as SNe \citep{kimm14}. This SN-driven escaping picture may be partly correct in that energetic explosions are required to blow out the neighboring ISM \citep[e.g.,][]{semenov21}. However, at least in mini-halos of mass $M_{\rm halo} \la 10^8\,\msun$,  radiation feedback creates low-density channels on a timescale of a few million years,  allowing a large fraction ($\sim 40 \%$) of the LyC radiation to escape to the IGM \citep{wise14,kimm17,xu16}. A recent analysis of star-forming disc galaxies also appears to suggest that local GMCs may be dispersed on a similar timescale of $\sim5$ \,{\rm Myr}, once massive stars are formed \citep{chevance20,kimjy21}.

The most prominent uncertainty in the modeling of reionization originates from our lack of understanding of the earliest stage of the propagation of ionizing radiation in GMCs. Although it is becoming possible to measure the escape fraction of individual HII regions in the local galaxy \citep{pellegrini12,doran13,mcleod19,mcleod20,della-bruna21}, the observational properties of star-forming clouds at high redshift are barely known. Moreover, simulating the evolution of GMCs is a complex task, owing to the simultaneous operation of several physical processes. Magnetic fields are known to delay gas collapse \citep[e.g.,][]{hennebelle14,girichidis18}, and SNe and pre-SN feedback can occur simultaneously for long-lived clouds containing multiple massive stars \citep[e.g.,][]{lopez14}. Thus, high-resolution simulations of GMCs with a variety of initial conditions and comprehensive physical processes are very much needed.

With these motivations, \citet{kimm19} performed radiation-hydrodynamic simulations of GMCs with a maximum resolution of 0.25 pc, particularly focusing on the effects of the total star formation efficiency (\tSFE) and spectral energy distribution of stellar populations. They found that the escape fraction of LyC photons (\fescLyC) is generally an increasing function of time, and that \fescLyC\ is higher if  the binary interaction of stellar populations is  taken into account or if a higher \tSFE\ or lower metallicity is used \citep[see also][for galactic-scale experiments]{yoo20}. Independent work by \citet{kimjg19} further demonstrated that \fescLyC\ decreases with an increase in the gas surface density, again confirming the increasing trend of \fescLyC\ with time \citep[c.f.,][]{howard18}. Because the simulation duration as well as initial cloud conditions are different, direct comparisons between simulations are difficult, but the luminosity-weighted \fescLyC\ tends to be very significant on cloud scales ($\sim 50\%$), which appears to be compatible with recent analyses of local HII regions \citep[e.g.,][]{mcleod20,choi20,della-bruna21}.

Studies of the propagation of LyC radiation in GMCs are also useful to improve our understanding of the properties of the neutral medium in and around galaxies. LyC radiation that does not escape from a cloud ionizes hydrogen, which then recombines with free electrons and generates Lyman-$\alpha$ (\Lya) photons with an energy of 10.2 eV. Because of its high transition probability, \Lya\ emission is one of the brightest lines in the spectrum of star-forming galaxies \citep{partridge67}, which renders it a  useful tool for studying the high-$z$ Universe \citep[e.g.,][]{dijkstra14,ouchi20}.  A neutral medium with a hydrogen column density as low as $\log N_{\rm HI} \sim 10^{14}\,{\rm cm^{-2}}$ can scatter \Lya, allowing us to probe the distribution of  HI in the ISM and circumgalactic medium (CGM) \citep[e.g.,][]{gronke17,smith19,song20,mitchell21}.  Furthermore, the relative ratio of the blue part to the red part of the \Lya\ spectrum has been shown to be a good indicator of galactic inflows/outflows \citep[e.g.,][]{zheng02,ahn03,dijkstra06,verhamme06,laursen09,barnes11}. Additionally, the fraction of \Lya\ emitters or the \Lya\ equivalent width distribution can be used to infer the reionization history of the Universe \citep[e.g.,][]{stark11,pentericci11,jung20,garel21}.  

Of those observational findings, the detection of extended  \Lya\ halos \citep{cantalupo14,borisova16,wisotzki16,leclercq17} is particularly noteworthy. To explain the presence of the giant \Lya\ nebulae, \citet{cantalupo14} claimed that there may exist an unresolved neutral, clumpy CGM. Recent numerical studies have indeed pointed out that neutral clouds in galactic outflows would be found to survive longer if hydrodynamic instabilities are properly simulated on sub-parsec scales \citep{mccourt18,gronke18,sparre19}. Several authors  further showed that an enhanced resolution of several hundred parsecs in the CGM region can better capture the thermal balance and  form a large amount of neutral hydrogen \citep{vandeVoort19,hummels19,bennett20}, implying that \Lya\ may be scattered to a greater extent in the CGM compared with the extent indicated by current theoretical predictions  \citep[e.g.,][]{mitchell21}. The increased  column density of neutral hydrogen is also likely to affect the observed \fescLyC\ and the velocity separation of double peaks in the \Lya\ spectra ($v_{\rm sep}$) in different ways, potentially providing us a useful metric to study the distribution of the ISM/CGM. On this basis, \citet{kimm19} compared the value of \fescLyC--$v_{\rm sep}$ for simulated GMCs with those for luminous compact galaxy samples and concluded that the lower \vsep\ predicted for the simulated GMCs should be attributed to the lack of interaction with the ISM/CGM. On the other hand, in their idealized simulations that mimicked GMC environments, \citet{kakiichi21} showed that the observed \fescLyC--$v_{\rm sep}$ could be reproduced if turbulent structures were well maintained. Given that the simulation setups of the two studies are idealized, theoretical experiments modeling the emergence of \Lya\ in realistic settings are thus required  to examine if  \fescLyC--$v_{\rm sep}$ can be used to study the distribution of neutral hydrogen in the ISM/CGM and  to physically interpret the observed \Lya\ properties of galaxies.

Because  \Lya\ emission is mainly produced by recombination \citep[e.g.,][]{kimm19}, all processes that ionize hydrogen, such as photoionization or collisional shocks, should be included in simulations of \Lya\ radiative transfer. Early works that did not solve the fully coupled radiation-hydrodynamic equations assumed  \Lya\ sources with constant line broadening \citep{verhamme12} or computed the emission by post-processing outputs of hydrodynamic simulations \citep{yajima14,smith19,byrohl21}. The finite resolution of galaxy-scale simulations ($\sim 10$--$100$ pc) was also a limiting factor. However, the use of the zoom-in technique in cosmological simulations and  simulations of ISM patches are now making it possible to investigate the evolution of \Lya\ equivalent widths \citep{smith19} or  the effects of cosmic ray-driven winds on \Lya\ profiles at sub-parsec resolutions \citep{gronke18b}. Nevertheless, there is a lack of simulations modeling \Lya\ sources directly from GMCs or the ISM \citep[][]{kimm19,kakiichi21}.
In order to measure the escape of LyC radiation and to understand the emerging characteristic spectra of GMCs, we perform a suite of radiation-magneto-hydrodynamic (RMHD) simulations with a maximum of 0.02--0.08 pc resolution, including stellar feedback due to ionizing radiation and SN explosions. Improving upon the work of \citet{kimm19}, we model star formation by self-consistently computing the formation and accretion on sink particles and include the effects of magnetic fields. Apart from spherical clouds, we also simulate filamentary or homogeneous clouds to examine the effect of morphology on the emergence of LyC and \Lya\ photons.

This paper is organized as follows. We describe the initial conditions and physical processes included in numerical simulations in Section 2. A Monte Carlo radiative transfer method for \Lya\ and a simple ray tracing method for LyC radiation are also presented. In Section 3, we analyze the dependence of escape fractions and emergent \Lya\ profiles on GMC properties, and in Section 4, we discuss the main parameter controlling the escape of LyC radiation and  compare the escape of LyC and \Lya\ photons with that of UV radiation at 1500\AA\ and Balmer $\alpha$ (H$\alpha$) photons. Finally, we summarize our primary findings and present our conclusions in Section 5.

\section{Methodology}
In this section, we describe our suite of RMHD simulations in which the evolution of GMCs is modelled with various initial conditions. We also present radiative transfer methods used to compute the propagation of LyC and \Lya\ photons.

\subsection{Numerical simulations}
We perform 18 idealized simulations of GMCs\footnote{These simulations are part of the {\sc Pralines} project (Bieri et al. {\sl in prep.}).} with the adaptive mesh refinement code, {\sc Ramses} \citep{teyssier02}. The magneto-hydrodynamic equations are solved using the Harten-Lax-van Leer-Discontinuity  Riemann solver with a constrained transport scheme \citep{fromang06}. The Poisson equation is evolved with the particle-mesh method \citep{guillet11}, and radiation transport is modelled using a moment-based method with the M1 closure relation for the Eddington tensor \citep{rosdahl13}. We adopt a courant factor of 0.8.  Physical processes considered in the simulations are identical to those of \citet{geen18}, except that SN explosions are also included in the current study. Interested readers are referred to \citet{geen18} for details of the initial conditions, and we recapitulate the main features below.

\begin{figure}
    \centering
    \includegraphics[width=8.5cm]{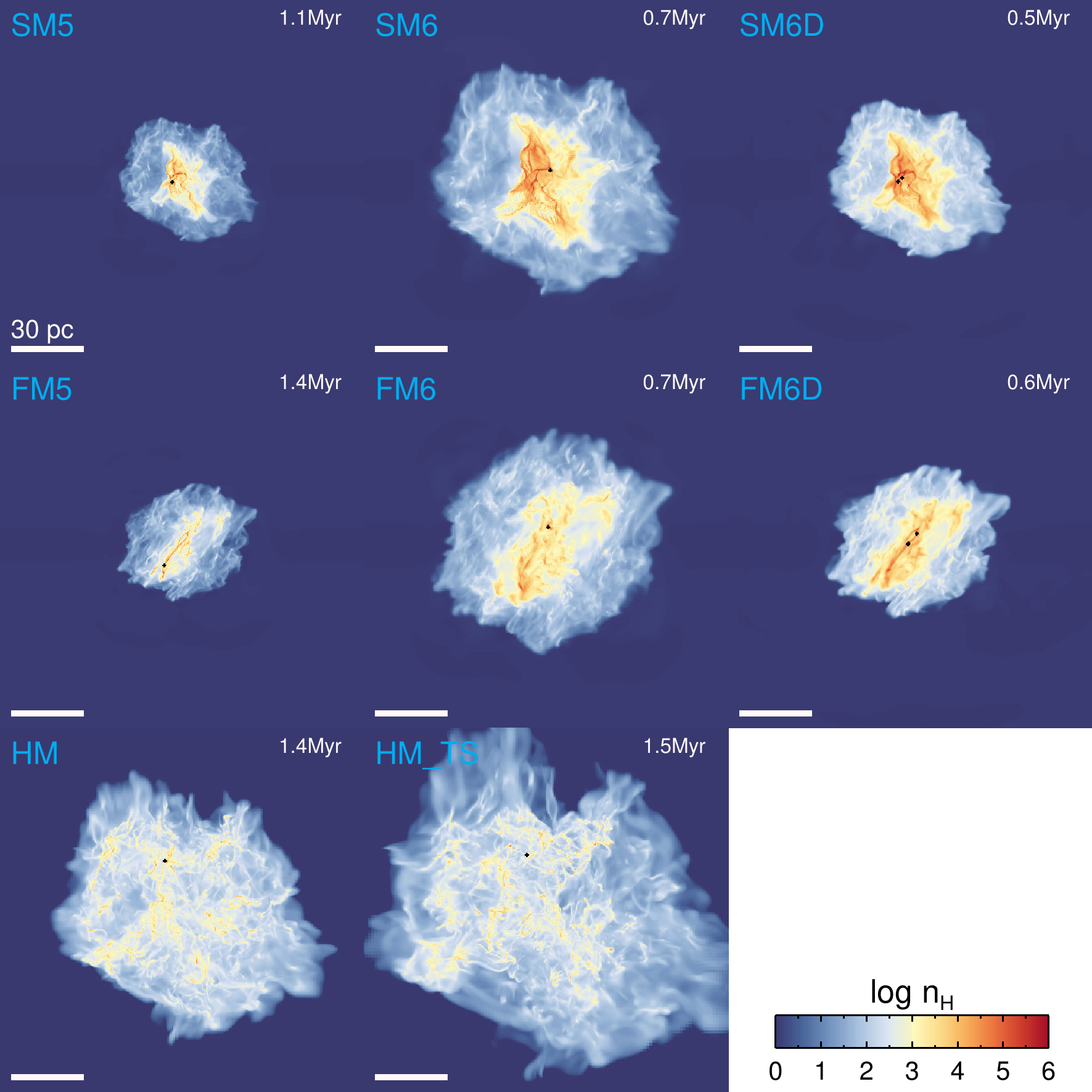}
    \caption{ Projected density distributions of the simulated clouds when the first star particle(s) are formed. The clouds have different morphologies (\texttt{S}: spherical, \texttt{F}: filamentary, \texttt{H}: homogeneous), masses (\texttt{M5}: $10^5$, \texttt{M6}: $10^6\,\msun$), surface densities (\texttt{D}: dense), and turbulence (\texttt{TS}: strong turbulence). The images measure 150 pc on a side, and the black dot indicates the position of the  star particles. }
    \label{fig:img_ic}
\end{figure}

\begin{figure}
    \centering
    \includegraphics[width=8.5cm]{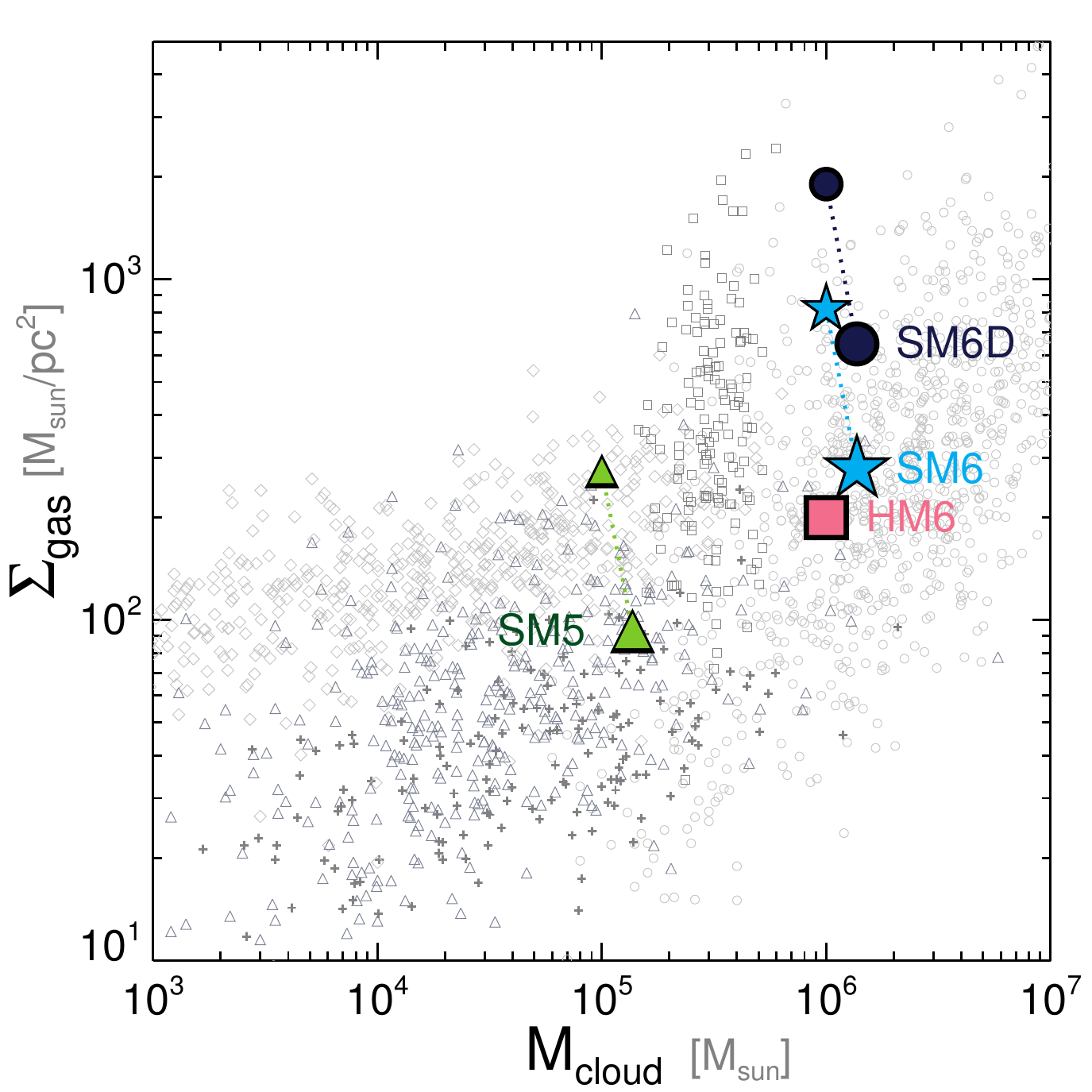}
    \caption{ Properties of the simulated GMCs. The cloud mass and surface density are displayed twice except for the homogeneous cloud (\texttt{HM6}), depending on whether the gas inside the inner (smaller) or outer layer (larger symbols) of the cloud is included. Filamentary clouds share the same initial conditions as spherical clouds (\texttt{SM}). The properties of the observed GMCs are shown as grey symbols: \citet[][crosses]{heyer09}, \citet[][diamonds]{roman-duval10}, \citet[][triangles]{wong11}, \citet[][circles]{colombo14}, \citet[][squares]{liu21} }
    \label{fig:cloud_prop}
\end{figure}

We simulate GMCs with different gas masses ($10^5$ [\texttt{M5}] or $10^6\,\msun$ [\texttt{M6}]), surface densities (typical versus dense [\texttt{D}]),  metallicities (low [\texttt{Z002}] or high [\texttt{Z014}]), resolutions (low [\texttt{LR}], high [\texttt{HR}]), input physics (\texttt{SN}), magnetic field strengths (\texttt{B}), turbulence strength (\texttt{T}), and morphology (spherical [\texttt{S}], filamentary [\texttt{F}], homogeneous [\texttt{H}]), as outlined in Table~\ref{tab:setting} and Figure~\ref{fig:img_ic}. The clouds initially comprise two layers of gas distributions with different densities, which are chosen by considering the total mass ($M_{\rm cl}=10^5$ or $10^6\,\msun$) and the typical average surface density of observed GMCs ($\Sigma\sim 100$--$200 \,\msun\,{\rm pc^{-2}}$) or the dense case (650$\,\msun\,{\rm pc^{-2}}$) (Figure~\ref{fig:cloud_prop}).   The region inside $r<20\,{\rm pc}$ ($10.8\,{\rm pc}$ for \texttt{M5}) is modelled as a Bonnor-Ebert sphere with scaling core radius of $R_c=6.6\,{\rm pc}$ (3.6 pc for \texttt{M5}), such that $\rho(r)\propto R_c^2/(R_c^2+r^2)$, which we refer to as an ``isothermal'' density profile.
The outer sphere has a constant hydrogen number density of $\nH=50\,\cmq$ up to 40 pc ($\nH=30\,\cmq$  up to 21.6 pc for \texttt{M5}). The background density is set to be $\nH\approx1\,\cmq$. Initial turbulent velocity fields with random phases are created using a Kolmogorov power spectrum, and we relax them for $0.5\,t_{\rm ff}$ without gravity, where $t_{\rm ff}$ is the free-fall timescale. Self-gravity is turned on after $0.5\, t_{\rm ff}$.  The initial conditions for filamentary clouds are generated by adopting identical velocity fields for the x- and y- directions. For  homogeneous clouds, we do not use double gas profiles, but simply adopt a sphere of uniform density $\nH=117\,\cmq$.

\begin{table*}
\small
    \centering
    \caption{Initial conditions and parameters used in the simulations. From left to right, the columns indicate (1) the simulation label, (2) cloud mass, (3) average gas surface density of the entire (inner) cloud, (4) radius of the entire (inner) cloud, (5) shape  (spherical (Sph), filamentary (Fil), and spherical with a uniform density (Hom)), (6) gas metallicity,  (7) size of the simulated box, (8) size of the finest AMR cells, (9) volume-weighted plasma beta ($\beta_P\equiv P_{\rm th}/P_{\rm mag}$) at $t_{\rm relax}$, (10) inclusion of Type II SN explosions, (11) one-dimensional velocity dispersion at $t_{\rm relax}$,  (12) virial parameter ($\alpha_{\rm vir,0}\equiv 5 \sigma_{1d,0}^2 R_{\rm cloud} / G M_{\rm cloud}$) of the entire (inner) cloud, and (13) time of the final snapshot. All simulations include the effect of photo-ionization heating. }
    \begin{tabular}{lccccccccccccc}
    \hline
    Name  & $M_{\rm cloud}$ &  $\Sigma_{\rm gas}$  & $r_{\rm cloud}$ & Shape  & $Z_{\rm gas}$   & $L_{\rm box}$  & $\Delta x_{\rm min}$ &  $\beta_P$ & SN & $\sigma_{\rm 1d,0}$ &  $\alpha_{\rm vir,0}$ & $t_{\rm final}$  \\
    &  [$\msun$] & [$M_\odot/{\rm pc^{2}}$] & [pc] & & & [pc] & [pc] & & & [\kms] &  & [Myr] \\    
(1) & (2)     & (3) &  (4)   &  (5) & (6) & (7)   & (8)     & (9) & (10)  & (11) & (12) & (13)               \\
       \hline
    
\texttt{SM5\_pZ002}             & $1.4\times 10^5$ & 93 (274) & 21.6 (10.8) & Sph  &  0.002 &  173 & 0.04 & 0.31 &  --  & 2.9 & 1.6 (1.1) & 8.4     \\
    
\texttt{SM5\_pZ014}              & $1.4\times10^5$ & 93 (274) & 21.6 (10.8) & Sph  &  0.014 &  173 & 0.04 &  0.31  &  -- &  2.9 & 1.6 (1.1) & 8.4   \\

\texttt{SM5\_pZ002\_BW}     & $1.4\times10^5$ & 93 (274) & 21.6 (10.8) & Sph  &  0.002 &  173 & 0.04 & 2.89  & --  & 2.9 & 1.6 (1.1) & 8.4      \\ 
    
\texttt{SM5\_pZ002\_BS}      & $1.4\times10^5$ & 93 (274) & 21.6 (10.8) & Sph  &  0.002 &  173 & 0.04 & 0.03  &-- & 2.9 & 1.6 (1.1) & 8.4     \\

\texttt{SM5\_pZ002\_HR}     & $1.4\times10^5$ & 93 (274) & 21.6 (10.8) & Sph  &  0.002 &  173 & 0.02 & 0.31   &--  &  2.9 & 1.6 (1.1) & 8.4     \\
    
\texttt{SM5\_pZ002\_LR}    & $1.4\times10^5$ & 93 (274) & 21.6 (10.8) & Sph  &  0.002 &  173 & 0.08 & 0.31   & --  &  2.9 & 1.6 (1.1) & 8.4  \\

\texttt{FM5\_pZ002}              & $1.4\times10^5$ & 93 (274) & 21.6 (10.8) & Fil &  0.002 &  173 & 0.04 & 0.31  & -- & 2.8 & 1.5 (1.0) & 8.4   \\
 
     \hline

\texttt{SM6\_pZ002}               & $1.4\times10^6$ & 278 (820)& 39.6 (19.8) & Sph &  0.002 &  317 & 0.08 & 0.11    & -- & 7.4 & 1.8 (1.2) & 8.3    \\

\texttt{SM6\_sZ002}       & $1.4\times10^6$ & 278 (820)& 39.6 (19.8) & Sph &  0.002 & 317 & 0.08 &  0.11   & \checkmark & 7.4 & 1.8 (1.2) & 8.3   \\
 
 \texttt{SM6\_sZ014}      & $1.4\times10^6$ & 278 (820)& 39.6 (19.8) & Sph &  0.014 &  317 & 0.08 & 0.11   & \checkmark & 7.4 & 1.8 (1.2) & 8.3    \\
 
\texttt{SM6\_sZ002\_HR} & $1.4\times10^6$ & 278 (820)& 39.6 (19.8) &  Sph &  0.002 &  317 & 0.04 & 0.11   & \checkmark & 7.4  & 1.8 (1.2) & 8.3  \\
  
\texttt{FM6\_sZ002}     & $1.4\times10^6$ & 278  (820)& 39.6 (19.8) &  Fil &  0.002 &  317 & 0.08 & 0.11   & \checkmark & 6.8 & 1.6 (1.1) & 8.3  \\

 \hline
 
 \texttt{SM6D\_pZ002}            & $1.4\times10^6$ & 647 (1905) & 26.0 (13.0) & Sph &  0.002 &  208 & 0.05 & 0.01   & --  & 7.4 & 1.2 (0.8) & 8.4   \\
 
 \texttt{SM6D\_pZ014}            & $1.4\times10^6$ & 647 (1905) & 26.0 (13.0) & Sph &  0.014 &  208 & 0.05 & 0.01   & -- & 7.4 & 1.2 (0.8) &8.3 \\
 
\texttt{SM6D\_sZ002}   & $1.4\times10^6$ & 647 (1905)& 26.0 (13.0) & Sph &  0.002 &  208 & 0.05 & 0.01   & \checkmark & 7.4 & 1.2 (0.8) &8.3   \\

\texttt{FM6D\_sZ002}   & $1.4\times10^6$ & 647 (1905) & 26.0 (13.0) &  Fil &  0.002 &  208 & 0.05 & 0.01 & \checkmark & 7.2 & 1.1 (0.8) & 8.3   \\
 
    \hline

\texttt{HM6\_sZ002}   & $10^6$ & 200 & 40 &  Hom &  0.002 &  320 & 0.08 & 0.24      & \checkmark &  5.5 & 1.4 &  8.3   \\ 
 
 \texttt{HM6\_sZ002\_TS} & $10^6$ & 200 & 40 & Hom &  0.002 & 320 & 0.08 &  0.11      & \checkmark & 10.0 & 4.6 & 8.3        \\
 \hline
    \end{tabular}

    \label{tab:setting}
\end{table*}

The simulated volume is covered with $128^3$ coarse cells (level 7), which are further refined up to level 11--13. The corresponding maximum resolution ranges from $\Delta x_{\rm min}=0.02$--$0.08$ pc (Table~\ref{tab:setting}). Note that the adopted  cell width is 3--10 times smaller than that used by \citet{kimm19}. Refinement continues until the maximum level is reached if the local Jeans length is resolved by fewer than 10 cells or the mass of a cell exceeds $0.27\,\msun$. Cells that contain a sink particle or are within the accretion radius of a sink particle ($4 \Delta x_{\rm min}$) are also maximally refined.

We model the formation of sink particles representing stars in dense regions where gas collapse occurs along the three axes, as described in \citet{bleuler14}. The clump finder first identifies dense clumps using a watershed algorithm, and places a sink particle of mass $0.001\,\msun$ in a virialized, dense cell  if $\nH\ge n_{\rm th,sink}$.  The critical density for sink formation is chosen as $n_{\rm th,sink}\ge 881\, \cmq / (\Delta x_{\rm min}/{\rm pc})^2$, motivated by the Jeans criterion. For example, $n_{\rm th,sink}$ in the fiducial \texttt{M6} clouds with $\Delta x_{\rm min}=0.08\,{\rm pc}$ is $1.47\times10^5\,\cmq$. Accretion onto the sink particle is modelled using a threshold such that 75\% of the mass above $n_{\rm th,sink}$ is transferred to the sink in each time step \citep{bleuler14}. Star particles are then created based on a pre-determined list of the masses of individual stars, which is randomly sampled assuming a Chabrier initial mass function \citep{chabrier03}. Note that we use the pre-determined list for reproducibility of the simulations and also to facilitate the comparison of simulations with different parameters. We assume that every time a sink particle accretes more than $120 \,\msun$, a massive star of mass between $8 \le m \le 120 \,\msun$ is formed and moves with the sink particle. The rest of the stellar mass is assumed to be non-radiating.

Radiative cooling due to primordial and metallic species is modelled on the basis of temperature, density, and metallicity. Metallicity-dependent gas cooling above $10^4\,{\rm K}$ is included by using piece-wise fits to \citet{sutherland93}. For metal cooling at lower temperatures, we consider fine structure transitions of CII and OI, as described in \citet{audit05}. 
By contrast, we solve the non-equilibrium chemistry for the primordial species (H, and He), which is coupled with the local radiation field produced by each star particle \citep{rosdahl13}, in order to accurately account for photoionization heating.  Stellar spectra are taken from the rotating stellar evolution model of the Geneva tracks \citep{ekstrom12,georgy13} and are used to calculate the time-dependent emissivity of ionizing radiation. In this study, we adopt three radiation bins ($13.6$--$24.59\, {\rm eV}$,   $24.59$--$54.42 \,{\rm eV}$, and $54.42\, {\rm eV}$--$\infty$).

When massive stars evolve off the main sequence and explode as SN, we inject a thermal energy of $10^{51}\,{\rm erg}$ into 27 neighboring cells of the sink particle. This method is often incapable of resolving the local cooling length and under-estimates the impact of explosions in dense environments. However, the inclusion of the radiation field and photoionization feedback decreases the density into which SNe explode \citep{krumholz07b}, and we verify that the local cooling radius is sufficiently resolved by more than three resolution elements by the time a SN explodes \citep{kim15}.

Magnetic fields are initially aligned in the $x$-direction, and then evolved following turbulent gas motions. The initial magnetic field strengths are set by considering the Alfv\'en crossing time $t_{A} \equiv r_c \sqrt{\rho_{\rm max,i}} / B_{\rm max,i}$, where $\rho_{\rm max,i}$ is the initial maximum density of the cloud and $B_{\rm max,i}$ is the initial maximum magnetic field strength. The volume-weighted magnetic field strengths of spherical clouds with low surface density (\texttt{SM5} and \texttt{SM6}) are $\approx7$ and $\approx12\mu G$, respectively, while for the \texttt{SM6D} models they are  $\approx45 \mu G$. A less massive cloud with weak (\texttt{BW}) or strong magnetization (\texttt{BS}) has a magnetic field strength of $\approx 2 \mu G$ or $ 23 \mu G$, respectively. Furthermore, filamentary clouds with the same mass and surface density are set to have nearly the same magnetic field strengths. \texttt{HM6} models with weak and strong turbulence (\texttt{HM6\_sZ002} and \texttt{HM6\_sZ002\_TS}) have a volume-weighted magnetic field strength of $\approx5\mu G$. The corresponding plasma beta ($\beta_P$), given by $P_{\rm th}/P_{\rm mag}$, is presented in Table~\ref{tab:setting}, where $P_{\rm th}$ and $P_{\rm mag}$ are the thermal and magnetic pressure, respectively. 

Simulations are run until $t\approx 8.4\,{\rm Myr}$, at which point the majority of gas is ionized and star formation is completed. Each run produces approximately $100$--$130$ snapshots at intervals of $\Delta t \la 0.1 \,{\rm Myr}$.

\begin{figure*}
    \centering
    \includegraphics[width=0.9\textwidth]{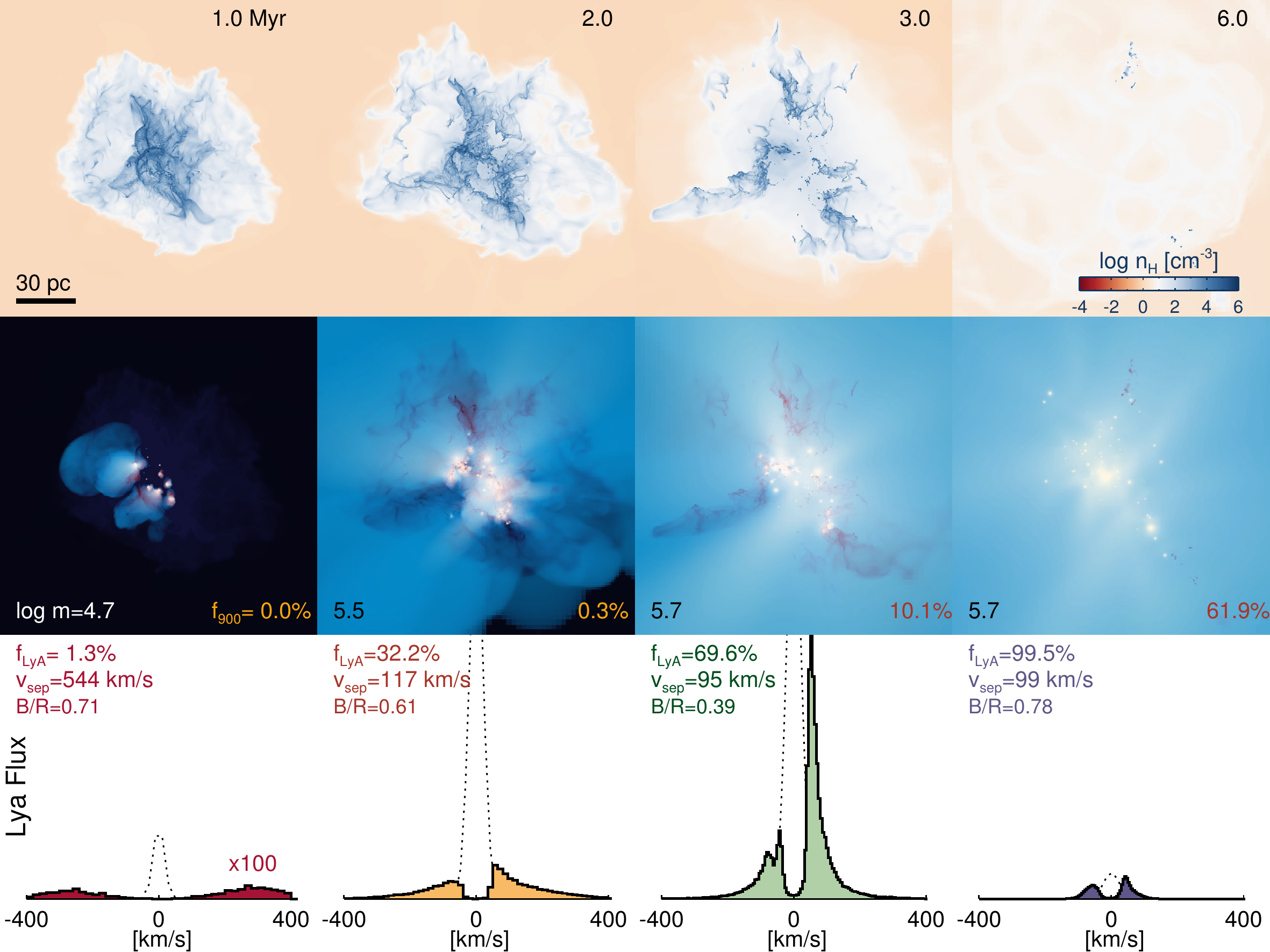}
    \caption{ Evolution of a GMC of mass $10^6\,\msun$ and metallicity of $0.002$ (\texttt{SM6\_sZ002}). The top panels depict projected distributions of the density, and the middle panels show composite images of the  gas density  (dark blue being  denser), the fraction of ionized hydrogen (blue means fully ionized), and HI-ionizing photon flux (brighter means more luminous) in the central region (158 pc on a side) of the simulation. The intrinsic (dotted) and emergent (solid) spectra of \Lya\ photons are shown in the bottom panels.  The emergent velocity profile of the leftmost column is magnified by a factor of 100 to improve readability. The simulation epoch is displayed in the upper right corner of the figures in the top row. The two values in the middle-row panels indicate the logarithmic total mass of stars and the escape fraction at $\lambda=900\,\AA$. In the bottom panels, we also include the escape fraction, the velocity separation of the two peaks,  and the flux ratio of the blue part to the red part of the \Lya\ spectrum.} 
    \label{fig:img1}
\end{figure*}

\subsection{Monte Carlo Ly$\alpha$ radiative transfer}
\label{sec:rascas}
To compute \Lya\ profiles, we post-process our simulation outputs with the Monte Carlo \Lya\ radiative transfer code {\sc Rascas} \citep{michel-dansac20}. We generate $10^5$ \Lya\ photon  packets in ionized regions via collisional excitation followed by de-excitation or recombinative radiation, as outlined in \citet{kimm19}.  The emission rates in each gas cell are calculated from the temperature, density, and  ionization fractions obtained from our RMHD simulations.  Note that LyC photons that escape from the simulated GMCs are not included as a source term in the calculation of \Lya\ radiative transfer. The initial frequency of the photons is determined from the gas motion, and we shift it by $x_{\rm ran} \Delta \nu$, where $\Delta \nu$ is the Doppler broadening due to the thermal motion and $x_{\rm ran}$ is a random number  drawn from the Gaussian distribution with the standard deviation of 1. \Lya\ photons are then resonantly scattered by neutral hydrogen and undergo diffusion in space and frequency. To reduce the computational cost, we use a core skipping algorithm \citep{smith15a}. We also include the recoil effect and transition due to a small amount of deuterium ($D/H=3\times10^{-5}$).

The most significant uncertainty in our modeling of \Lya\ radiative transfer comes from the determination of the amount of dust, for which we follow the method of \citet{laursen09}, as   
\begin{equation}
    n_{\rm dust} = \left( n_{\rm HI} + \fion \, n_{\rm HII} \right) Z/Z_{\rm ref}.
\end{equation}
Here $n_{\rm dust}$ is the pseudo number density of dust, $n_{\rm HI}$ and $n_{\rm HII}$ are the number density of neutral and ionized hydrogen, and \fion\ is a free parameter that controls the amount of dust in ionized media; we take $\fion=0.01$ \citep{laursen09}. $Z_{\rm ref}$ is a parameter that determines the overall dust mass relative to hydrogen. We take $Z_{\rm ref}=0.005$ with an albedo ($\mathcal{A}$) of 0.32 for low-metallicity runs, while  $Z_{\rm ref}=0.02$ and $\mathcal{A}=0.325$ are used for metal-rich cases \citep{weingartner01}. Dust absorption cross-sections for the low- and high-metallicity runs are taken from the dust models of the Small Magellanic Cloud (SMC) and Milky Way (MW) presented by \citet{weingartner01}, respectively.

\begin{figure}
    \centering
    \includegraphics[width=\linewidth]{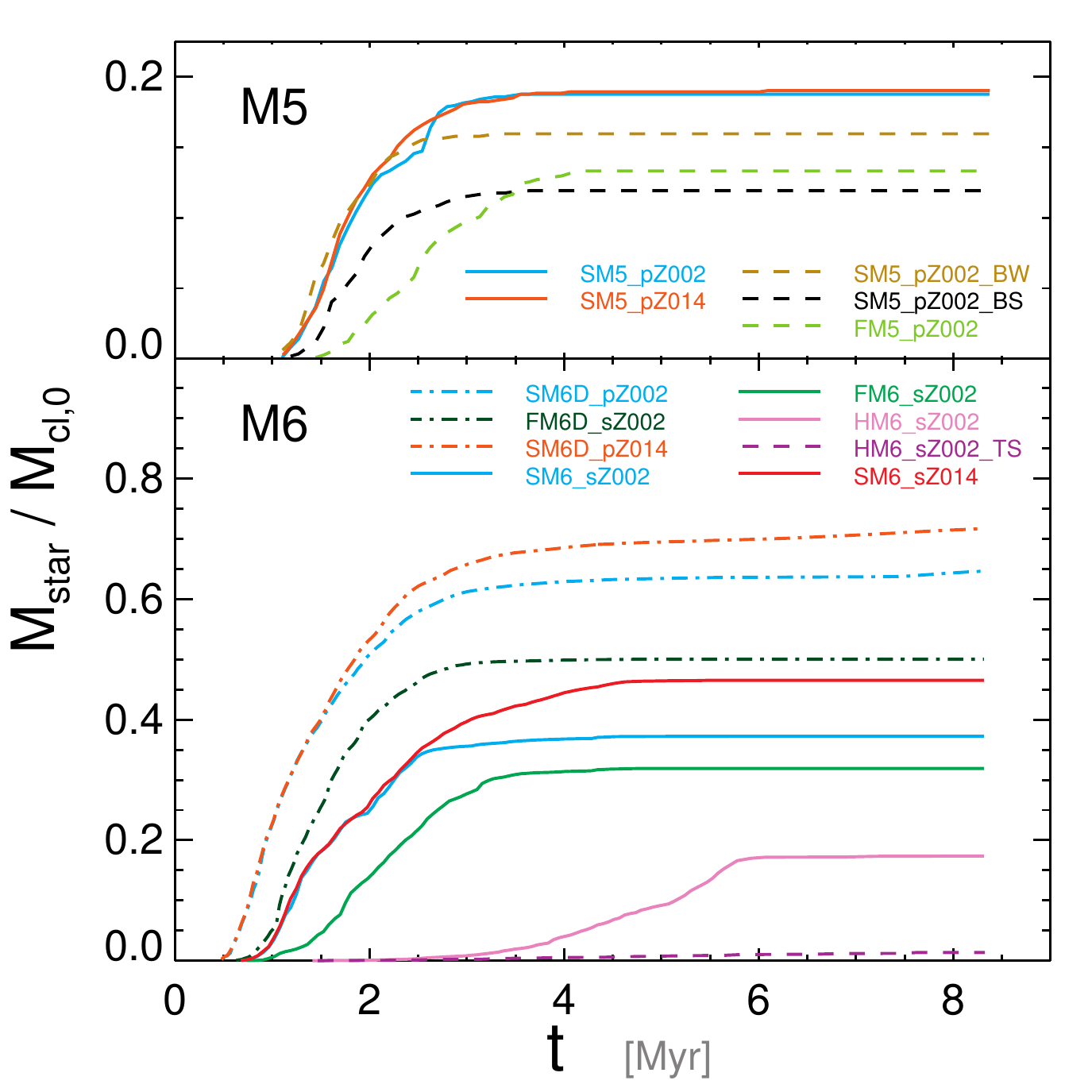}
    \caption{ Star formation histories of the simulated GMCs. Different color-codes and line styles correspond to runs with different initial conditions and input physics, as indicated in the legend. In most of the clouds, star formation ends within about $3\,{\rm Myr}$, except for the homogeneous spherical clouds in which gas collapses slowly because of its lower surface density. }
    \label{fig:SFH}
\end{figure}

\subsection{Measurement of LyC escape}
\label{sec:ray_tracing}
We measure the escape fraction of LyC radiation by computing the optical depth in 12,288 directions  from each star particle using the {\sc Healpix} algorithm with $N_{\rm side}=32$ \citep{gorski05}. The optical depth associated with primordial species is calculated by assuming that the ionizing radiation is absorbed by neutral hydrogen, singly ionized helium \citep{osterbrock06}, and neutral helium \citep{yan01}. Attenuation due to dust is modeled using the SMC-type absorption coefficient for the runs with $Z=0.002$ or MW-type coefficient for the runs with $Z=0.014$ \citep[][]{weingartner01}. We do not include absorption by molecular hydrogen as its formation and destruction are not modeled explicitly in this study. Moreover, neglecting  absorption by molecular hydrogen is unlikely to change the escape fraction of LyC photons significantly, as its optical depth is an order of magnitude smaller than that of neutral hydrogen  \citep[see Figure 5 of][]{kimm19}.

\section{Results}
In this section, we analyze the general evolutionary features of our fiducial cloud, present the systematic analysis on the escape of LyC and \Lya\ photons, and discuss the shape of \Lya\ spectra predicted from GMCs with different physical properties. 

\subsection{General evolutionary features}
Figure~\ref{fig:img1} (top row) shows the general evolution of the density distribution and emergent \Lya\ profiles of our fiducial cloud of mass $10^6\,\msun$ and metallicity of $Z=0.002$ (\texttt{SM6\_sZ002}). The initially driven turbulence in the GMC leads to the formation of filamentary structures, which then collapse to form sink particles (we re-iterate that these do not represent individual stars). Since there is no continuous energy input to act against gravity, gas accretes efficiently onto sink particles and a number of massive stars emerge within a short timescale of 1--2 Myr. By $t=3\,{\rm Myr}$, approximately 50 per cent of the cloud mass is converted into stars (Figure~\ref{fig:SFH}), and strong UV radiation photoheats and expels gas from the cloud,  as can be gleaned from the composite images of the density, the HII fraction, and HI-ionizing flux (middle row). 

As previously noted in the literature \citep{dale12,howard18,kimm19,he20,yoo20,kimjg21}, the escape of LyC photons is regulated by the evolution of the GMCs. In the early phase of star formation, no LyC photon escapes from the clouds. However, a significant fraction ($\ga 10\%$) of the ionizing radiation leaks from the cloud, once the cloud is disrupted because of  photoionization feedback. By $t=8.3\,{\rm Myr}$, 42.0 per cent of the ionizing radiation generated until then escapes from the fiducial cloud with a total \tSFE\ of 0.51. If we further count photons that would be produced during $8.31<t\le20\,{\rm Myr}$ and assume that the escape fraction is kept the same as the value at the last snapshot (i.e. $\fescLyC=64\%$), the luminosity-weighted escape fraction would increase only slightly  to $\left<f_{\rm LyC}\right>_{\rm 20\,Myr}=42.3\%$, as only a small fraction ($\approx 2\%$) of the total number of LyC photons is generated after $t=8.31\,{\rm Myr}$ in the stellar evolution model adopted in this study. 

Similarly, the escape fraction of \Lya\ photons ($\fescLya$) increases with the disruption of the cloud, but the escape is systemically more efficient than that of LyC radiation. This is not surprising, given that \Lya\ photons are not destroyed but scattered by neutral hydrogen. 
 We also note that the typical properties of \Lya-emitting gas change rapidly within the timescale of a few Myr.
During the early phases of cloud evolution, most of the \Lya\ photons are produced in dense ($\nH\approx10^{3.5-4.5}\,\cmq$) and warm ($T\approx 10^{4.1}\,{\rm K}$) gas.  By contrast, the density is reduced to $\nH\sim5$--$10\,\cmq$ when most of the GMC is dispersed. More interesting is the large number of scattering events which results in a significant velocity shift in the \Lya\ profile. As evident in the bottom panels of Figure~\ref{fig:img1}, the separation of the two velocity peaks ($v_{\rm sep}$) is as large as $\sim 500 \,\kms$ in the early phase. Once the dense clumps enshrouding young massive stars are ionized and destroyed, \Lya\ photons are scattered less and  $v_{\rm sep}$ decreases to $\approx 100\,\kms$. Expanding motion of the cloud rapidly develops because of photoionization feedback \citep[e.g.,][]{rosdahl15RT,geen17,kimjg18,grudic21a}, and back-scattered \Lya\ photons are preferentially observed \citep{verhamme06,dijkstra06}. Even when the cloud becomes optically thin to LyC radiation ($t=6\,{\rm Myr}$), \vsep\ does not fall below $\sim 100\,\kms$, as the \Lya\ photons are still scattered by the residual neutral hydrogen present in the expanding cloud.

In the simulation, starting from the first SN that  explodes at $t=4.5\,{\rm Myr}$, a total of 438 SNe explode,  blowing away the ambient medium. These explosions not only provide thermal energy but also re-distribute mass from stars to gas, sporadically enhancing the neutral fraction of hydrogen via radiative cooling. Moreover, our adopted stellar evolution model predicts a relatively short period ($\sim 5\,{\rm Myr})$ of bright phases in LyC because rejuvenation through binary interaction is neglected. Consequently, a small fraction ($\sim 0.1$--$1\%$) of neutral hydrogen persists in the dispersed bubbles and 40 per cent of LyC photons are still absorbed at $t\approx 8 \,{\rm Myr}$. We expect  the absorption by residual neutral hydrogen to be suppressed once a sufficient number of SNe explode and blow out most of the gas in the GMC.

We note that the simulated cloud shows the brightest \Lya\ emission in the intermediate stage, during which only a portion of the cloud is disrupted (third column  in Figure~\ref{fig:img1}).  Although \fescLya\ is close to unity in later stages, the recombinative emission channel is not efficient because the cloud is already dispersed and few LyC photons are converted into \Lya. However, if sufficient LyC photons continue to escape from the cloud, \Lya\ photons will be produced in the ISM, leading to a  significant change in the emergent \Lya\ spectrum, as discussed in \citet[][Figure 14]{kimm19}.

\begin{table*}
    \centering
    \caption{ Escape fractions at various wavelengths. From left to right, the columns indicate the name of the simulation, (1) simulation time in Myr, (2) \tSFE, (3) escape fraction of LyC radiation, (4) extrapolated escape fraction obtained by assuming that \fescLyC\ in the final snapshot is maintained constant until 20 Myr, (5) escape fraction of photons at [880, 912] \AA, (6) escape fraction of \Lya\ photons, (7) escape fraction of photons at [1475, 1525] \AA, (8) escape fraction of H$\alpha$ photons, (9) intrinsic UV magnitude at 1500 \AA, (10) intrinsic Ly$\alpha$ luminosity in units of $\rm erg\,s^{-1}$, (11) intrinsic H$\alpha$ luminosity in units of $\rm erg\,s^{-1}$, and (12) destruction timescale ($t_{\rm desc}$) in Myr. The escape fractions are luminosity-weighted over the entire simulation duration, whereas $\left<M_{1500}\right>$, $\left<L_{\rm Ly\alpha}\right>$, and $\left<f_{\rm H\alpha}\right>$ are the average during $t_{\rm desc}$ from the onset of star formation. The parameter $t_{\rm desc}$ is defined by the moment at which a gas mass denser than $100\,\cmq$ is reduced by 90\% relative to its initial value.}
    \begin{tabular}{lcccccccccccc}
    \hline
    Name  & $t_{\rm final}$ &  \tSFE\ & $\left<f_{\rm LyC}\right>$ & $\left<f_{\rm LyC}\right>_{\rm 20}$ & $\left<f_{\rm 900}\right>$ & $\left<f_{\rm Ly\alpha}\right>$ & $\left<f_{\rm 1500}\right>$ & $\left<f_{\rm H\alpha}\right>$ & $\left<M_{1500}\right>$ &  $\left<L_{\rm Ly\alpha}\right>$ &  $\left<L_{\rm H\alpha}\right>$ & $t_{\rm desc}$ \\
     & (1) & (2) & (3) & (4) & (5) & (6) & (7) & (8) & (9) & (10) & (11) & (12) \\
         \hline
\texttt{SM5\_pZ002     } &  8.4 & 0.188 & 0.690 & 0.690 & 0.629 & 0.711 & 0.946 & 0.833 & -10.9 &  39.6 &  38.6 & 2.7 \\
\texttt{SM5\_pZ014     } &  8.4 & 0.190 & 0.574 & 0.569 & 0.537 & 0.553 & 0.878 & 0.643 & -10.9 &  39.7 &  38.8 & 3.0 \\
\texttt{SM5\_pZ002\_BW } &  8.4 & 0.159 & 0.656 & 0.657 & 0.598 & 0.738 & 0.948 & 0.879 & -10.7 &  39.6 &  38.7 & 2.4 \\
\texttt{SM5\_pZ002\_BS } &  8.4 & 0.119 & 0.524 & 0.519 & 0.430 & 0.719 & 0.931 & 0.908 & -10.5 &  39.5 &  38.6 & 3.0 \\
\texttt{SM5\_pZ002\_HR} &  8.4 & 0.174 & 0.659 & 0.660 & 0.595 & 0.725 & 0.939 & 0.841 & -10.7 &  39.6 &  38.6 & 2.7 \\
\texttt{SM5\_pZ002\_LR} &  8.4 & 0.140 & 0.620 & 0.620 & 0.544 & 0.765 & 0.954 & 0.927 & -10.6 &  39.5 &  38.6 & 2.9 \\
\texttt{FM5\_pZ002    } &  8.4 & 0.133 & 0.635 & 0.641 & 0.545 & 0.844 & 0.957 & 0.939 & -10.3 &  39.4 &  38.5 & 2.9 \\
\hline
\texttt{SM6\_pZ002    } &  8.3 & 0.374 & 0.441 & 0.444 & 0.404 & 0.523 & 0.843 & 0.728 & -14.1 &  41.1 &  40.1 & 3.1 \\
\texttt{SM6\_sZ002    } &  8.3 & 0.373 & 0.420 & 0.423 & 0.378 & 0.542 & 0.843 & 0.740 & -14.1 &  41.1 &  40.1 & 3.1 \\
\texttt{SM6\_sZ014    } &  8.3 & 0.465 & 0.342 & 0.360 & 0.358 & 0.299 & 0.653 & 0.427 & -14.3 &  41.3 &  40.3 & 4.2 \\
\texttt{SM6\_sZ002\_HR} &  8.3 & 0.401 & 0.327 & 0.332 & 0.295 & 0.436 & 0.755 & 0.632 & -14.3 &  41.2 &  40.2 & 3.5 \\
\texttt{FM6\_sZ002    } &  8.3 & 0.319 & 0.466 & 0.468 & 0.399 & 0.556 & 0.866 & 0.751 & -13.8 &  41.0 &  40.0 & 3.6 \\
\hline
\texttt{SM6D\_pZ002   } &  8.4 & 0.647 & 0.192 & 0.192 & 0.165 & 0.360 & 0.563 & 0.623 & -14.7 &  41.4 &  40.4 & 3.0 \\
\texttt{SM6D\_pZ014   } &  8.3 & 0.717 & 0.258 & 0.260 & 0.255 & 0.218 & 0.514 & 0.468 & -14.6 &  41.4 &  40.5 & 3.0 \\
\texttt{SM6D\_sZ002   } &  8.3 & 0.631 & 0.302 & 0.311 & 0.207 & 0.354 & 0.597 & 0.584 & -14.7 &  41.4 &  40.4 & 3.5 \\
\texttt{FM6D\_sZ002   } &  8.3 & 0.500 & 0.363 & 0.370 & 0.328 & 0.450 & 0.793 & 0.718 & -14.4 &  41.3 &  40.3 & 2.9 \\
\hline
\texttt{HM6\_sZ002    } &  8.3 & 0.174 & 0.576 & 0.633 & 0.497 & 0.657 & 0.879 & 0.799 & -12.1 &  40.3 &  39.3 & 4.6 \\
\texttt{HM6\_sZ002\_TS} &  8.3 & 0.013 & 0.391 & 0.448 & 0.300 & 0.852 & 0.873 & 0.954 &  -9.6 &  39.3 &  38.3 & 4.5+ \\
   \hline
    \end{tabular}

    \label{tab:res}
\end{table*}

\subsection{Effects of physical properties on the escape of LyC and Ly$\alpha$ }

We now investigate how different physical properties of a cloud affect the propagation of LyC and \Lya\ photons. We first show the star formation histories  of our various clouds, which are essentially driven by free-fall collapse, in Figure~\ref{fig:SFH}. We then compare the escape fractions for different cloud masses, surface densities,  metallicities, magnetic fields, morphologies, and turbulent strengths in Figures~\ref{fig:fescLyC_all}--\ref{fig:fescLyA_all}. The effects of resolution are discussed in Appendix~\ref{appendix:resolution}.  The luminosity-weighted escape fractions are summarized in Table~\ref{tab:res} and shown in Figure~\ref{fig:fesc_summary} for a convenient overview.

\begin{figure*}
    \centering
    \includegraphics[width=0.95\textwidth]{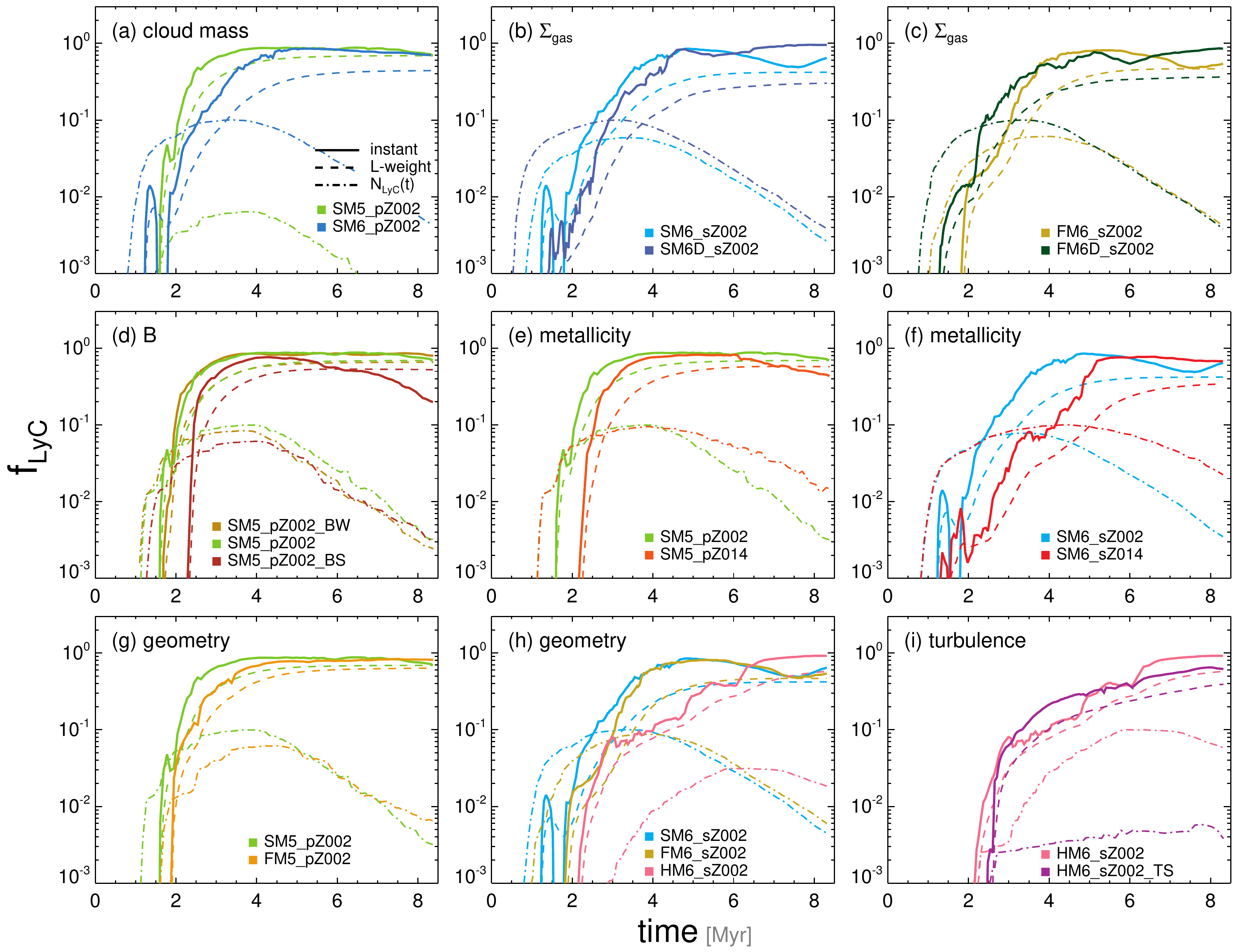}
    \caption{ Escape fraction of LyC photons (\fescLyC) in GMCs simulated with different input physics and parameters. The solid and dashed lines indicate instantaneous and luminosity-weighted escape fractions, respectively, and the  dot-dashed lines show the intrinsic number of LyC photons ($N_{\rm LyC}$). Note that the maximum $N_{\rm LyC}$ is normalized to 0.1 in each panel for better readability. The different panels compare the escape fraction from different runs, which are indicated in the legend. The escape fraction generally increases with time since the GMC gets disrupted.  }
    \label{fig:fescLyC_all}
\end{figure*}

\begin{figure*}
    \centering
    \includegraphics[width=0.95\textwidth]{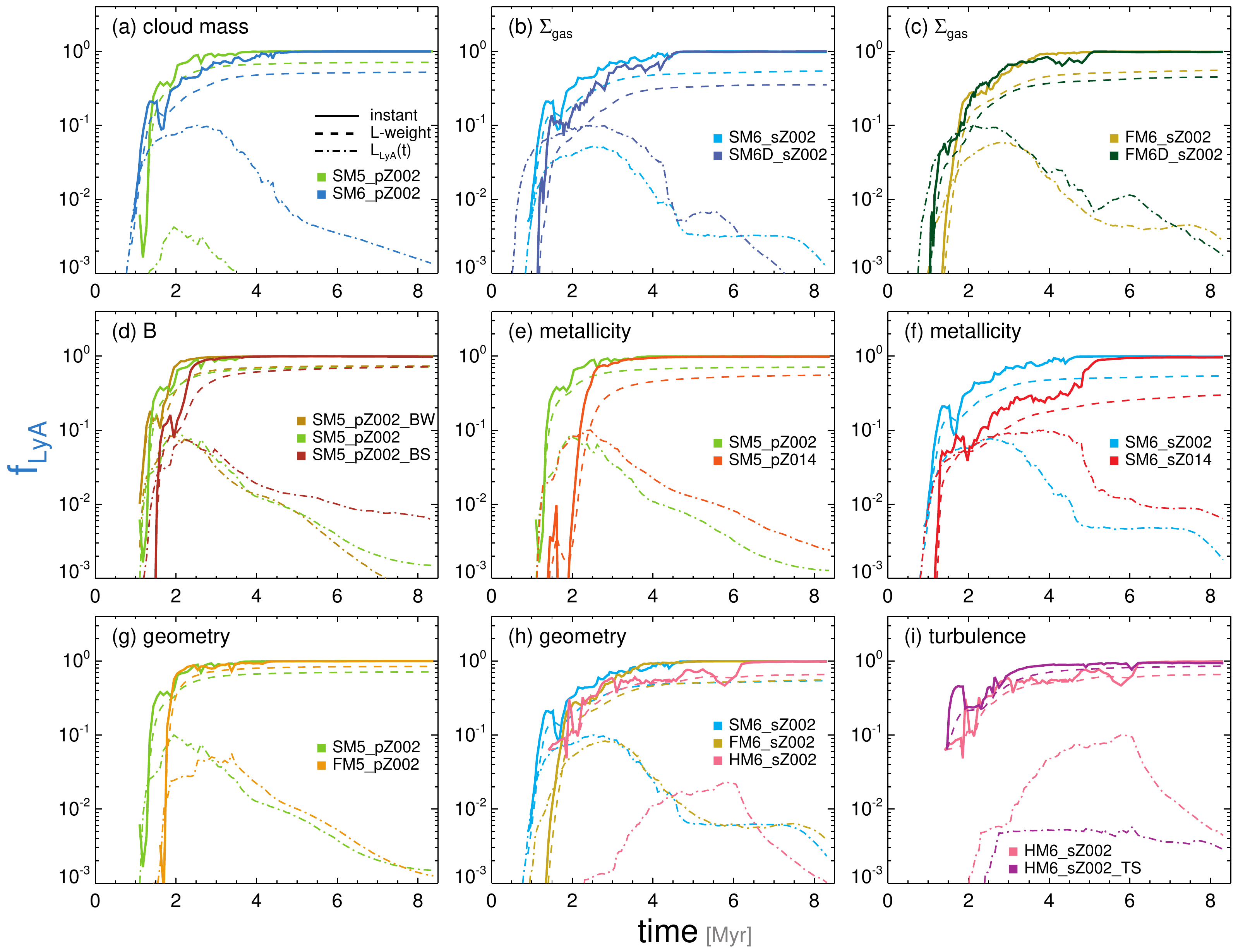}
    \caption{ Same as Figure~\ref{fig:fescLyC_all}, but for  \Lya\ photons. Intrinsic \Lya\ luminosities are computed from recombinative and collisional processes by taking the gas density and ionized fraction from each computational cell in the simulation. The maximum luminosity is normalized to 0.1 for better readability. Note that \Lya\ luminosities drop more rapidly than LyC luminosities (Figure~\ref{fig:fescLyC_all}), since recombinative radiation ($L \propto n_e n_{\rm HII}$) weakens as the GMCs get dispersed. }
    \label{fig:fescLyA_all}
\end{figure*}

\begin{figure*}
   \centering
   \includegraphics[width=\linewidth]{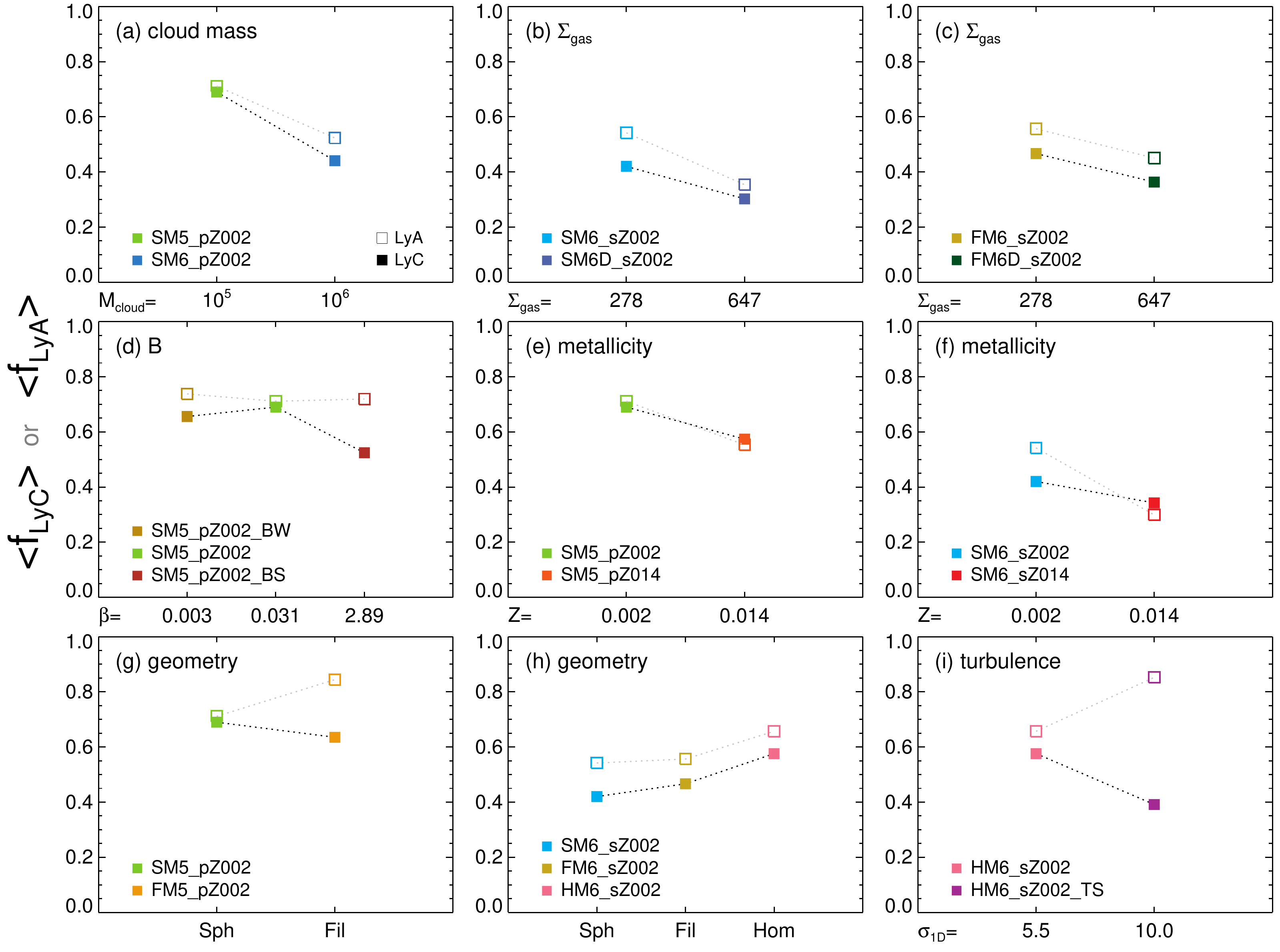} 
   \caption{Luminosity-weighted escape fractions of LyC (filled squares) and \Lya\ photons (empty squares). The escape fractions are averaged over the simulation duration ($\approx 8.4\,{\rm Myr}$). Different color-codes correspond to runs with various physical properties or resolutions, as indicated in the legend. }
   \label{fig:fesc_summary}
\end{figure*}

\subsubsection{Cloud mass}

Our simulated GMCs with a mass of $10^5\,\msun$ tend to have a relatively low \tSFE\ of $\approx 0.19$, independent of the metallicity (\texttt{SM5\_pZ002} or \texttt{SM5\_pZ014}). By contrast, more massive clouds (\texttt{SM6\_sZ002} and \texttt{SM6\_sZ014}) exhibit a higher \tSFE\ of $\approx 0.3$--$0.5$. The mass dependence partly arises because it takes more time to photoionize and destroy more massive clouds for a given \tSFE\ and density, since they are larger in size \citep[e.g.,][see their Appendix A]{kimm19}. The higher \tSFE\  also result from the higher gas surface density considered for massive clouds ($\Sigma=278$ versus $93\,\msun \,{\rm pc^{-2}}$); this is discussed in the following subsection (Sec.~\ref{sec:surface_density}). On the other hand, the inclusion of SN explosions does not change a \tSFE\ dramatically (\texttt{SM6\_pZ002}  versus \texttt{SM6\_sZ002}), as they occur after photoionization feedback destroys the cloud \citep[see also][]{grudic21a}. 

Despite the higher \tSFE, a smaller fraction of LyC and \Lya\ photons escape from massive GMCs. As shown in Figure~\ref{fig:fescLyC_all}-(a), the escape fractions reach almost unity in both the $10^5\,\msun$ and $10^6\,\msun$ clouds, but the rise in \fescLyC\ is much slower in the massive one because of the slow disruption. In total, our fiducial GMC (\texttt{SM6\_pZ002}) shows a luminosity-weighted escape fraction of $\fescLyCL=44.1\%$, whereas $69.0\%$ of LyC radiation leaks from the less massive cloud (\texttt{SM5\_pZ002}). Similarly, we obtain $\fescLyaL= 52.3\%$ and $71.1\%$ for these clouds, respectively, ignoring the  \Lya\ photons that would be potentially produced in the ISM by the leaking LyC radiation (Figure~\ref{fig:fescLyA_all}-(a)).

Our results are consistent with previous findings. 
 \citet{dale12} computed the escape fraction of LyC radiation from metal-rich clouds with various sizes and turbulence for 2--3$t_{\rm ff}$. Their run `D' was similar to our SM5 series in that the velocity dispersion was low and feedback was efficient. In their run, the resulting escape fraction of LyC radiation within $\approx 20\,{\rm Myr}$ was 70\%, similar to our predictions. In \citet{dale13}, the massive cloud of mass $10^6\,\msun$ (UZ) showed an escape fraction of 23\%. However, this measurement was performed within 3 Myr, which implies that the final escape fraction could be higher, consistent with our predictions. \citet{kimm19} and \citet{he20} also found that the escape fractions of LyC and \Lya\ photons were higher in less massive clouds. 

\subsubsection{Gas surface density}
\label{sec:surface_density}

Now we turn to the impact of the gas surface density on the escape of LyC and \Lya\ radiation. As shown in Figure~\ref{fig:cloud_prop}, our fiducial massive GMCs have a surface density ($\Sigma_{\rm gas}$) of $278\,\msun \, {\rm pc^{-2}}$, motivated by observations of GMCs in M51  \citep[e.g.,][]{colombo14}. However, GMCs with a higher surface density are also observed \citep[e.g.,][]{liu21,Dessauges-Zavadsky19} and hence we compare to simulations with a higher surface density of $\Sigma_{\rm gas}=647\,\msun \, {\rm pc^{-2}}$. 

We find that star formation commences earlier in denser GMCs (\texttt{SM6D\_sZ002} or \texttt{FM6D\_sZ002}) compared with GMCs with the fiducial surface density (Figure~\ref{fig:fescLyC_all}-(b) and (c)). While the denser clouds collapse more rapidly because of their shorter free-fall timescale, it takes longer for the LyC photons to escape from the cloud.  Since the cloud mass is set to be the same in the runs with different surface densities,  denser clouds are smaller in size, and the high density around young stars prevents the ionization front from expanding efficiently. As a result, despite extreme \tSFE\  of (0.631, 0.500), only (30.2\%,  36.3\%) of the LyC photons escape from the (\texttt{SM6D\_sZ002}, \texttt{FM6D\_sZ002}) runs, respectively. Such negative dependence on the gas surface density was also observed by \citet{kimjg19} from an analysis of RHD simulations of turbulent GMCs of mass $10^4$--$10^6\,\msun$ \citep[see also][for smaller and higher density clouds]{he20}. 

Similarly, the escape of \Lya\ photons is less efficient in dense clouds. While more than half of the \Lya\ photons leak from the clouds with the fiducial  $\Sigma_{\rm gas}$, \fescLyaL\ is reduced to $\approx 1/3$  in the dense cloud cases. Although not shown in Figures~\ref{fig:fescLyC_all} and \ref{fig:fescLyA_all}, we also compute the evolution of a dense, metal-rich cloud without SNe (\texttt{SM6D\_pZ014}), where the number of stars formed is the largest ($\tSFE=0.717$). In spite of the high \tSFE, the cloud shows the lowest \fescLyaL\ of 21.8\%, which is even smaller than that  \fescLyaL\ of the \texttt{SM6D\_sZ002} run by a factor of $\approx 1.5$.  These low \fescLyaL\ results are intriguing, given that observations of local starbursts sometimes reveal a high \fescLya\ exceeding $50\%$ \citep[e.g.,][]{verhamme17}. Although a direct inference should be drawn with caution, our experiments suggest that such a high \fescLya\ is better explained if metal-poor galaxies comprises  diffuse GMCs (e.g., $\Sigma_{\rm gas}= 278 \,\msun \, {\rm pc^{-2}}$), or if their typical GMC mass is low (e.g., $10^5\,\msun$).  

\subsubsection{Metallicity}


A comparison of the star formation histories  between the metal-poor (\texttt{Z002}) and metal-rich (\texttt{Z014}) runs shows that local dense clumps in our GMCs undergo free-fall collapse and form nearly the same number of stars at $t\la 2.5 \,{\rm Myr}$  regardless of gas metallicity (Figure~\ref{fig:SFH}). In particular, less massive clouds are dispersed on a comparable timescale, and thus the total number of stars formed by the end of the simulation ($t=8.4\,{\rm Myr}$) is also very similar in the \texttt{SM5\_pZ002} and \texttt{SM5\_pZ014} runs. Yet, a smaller fraction of LyC photons escapes from  \texttt{SM5\_pZ014} compared with the metal-poor cloud (57.4\% versus 69.0\%, Figure~\ref{fig:fescLyC_all}-(e)). This can be attributed to additional cooling due to metal agents, which mitigates the effect of photoionization heating and thereby results in a slower  propagation of ionization fronts \citep[][]{kimm19,he20,fukushima20}.  In the case of more massive clouds (\texttt{SM6}), fewer stars are formed between $2.5\la t<8.4\,{\rm Myr}$ in the metal-poor run (\texttt{SM6\_pZ002}) as radiation feedback starts to control gas collapse and suppress star formation in the late phases.  In spite of the lower \tSFE\ , the resulting \fescLyCL\ is still significantly higher in metal-poor cases (42.0\% versus 34.2\%). A similar conclusion was reached in a previous galactic-scale study of LyC escape using idealized disk simulations \citep{yoo20}.

The effect of metallicity is somewhat more pronounced in \Lya\  (Figure~\ref{fig:fescLyA_all}-(e) and (f)),  as \fescLya\ emerging from a given structure is higher than \fescLyC\ \citep{dijkstra16,kimm19}. In the metal-poor GMCs   (\texttt{SM5\_pZ002} and \texttt{SM6\_sZ002}),  \fescLyaL\ is indeed greater (71.1\% and 55.3\%) than \fescLyCL\ (69.0\% and 42.0\%). Interestingly, the metal-rich GMCs (\texttt{SM5\_pZ014} and \texttt{SM6\_sZ014}) exhibit an opposite trend, which can be attributed to different production mechanisms.  During LyC dark phases, \Lya\ photons are generated in the vicinity of young stars, but once the clouds are disrupted and fully ionized, only a few \Lya\ photons are produced per LyC photon (compare the dot-dashed lines in Figures~\ref{fig:fescLyC_all} and \ref{fig:fescLyA_all}). At this point, although \fescLya\ is close to unity, it does not contribute significantly to the total budget of \Lya\ photons and hence to the increase in \fescLyaL. As a result, \fescLyaL\ (55.5\% and 29.9\%) is slightly lower than \fescLyCL\  (57.4\% and 34.2\%) on GMC scales.  However, if we include the contribution from the \Lya\ photons that are produced by the escaping LyC radiation, \fescLyaL\ would become larger than \fescLyCL\ on galactic scales. 

\subsubsection{Magnetic field strength}

Figure~\ref{fig:fescLyC_all}-(c) shows the effect of magnetic fields in the cloud of mass $10^5\,\msun$. We compare three runs (\texttt{SM5\_pZ002\_BS}, \texttt{SM5\_pZ002}, and \texttt{SM5\_pZ002\_BW}) with different plasma beta parameter ($\beta_P=0.03$, $0.31$, and $2.89$, respectively), where $\beta_P$ is ratio of the thermal pressure  ($P_{\rm th}$) to the magnetic pressure ($P_{\rm mag}$). Note that the mass-weighted magnetic pressure in the fiducial case (\texttt{SM5\_pZ002}) is smaller than the turbulent pressure by an order of magnitude, whereas the two pressures are more comparable ($\approx$1:4) in the strongly magnetized GMC (\texttt{SM5\_pZ002\_BS}).

In the low-$\beta_P$ model, gas collapse occurs slowly because of the strong magnetic pressure acting against gravity \citep[e.g.,][]{hennebelle14,girichidis18}. Star formation is delayed by $\approx 0.3 \,{\rm Myr}$, and the number of stars formed is the smallest ($\tSFE=0.119$) among the three runs. The escape of ionizing photons is delayed even more significantly ($\approx 1\,{\rm Myr}$), compared with the cases with weaker magnetic fields. This is because LyC photons are enshrouded by a larger amount of neutral gas by the time  stars are formed and the feedback hence becomes less efficient at breaking out. For example, the luminosity-weighted \fescLyC\ of the stars younger than 0.1 Myr in \texttt{SM5\_pZ002} is 5.4\%, but in \texttt{SM5\_pZ002\_BS} it is only 2.5\%. We find that part of this difference is already established on small scales;  when the escape fraction is measured at 5 pc from each star, \fescLyCL\ is 14.4\% and 9.2\%, respectively.  Although not very significant in terms of photon budget, the escape of LyC photons in the late phase ($t\ga 5\,{\rm Myr}$) is also noticeably reduced in the low-$\beta_P$ case owing to the combined effects of a lower \tSFE\ and the higher density of gas remaining in the GMC. Consequently, the neutral fraction of hydrogen increases and a smaller fraction (52\%) of LyC photons leaks from their birth cloud (\texttt{SM5\_pZ002\_BS}) compared with the other runs (66--69\%).

Comparison of the runs with weak and fiducial magnetic field strengths reveals a more complex behavior. Until $t \approx 2.5\,{\rm Myr}$, star formation proceeds more efficiently in the run with a weaker B field (\texttt{SM5\_pZ002\_BW}) than the  \texttt{SM5\_pZ002} case. However, this trend is reversed at $t \ge 2.5\,{\rm Myr}$ and the  \texttt{SM5\_pZ002} run forms more stars since photoionization feedback comes into play earlier than in the other run. Although the differences are not dramatic, the run with a moderate $B$ field forms the largest number of stars and shows the highest escape fraction (69\%) among the three cases. Note that this is not the case for the escape of \Lya\ photons, owing to differences in the \Lya\ bright phase during cloud evolution.

Our results are consistent with those of RMHD simulations of turbulent clouds performed by \citet{kimjg21}.  By simulating five different turbulent seeds per cloud with different magnetic field strengths, the study concluded that strong magnetic fields tend to suppress star formation and the escape fraction. In their study, strongly magnetized clouds with a mass-to-magnetic flux ratio of $\mu_{B,0}=0.5$ exhibited escape fractions that are twice smaller than those in weakly magnetized cases ($\mu_{B,0}\ga 2$), where $\mu_{B,0}\equiv 2 \pi \sqrt{G} \Sigma_{\rm gas} / B_{0}$. Similarly, our \texttt{SM5\_pZ002\_BS} run also yields the smallest \fescLyCL.  However, the difference is less noticeable because the magnetization of our simulated clouds is weaker  ($\mu_{B,0} \approx$1.4, 2.3, and 14\footnote{The dimensionless mass-to-magnetic flux ratio is higher in the inner region of the cloud where the isothermal density profile is employed.}) than the strongest case in \citet{kimjg21}. Moreover, gravitationally, our simulated clouds are bound better than those of \citet{kimjg21}. The initial virial parameter of the \texttt{SM5} series is  $\alpha_{\rm vir,0} = 5 \sigma_{1d,0}^2 R_0 / G M_{\rm 0} \approx 1.1$ or $1.6$, depending on whether $R_0$ and $M_0$ are measured for the inner (10.8 pc) or all (20 pc) layers of the clouds, which is smaller than the value ($\alpha_{\rm vir,0} = 2$) adopted by \citet{kimjg21}. Thus, our simulations are likely to have probed the regime where the effects of magnetic fields are less significant.

\begin{figure*}
    \centering
    \includegraphics[width=\textwidth]{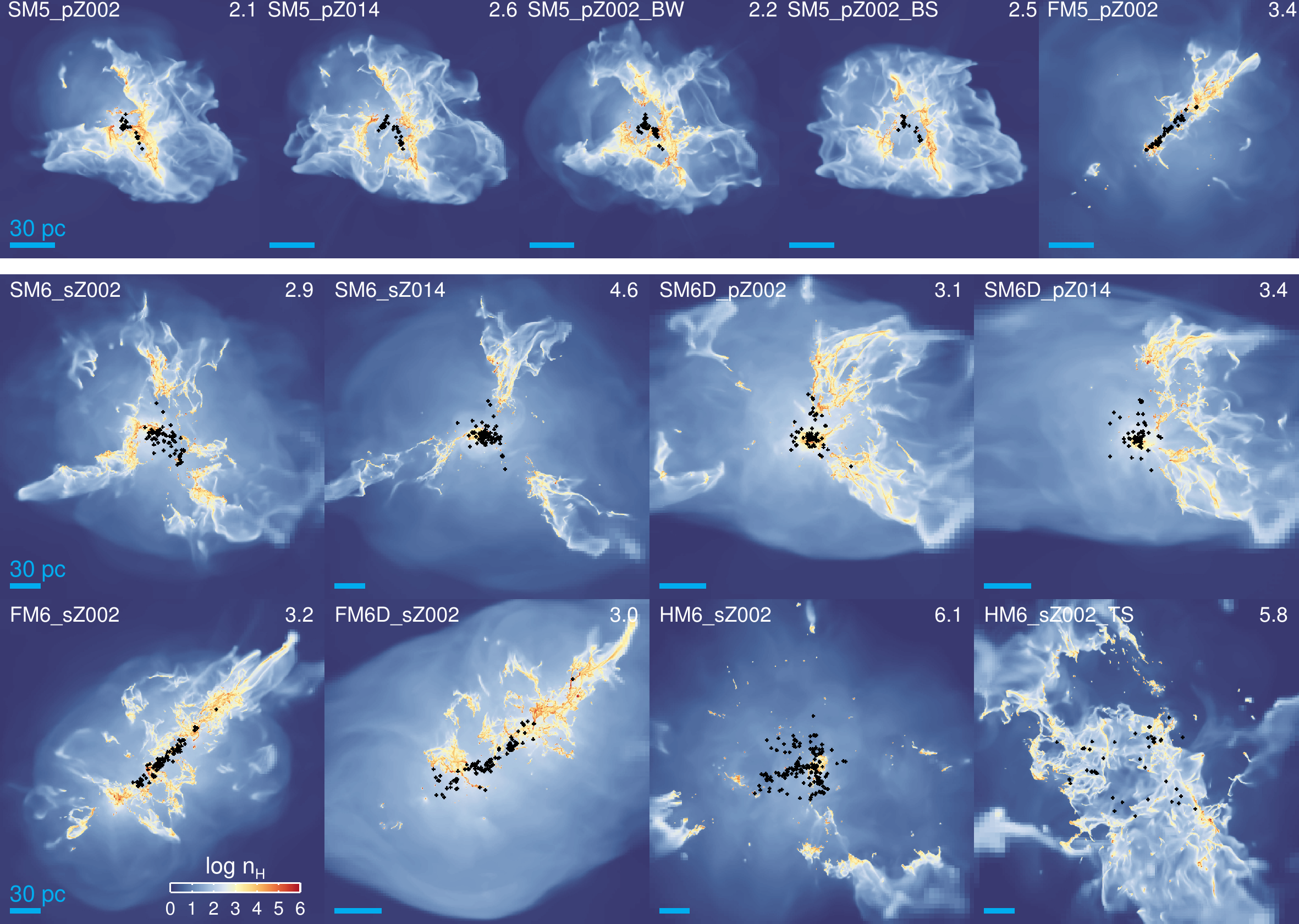}
    \caption{ Projected density distributions of GMCs when the escaping \Lya\ luminosity is the maximum. The time of each snapshot (in Myrs) is specified at the top right corner. Black dots denote stars with $>8\,\msun$, and the cyan bar measures 30 pc. The simulated clouds show the brightest  \Lya\ emission before they are completely dispersed. An exception is \texttt{HM6\_sZ002}, where gas collapses at the center contributes significantly to the total \Lya\ luminosity.   }
    \label{fig:img_all}
\end{figure*}

\subsubsection{Morphology}
While previous studies have focused on the escape of LyC and \Lya\ photons from spherical GMCs, local star-forming clouds often exhibit filamentary structures. Furthermore, GMCs do not necessarily have an isothermal density profile, which is imposed in our spherical cases. Therefore, we examine the effects of morphology on \fescLyC\ and  \fescLya\ in panels (g) and (h) in Figures~\ref{fig:fescLyC_all} and \ref{fig:fescLyA_all}.  Note that filamentary shapes in the \texttt{FM} series are generated using the same initial density distribution employed in the spherical cases (\texttt{SM} series), and thus, any change in the physical properties of star-forming clouds with filamentary structures can be attributed to the morphological difference. 

Unlike the spherical case where most of the star particles are formed in the central region, star formation occurs rather slowly along filamentary structures in the \texttt{FM} runs (Figure~\ref{fig:img_all}). Owing to the elongated geometry, young stars are less enshrouded by dense gas than in the spherical case, and therefore, they over-pressurize the neighboring region more easily. For example, we find that $3.3\times10^5\,\msun$ of stellar mass is needed to ionize 25\% of the total hydrogen in the \texttt{SM6\_sZ002} run, whereas roughly half ($1.8\times10^5\,\msun$) of this stellar mass is needed to achieve the same level of ionization in  \texttt{FM6\_sZ002}. The resulting \tSFE\ is decreased from 0.188 (\texttt{SM5\_pZ002}) to 0.133 (\texttt{FM5\_pZ002}) in the $10^5\,\msun$ GMCs, and from 0.465 (\texttt{SM6\_sZ002}) to 0.319 (\texttt{FM6\_sZ002}) in the massive GMCs. Naively, one may think that the filamentary structure would lead to a high escape fraction of ionizing photons, but no clear trend is found in \fescLyCL\ between the two geometries. While the filamentary cloud shows a higher \fescLyCL\ when it is massive, the opposite is true for the less massive cloud. This is likely because a small GMC is more susceptible to photoionization feedback, and because in a spherical cloud, the effects of a higher \tSFE\ dominate over the geometrical effect that allows for more efficient propagation along low-density channels. On the other hand, \fescLyaL\ is always larger in filamentary cases, indicating that resonant scattering is more sensitive to the cloud geometry. 

The most dramatic difference is seen in the run where the gas is homogeneously distributed (\texttt{HM6\_sZ002}). Because this homogeneous spherical cloud starts without any central concentration in density, stars are formed late ($t > 2\,{\rm Myr}$), with the peak being around $t=6\,{\rm Myr}$. \tSFE\ is also significantly lower (0.174) than the \texttt{SM6} runs, as photoionization feedback from the early populations effectively controls the gas collapse and star formation in neighboring regions  within the GMC. Despite the low \tSFE, their \fescLyCL\ is  higher (57.6\%) than that of the spherical GMC (\texttt{SM6\_sZ002}, 42.0\%), as the late-forming stars can take advantage of the low-density channels created by the early generations of stars. 

The physical properties of the \texttt{HM6\_sZ002} cloud are similar to those of the \texttt{M1E6R45} cloud by \citet{kimjg19}, and therefore, it is of interest to compare the results for the two runs. Although our virial parameter ($\alpha_{\rm vir,0}=1.4$) is slightly lower than theirs ($\alpha_{\rm vir,0}=2$), the mass and radius are nearly identical and both cases  assume a uniform density profile. The resulting \tSFE\ are comparable (0.174 versus 0.22), despite differences in the numerics and turbulent structures. \citet{kimjg19} found that within the first 3 Myr from the onset of star formation, \fescLyCL\ was 10\%,  which is very similar to our \fescLyCL\ if the escape fraction is integrated for the first 3 Myr, demonstrating overall consistency. 

\subsubsection{Turbulence}
As demonstrated by \citet{kimjg21},  the presence of strong turbulence regulates star formation and thereby alters the escape of LyC radiation \citep[see also ][]{safarzadeh16,kakiichi21}. In a similar spirit, we examine the effects of turbulence by imposing a turbulence that is approximately twice ($\sigma_{\rm 1d,0}=10.0\,\kms$)  stronger than \texttt{HM6\_sZ002}. The corresponding virial parameter of the  \texttt{HM6\_sZ002\_TS} run is $\alpha_{\rm vir,0}=4.6$. Note that we use the homogeneous cloud to facilitate comparison with other studies and also because its uniform density makes it easier to interpret the impact of turbulence compared with clouds with an isothermal density profile. 

Figures~\ref{fig:fescLyC_all}-(i) and \ref{fig:fescLyC_all}-(i) show the impact of turbulence. Because the cloud is no longer gravitationally well bound, star formation proceeds very slowly, and only a small amount of gas is turned into stars. Compared with the \texttt{HM6\_sZ002} run ($\tSFE=0.174$), the total \tSFE\ is reduced by an order of magnitude in the \texttt{HM6\_sZ002\_TS} run ($\tSFE=0.013$). However, intriguingly, a significant fraction of LyC photons escapes from the GMC (\fescLyCL=39.1\%). This fraction is smaller than that in the weaker turbulence case (\fescLyCL=57.6\%), but still substantial and comparable to those for spherical clouds with high \tSFE. The large \fescLyCL\ can be attributed to the extended spatial distribution of young stars that are less enshrouded by the GMC gas compared with stars formed in spherical clouds (Figure~\ref{fig:img_all}).  An opposite example is the \texttt{M6\_SFE1\_sng} run from \citet{kimm19}, where star particles with a mass of 1\% of the GMC were placed randomly in the central dense region of $r\la 5\,{\rm pc}$.  In this model, only 0.4\% of the LyC photons managed to escape from the GMC, which emphasizes the importance of the environment of star forming sites. 

Despite the low \tSFE, the escape of \Lya\ photons is the most efficient in the GMC with strong turbulence. A total of 85.2\% of \Lya\ photons escape from the \texttt{HM6\_sZ002\_TS} run, while a smaller fraction leaves the homogeneous cloud with weaker turbulence (65.7\%). We note that  \fescLyaL\ is  even larger than any other clouds with different geometries and high \tSFE. As evident in  Figure~\ref{fig:fescLyA_all}-(i), the \Lya\ escape fraction is high even before a significant number of stars form and disrupt the cloud ($t\sim2\,{\rm Myr}$), as the turbulent motions naturally generate low-density channels through which \Lya\ photons can propagate. Once the LyC radiation permeates the low-density regions, almost all  \Lya\ photons escape from the GMCs, even before SNe explode \citep[see also][]{kakiichi21}.

\begin{figure*}
    \centering
    \includegraphics[width=0.9\textwidth]{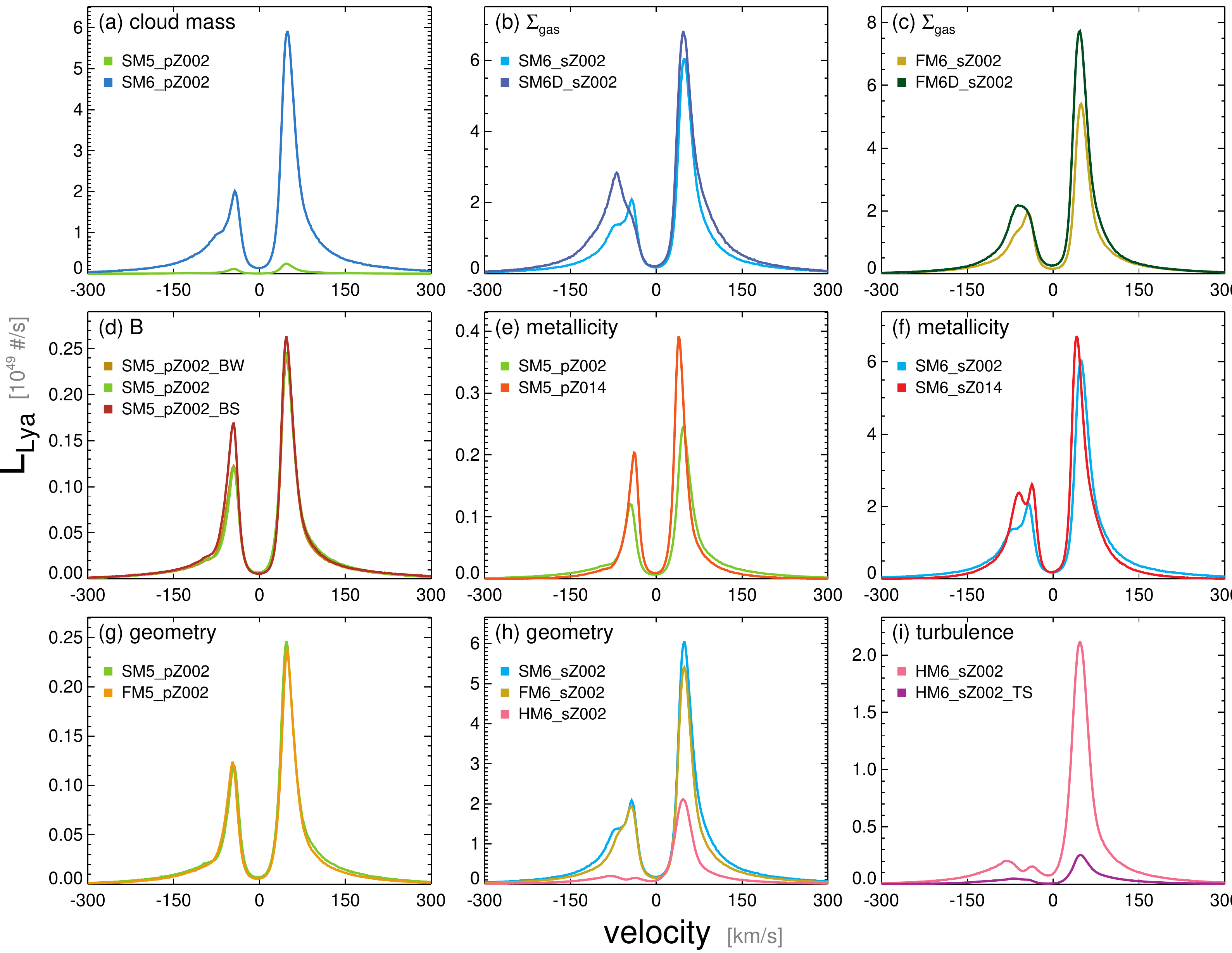}
    \caption{ Time-averaged velocity spectra of \Lya\ photons from simulated GMCs. The abscissa represents the angle-averaged escaping \Lya\ luminosity summed over each bin in units of $10^{49}\,{\rm \#}/s$, where ${\rm \#}$ is the number of \Lya\ photons. The bin size is $2 \,\kms$. Different color-codes correspond to results from simulations with different physical parameters or resolutions, as indicated in the legend.  }
    \label{fig:lya_prof_all}
\end{figure*}

\subsection{Properties of Ly$\alpha$ spectra}

As shown in Figure~\ref{fig:img1}, the shape of \Lya\ profiles changes dramatically over the lifetime of GMCs. In this subsection, we discuss our investigation of the dependence of \Lya\ shapes on cloud properties and our qualitative examination of line properties, such as the velocity separation of the double peaks ($v_{\rm sep}$) and the ratio of the blue part to the red part of the \Lya\ spectrum ($L_{\rm blue}/L_{\rm red}$).

\subsubsection{Similar $v_{\rm sep}$ in emergent \Lya}
In Figure~\ref{fig:lya_prof_all}, we present the luminosity-weighted \Lya\ profiles of the simulated GMCs by stacking the results between the first snapshot that shows star formation and the last snapshot of the simulation that the GMC is mostly dispersed. The \Lya\ lines show the well-known double-peak profiles, with a more pronounced spectrum redward of \Lya. The asymmetric feature is ubiquitously seen, indicating the presence of neutral outflows resulting from the disruption of clouds. The typical separation of the two velocity peaks is about $90$--$100\,\kms$, and it shows little dependence on the magnetic field strength (panel (d)) or morphology (panels (g) and (h)). On the other hand, GMCs with a higher surface density tend to exhibit a slightly larger $\vsep\approx120\,\kms$, owing to the larger optical depth (panels (b) and (c)). By contrast, more metal-rich runs show a slightly smaller $\vsep$ of $\approx 80\,\kms$, as \Lya\ photons are destroyed by dust before too many scattering events occur (panels (e) and (f)).
\begin{figure}
    \centering
    \includegraphics[width=0.5\textwidth]{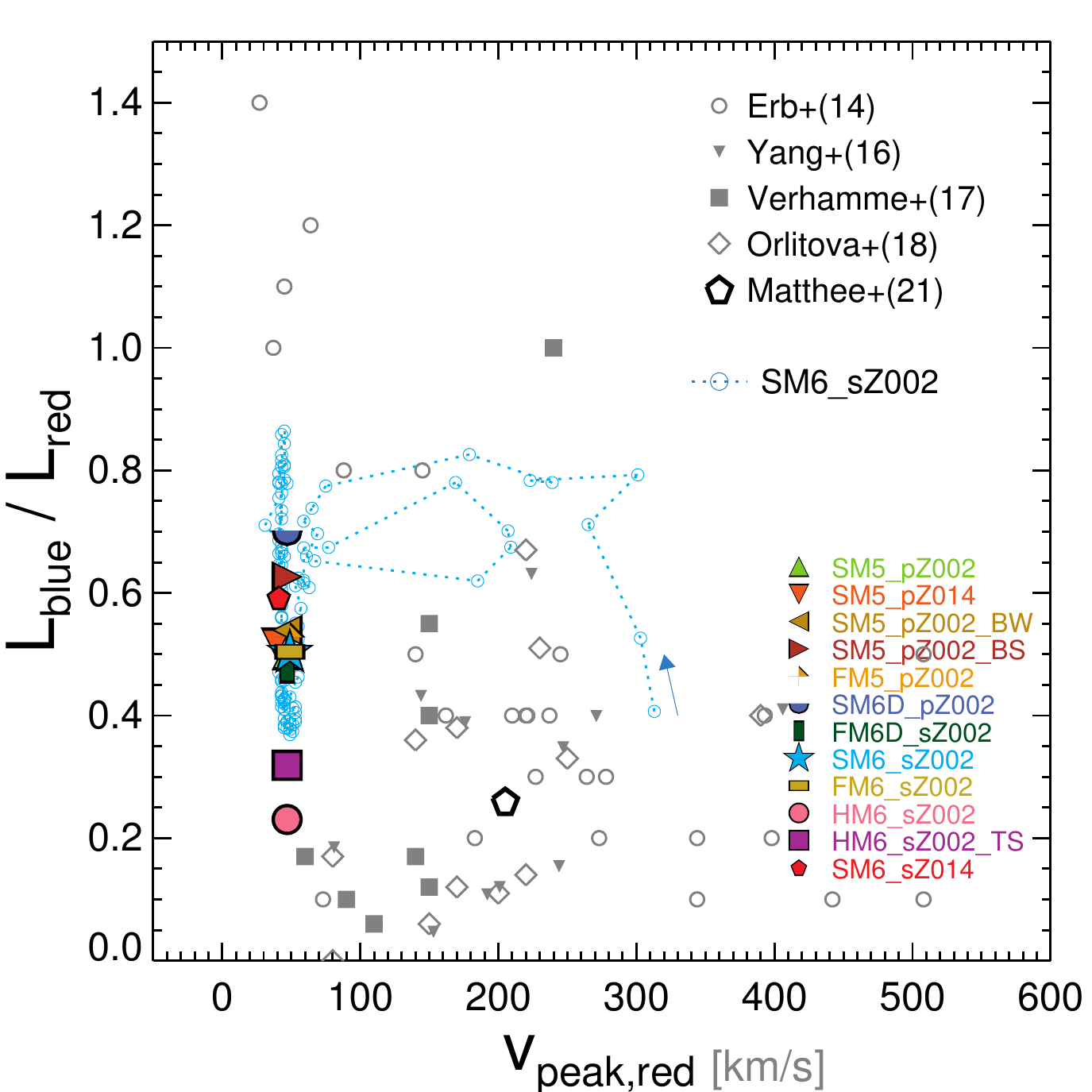}
    \caption{ Plot of the location of the red peak ($v_{\rm peak,red}$) versus the relative flux ratio between the blue part and red part of the \Lya\ spectrum ($L_{\rm blue}/L_{\rm red}$). Blue empty circles connected by dotted lines show an example of the time evolution of a metal-poor massive cloud (\texttt{SM6\_sZ002}).  Colored symbols indicate values obtained from the stacked \Lya\ spectrum in each simulation, and observational data are marked in gray color with various symbols, as indicated in the legend.  }
    \label{fig:Lratio}
\end{figure}

\begin{figure}
    \centering
    \includegraphics[width=0.5\textwidth]{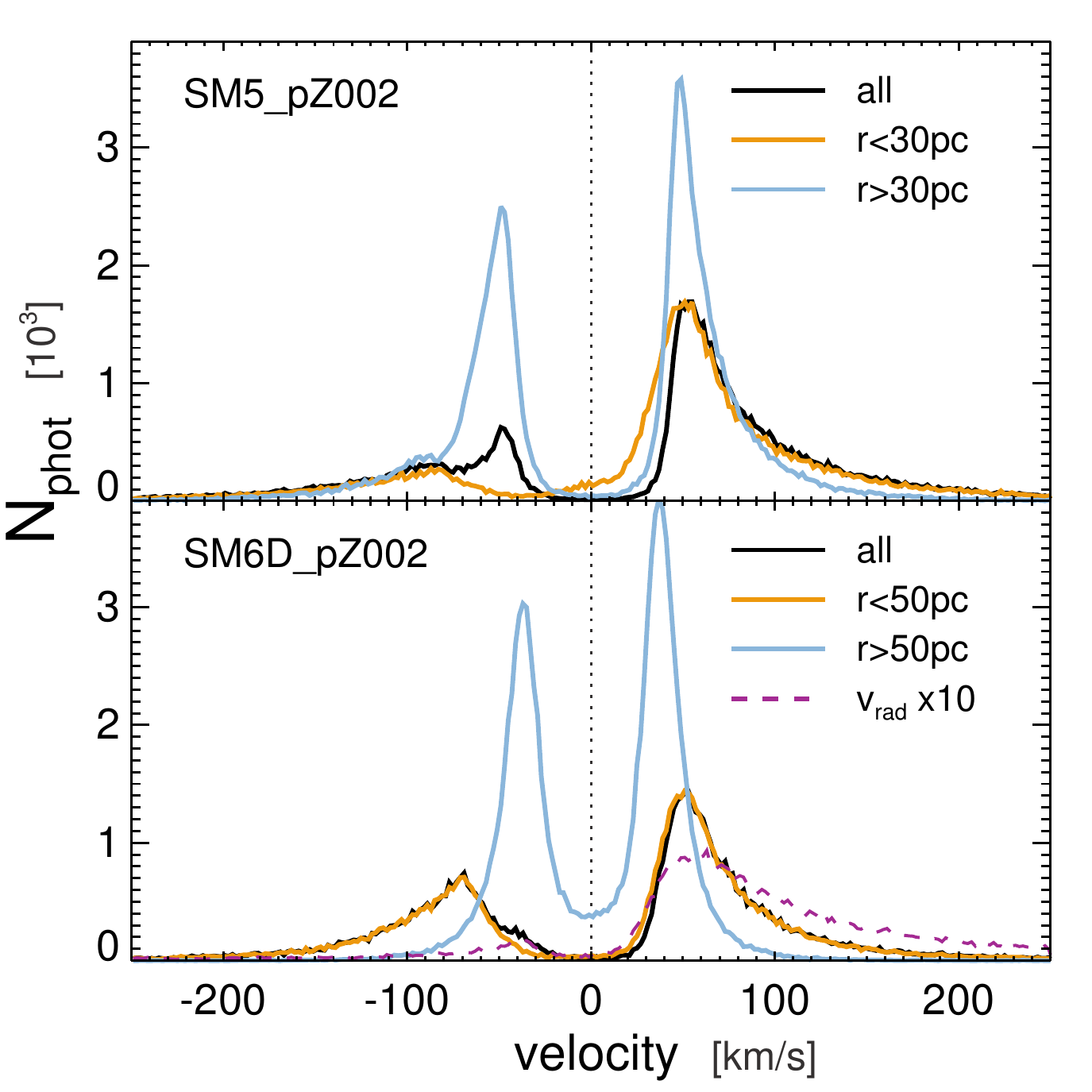}
    \caption{ Effects of scattering due to the inner or outer cloud gas in the simulations. The upper panel shows an example of \Lya\ spectra from the lower-mass metal-poor cloud (\texttt{SM5\_pZ002}), while the lower panel exhibits spectra from a massive cloud with high surface density (\texttt{SM5D\_Z002}) during the \Lya\ bright phase (i.e. the same snapshot as in Figure~\ref{fig:img_all}). Colored solid lines indicate the spectra in the absence of scattering due to the diffuse ISM region at $r>30$ or $50\,{\rm pc}$ (orange) or the dense GMC region at  $r<30$ or $50\,{\rm pc}$ (blue), where $r$ is the three dimensional radius from the \Lya\ luminosity center. The black lines correspond to the \Lya\ spectra processed by the entire gas in the simulated volume. The purple dashed line corresponds to the case in which the radial outflow velocity of the cloud is artificially augmented by a factor of 10.}
    \label{fig:vprof_ex}
\end{figure}

In order to qualitatively investigate the line features, we compute the location of the red peak by binning the stacked \Lya\ spectrum with $\Delta v=2\,\kms$ and show it as a function of the relative flux ratio between the blue part and red part of the spectrum ($L_{\rm blue}/L_{\rm red}$) in Figure~\ref{fig:Lratio}. The blue empty circles connected by dotted lines show an example of the time evolution of the metal-poor massive cloud of \texttt{SM6\_sZ002}.  Note that the earliest stages during which \Lya\ photons are fully absorbed by dust are not shown.  The evolutionary sequence begins around $L_{\rm blue}/L_{\rm red} \sim 0.4$--$0.8$ when  gas outflows start to develop locally. The velocity offset of the red peak is substantial ($v_{\rm peak,red}\sim 200$--$300\,\kms$) in this \Lya-faint phase, but it quickly decreases and settles at $v_{\rm peak,red}\approx 50\,\kms$. This happens mainly because the average column density of the volume-filling neutral hydrogen in later stages of the evolution is maintained around $N_{\rm HI}\sim 10^{18}\,{\rm cm^{-2}}$. As mentioned in the literature, the peak of the velocity offset is formed at $v_{\rm peak}\approx \pm 1.06 \left(a_{\rm V}  \tau_0 \right)^{1/3} \,b$ when photons are injected into the center of a homogeneous static slab \citep{neufeld90,dijkstra06,verhamme06}, where $a_{\rm V}=4.7\times10^{-4} \, T_4^{-1/2}$ is the Voigt parameter, $\tau_0=\sigma_0 N_{\rm HI}$ is the optical depth to \Lya\ at the line center, $T_4=T/10^4\,{\rm K}$ is the temperature in units of $10^4\,{\rm K}$, and $\sigma_0=5.88\times10^{-14}\,{\rm cm^2}\,T_4^{-1/2}$ is the cross-section. Here $b=\sqrt{v_{\rm th}^2 + \sigma_{\rm turb}^2}$ is the Doppler parameter, where $v_{\rm th}=12.9\,\kms\,T_4^{1/2}$ is the thermal velocity and $\sigma_{\rm turb}$ is the turbulent velocity. The resulting $\vsep$ may be written as
\begin{equation}
v_{\rm sep} = 82.7 \,{\rm km\,s^{-1}} \sqrt{1+\mathcal{M}^2}\, \left(\frac{N_{\rm HI}}{10^{18}\,{\rm cm^{-2}}}\right)^{1/3} \left(\frac{T}{10^4\,{\rm K}}\right)^{1/6}
\label{eq:vsep}
\end{equation}
where $\mathcal{M}\equiv \sigma_{\rm turb}/v_{\rm th}$ is the turbulent Mach number. 
Thus, \Lya\ photons that are scattered in a medium with $T=20,000\,{\rm K}$ and $\sigma_{\rm turb}\approx 0- 10\,\kms$ would easily produce $v_{\rm peak,red}\approx 50 \,\kms$, as evident in Figure~\ref{fig:Lratio}.

Intriguingly, $v_{\rm peak,red}$ from the stacked \Lya\ spectrum (star symbols in Figure~\ref{fig:Lratio}) is very similar ($v_{\rm peak,red}\approx 50\,\kms$), regardless of the physical properties of the clouds. This is unexpected, given that the average hydrogen column density varies by an order of magnitude from $\left< \log N_{\rm H} \right> \approx 22.3$ in \texttt{SM5\_pZ002} to $\approx 23.2$  in \texttt{SM6D\_pZ002}. Here, we measure $\left< \log N_{\rm H}\right>$ from  the simulation center in the radial direction using the Healpix algorithm along 196,608 sight-lines \citep{gorski05}. However, the difference in the column density of {\em neutral} hydrogen is less dramatic as \tSFE\ tends to be higher by a factor of 4--5 in the runs with larger $\left< \log N_{\rm H} \right>$ and therefore more ionizing photons are available per unit gas mass.

To corroborate this, we present in Figure~\ref{fig:vprof_ex} example spectra from the \texttt{SM5\_pZ002} and \texttt{SM6D\_pZ002} runs  when the largest number of \Lya\ photons escape from the GMC ($t\approx 3\,{\rm Myr}$). The gas distributions at the same epoch are shown in Figure~\ref{fig:img_all}. Note that during this \Lya\ bright phase, there exist dense filaments that continue to form stars, but a large fraction of the GMC is covered with low-density ionized hydrogen. Figure~\ref{fig:vprof_ex} shows that $v_{\rm red,peak}$ is primarily determined by the cloud gas located within $r \la 30$ (top) or $50\,{\rm pc}$ (bottom), where the $N_{\rm HI}$ distributions may be characterized by double log-normal profiles with one centered on  $\sim 10^{18}\,{\rm cm^{-2}}$ and the other on $\sim 10^{22}\,{\rm cm^{-2}}$. The covering fraction of the latter component ($\log N_{\rm HI}>19$) is large ($\sim 30$--$50\%$) in these snapshots, but insufficient to trap \Lya\ photons. Thus, \Lya\ could easily travel away from the dense regions and be scattered in the low density medium at $r \la 30$ or $50\,{\rm pc}$, forming the red peak at $\approx 50\,\kms$.  Likewise, the velocity offset of the blue peak is mostly determined by the gas at $r \la 30$ or $50\,{\rm pc}$, but it appears further away from the line center  ($v_{\rm peak,blue} \approx -70$--$80\,\kms$), similar to the feature associated with zero back-scattering in expanding shells \citep[e.g., Figure 12 in][]{verhamme06}.  
The gas outside the GMC ($r>30$ or $50$ pc) then re-distributes the frequency of any residual photons with $|v| \la 30 \,\kms$. For example, scattering in the inner region gives rise to a blue peak  at $v\approx -80\,\kms$ in the \texttt{SM5\_pZ002} run, but more \Lya\ photons pile up at $v\approx -50\,\kms$ after being scattered by the background medium.  This effect is not very prominent in the \texttt{SM6D\_Z002} run, as there are fewer \Lya\ photons close to the line center after the interaction with the GMC. The exact location of the new blue peak found in the  \texttt{SM5\_pZ002} run should depend on the assumption about the background ISM, but it is clear from Figure~\ref{fig:vprof_ex} that a similar \vsep\ originates from the scattering with neutral hydrogen in the GMCs.

\begin{figure*}
    \centering
    \includegraphics[width=0.45\linewidth]{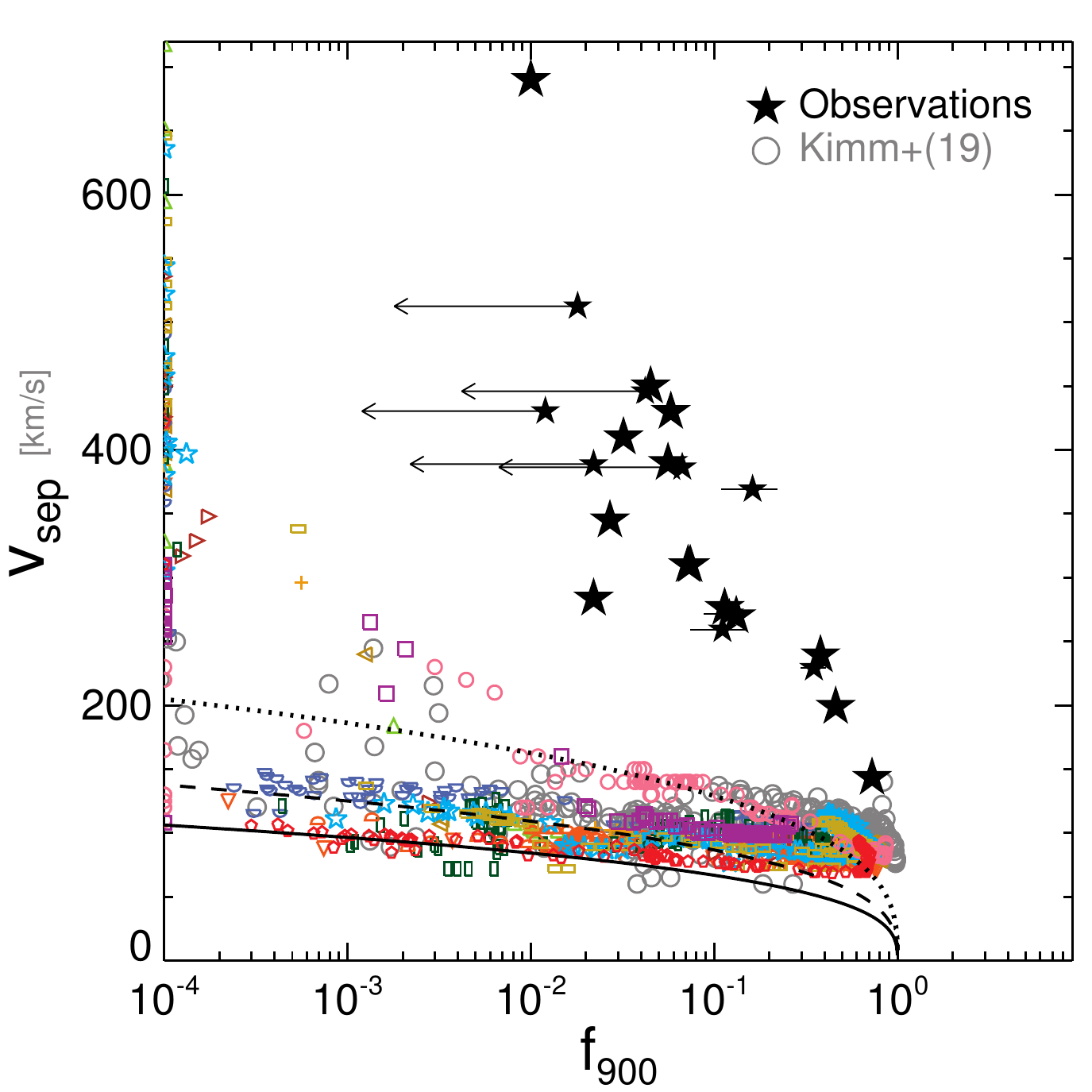}
    \includegraphics[width=0.45\linewidth]{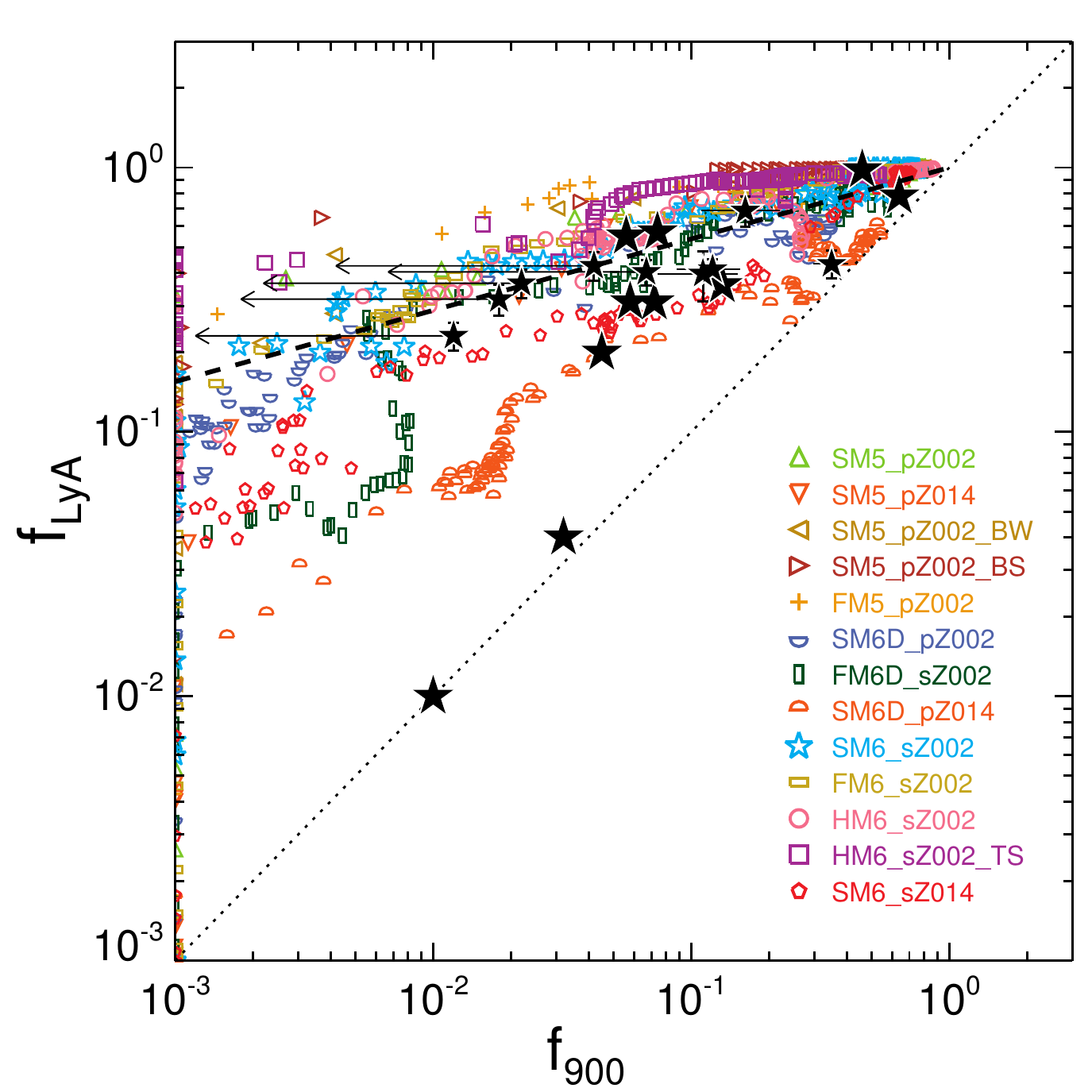}
    \caption{Relationship between the velocity separation of the \Lya\ double peak (\vsep) and the escape fraction at $\lambda\approx 900\,\AA$ ($f_{900}$, left panel). Results for our simulated GMCs are shown as colored symbols, as indicated in the legend of the right panel. The gray empty circles indicate earlier results obtained from RHD simulations of GMCs \citep{kimm19}. The black star symbols correspond to {\em galactic-scale} measurements taken from \citet{vanzella15,izotov16a,izotov16b,verhamme17,vanzella18,izotov18}. The smaller star symbol with error bars indicates the results obtained from a low-mass galaxy sample by \citet{izotov21}. The black lines show simple analytic estimates obtained by assuming $T=20,000\,{\rm K}$ for various turbulent velocities (solid, dashed, and dotted lines: $\sigma=0$, $15$, and $30\,\kms$, respectively). The right panel shows the comparison between $f_{900}$ and \fescLya\ on cloud scales.  The dashed line indicates a simple fit to our results, $\fescLya\approx f_{900}^{0.27}$, at $f_{900}>10^{-3}$.}
    \label{fig:vsep}
\end{figure*}

\subsubsection{ Flux ratio}

Figure~\ref{fig:Lratio} shows that the typical flux ratio between the blue part and red part of the \Lya\ profiles is less than unity ($L_{\rm blue}/L_{\rm red}\approx 0.4$--$0.6$). As shown in Figure~\ref{fig:fescLyA_all}, simulated clouds are bright in \Lya\ at $t\sim2$--$4\,{\rm Myr}$ during which roughly 50--70\% of the initial neutral hydrogen is ionized. Because outflows of the cloud gas are driven by photoionization heating, they are generally slow ($\sim 10\,\kms$), and \Lya\ properties of the simulated clouds occupy a narrower region in the $v_{\rm peak,red}$--$L_{\rm blue}/L_{\rm red}$ plane, compared with the observations of actively star-forming galaxies \citep{erb14,yang16,verhamme17,orlitova18}. 

The relative flux ratio can be further reduced if the outflow velocities are increased. For illustrative purposes, we present a case in which the radial velocity of each outflowing gas cell is artificially augmented by a factor of 10 in the \texttt{SM6D\_pZ002} run (purple dashed line in Figure~\ref{fig:vprof_ex}).  Here, the velocity center is chosen as the \Lya\ luminosity center, and the mass-weighted outflow velocity is increased from 8.2 \kms\ to 82 \kms. We find that the flux ratio is reduced to $\approx0.15$,  stressing the importance of the velocity structure in the formation of $L_{\rm blue}/L_{\rm red}$, while the location of the red peak is little changed.

It is also worth noting that optically thick outflows with a large covering fraction can lead to a low flux ratio. By performing a simple Monte Carlo \Lya\ radiative transfer of a uniform medium with a central source, we confirm that $L_{\rm blue}/L_{\rm red} \la 0.1$ if $N_{\rm HI}=10^{18}\,{\rm cm^{-2}}$ and the outflow velocity is $10\,\kms$. A more realistic setting of the homogeneous turbulent cloud of \texttt{HM6\_sZ002} also reveals a low $L_{\rm blue}/L_{\rm red}$ of $\approx 0.2$, despite the fact that the average outflow velocity of neutral gas is not significant ($\sim10$--$20\,\kms$). In the \texttt{HM6\_sZ002} run, $L_{\rm blue}/L_{\rm red}$ is kept relatively high ($\ga 0.4$) in the early phase ($t\la 4\,{\rm Myr}$), but the slow star formation and SN explosions drive gentle, optically thick outflows that subtend a large solid angle. The flux ratio drops to the minimum of $\sim 0.1$ at $t>6\,{\rm Myr}$ during which the majority of LyC photons are produced in the  \texttt{HM6\_sZ002} run. By contrast, in other spherical or filamentary clouds, strong SN bursts lower the density of the GMC and suppress \Lya\ emission at $t\ga 6\,{\rm Myr}$. The resulting flux ratio becomes close to unity, as shown as empty blue circles of $L_{\rm blue}/L_{\rm red} > 0.6$ in Figure~\ref{fig:Lratio}.
Our numerical experiments thus suggest that simultaneously reproducing the galactic-scale properties of  $v_{\rm peak,red}\sim 100$--$200\,\kms$ and $L_{\rm blue}/L_{\rm red}\sim 0.1$--$0.5$ is likely to require optically thick, fast outflows in the ISM/CGM and/or neutral outflows with large covering fractions.

\subsubsection {Connection between \Lya\ and LyC photons}
Another common feature of the simulated GMCs is the anti-correlation between the velocity separation and the escape fraction of photons in the wavelength range of [880$\AA$, 912$\AA$] ($f_{900}$) (Figure~\ref{fig:vsep}). This is expected because $f_{900}$ decreases with an increase in $N_{\rm HI}$, while the number of \Lya\ scatterings is large in optically thick media. The black lines in Figure~\ref{fig:vsep} exhibit simple analytic estimates obtained from Equation~\ref{eq:vsep} by assuming $T=20,000\,{\rm K}$ for various values of the turbulent velocity (solid, dashed, and dotted lines: $\sigma_{\rm turb}=0$, $15$, and $30\,\kms$, respectively). Here the analytic $f_{900}$ is approximated as $ \exp\left(-\sigma_{\rm 900} N_{\rm HI}\right)$, where $\sigma_{\rm 900}=6.09\times10^{-18} \,{\rm cm^{-2}}$ is the absorption cross-section at $900\,\AA$. The simulated clouds are largely consistent with the case with $\sigma_{\rm turb}=15\,\kms$, which is indeed the typical velocity dispersion of neutral hydrogen in the simulations ($\sigma\sim 10$--$15\,\kms$). There is also a distinctive trend at $f_{900}\ga 0.5$ in the runs of massive clouds, which can be attributed to a strong turbulence ($\sigma \sim 20$--$30\,\kms$) driven by photoionization heating and SN explosions. 

Figure~\ref{fig:vsep} further shows that the clouds simulated with different physical conditions follow a similar locus in the \vsep-$f_{900}$ plane, despite differences in \tSFE. Again, this happens because the neutral column density of the low-density channels through which LyC photons escape is similar ($N_{\rm HI}\sim 10^{18}\,{\rm cm^{-2}}$) on GMC scales. As the clouds become ionized and are disrupted, \Lya\ photons propagate through optically thin, volume-filling gas with  $1\la n_{\rm H} \la  30\,\cmq$ and a  neutral fraction of $10^{-5} \la x_{\rm HI} \la 10^{-4}$. The low but non-negligible neutral fraction is set by radiation-hydrodynamics calculations, and therefore, \vsep\ of the clouds for a given $f_{900}$ is more sensitive to the disruption phase of the GMCs rather than their initial properties, such as turbulence or metallicity. We also note that the same trend has been observed in RHD simulations of GMCs \citep{kimm19} where massive star particles are placed at random, instead of self-consistently modelling accretion onto sink particles as done in this study, supporting the aforementioned picture.

These findings suggest that the \vsep-$f_{900}$ sequence is reasonably well defined on GMC scales. Any positive offset from the sequence is likely to indicate the need for additional scattering with neutral hydrogen. Indeed,  we find that \vsep\ from simulated GMCs is systematically smaller for a given $f_{900}$ than measured from luminous compact galaxy samples \citep{vanzella15,izotov16a,izotov16b,verhamme17,vanzella18,izotov18,izotov21}. Introducing  further scattering  by neutral ISM/CGM naturally broadens the \Lya\ spectra, but the column density should be sufficiently low so that $f_{900}$ is not reduced significantly.

\begin{figure}
    \centering
    \includegraphics[width=\linewidth]{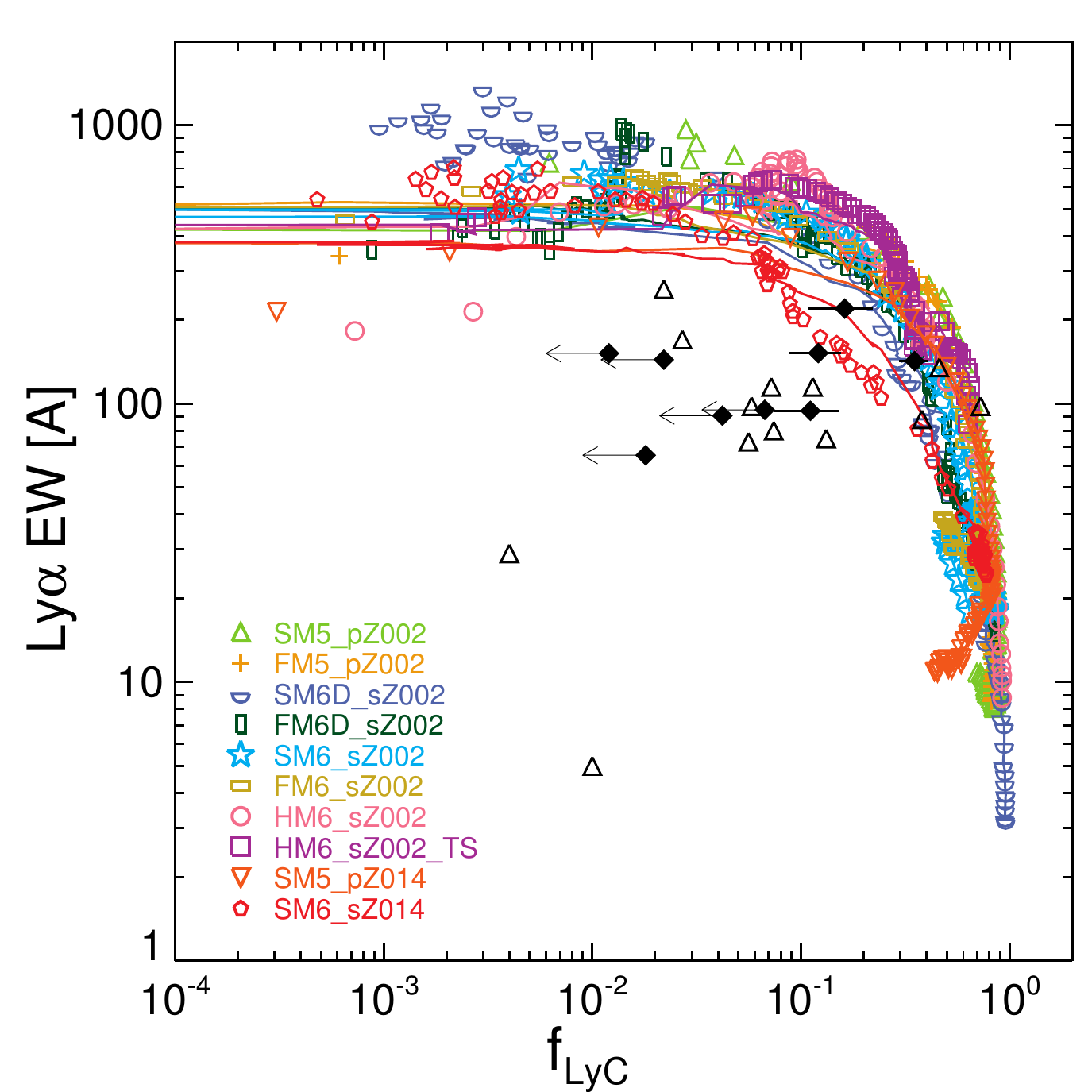}
    \caption{ Correlation between the LyC escape fraction and \Lya\ EW in GMCs. Different symbols show the attenuated \Lya\ EW in different simulations, as indicated in the legend.  The intrinsic \Lya\ EWs are shown as solid lines, and black empty triangles and filled diamonds represent galactic measurements from \citet{gazagnes20} and \citet{izotov21}, respectively. Note that the attenuated \Lya\ EWs  are often larger than the intrinsic EWs. 
    }
    \label{fig:ew}
\end{figure}

As previously noted, we find that clouds with efficient LyC escape show high \fescLya\ \citep[][see also \citealt{yajima14} for galactic scale simulations]{dijkstra16,kimm19} (Figure~\ref{fig:vsep}, right panel). Although there is  substantial scatter, \fescLya\ may be approximated as $ f_{\rm 900}^{0.27}$  in the relatively bright regime ($f_{\rm 900}> 10^{-3}$), which is in reasonable agreement with escape fraction measurements of luminous compact galaxies. In principle, LyC photons are significantly absorbed by neutral hydrogen at $N_{\rm HI}\ga 10^{17}\,{\rm cm^{-2}}$, while only half of \Lya\ photons are destroyed by dust even at $N_{\rm HI}\ga 5\times 10^{19}\,{\rm cm^{-2}}$ in the metal-poor ($Z=0.002$) slab \citep{neufeld90,hansen06,verhamme06}. Therefore, the correlation evident in the right panel of Figure~\ref{fig:vsep} cannot be easily explained by considering different cross-sections of LyC and \Lya\ photons. Rather, it is mainly shaped by the covering fraction of optically thick gas \citep[e.g.][]{hansen06,dijkstra16}. One may wonder whether the correlation is affected by different source positions between \Lya\ and LyC, but we confirm that the typical densities of the \Lya\ emitting gas and the host cell of star particles are very similar ($\nH\approx 10^{3-4}\,\cmq$) during the \Lya\ bright phase ($t\la 4\,{\rm Myr}$)\footnote{At $t\ga 5\,{\rm Myr}$, \Lya\ is produced in denser environments ($\nH\sim 5$--$10\,\cmq$), compared to LyC ($\nH \sim 0.001$--$0.01\,\cmq$), but at this stage not only \Lya\ but also LyC photons show the escape fractions of unity.}. The question that then arises is how GMCs and galaxies follow similar sequences, despite having marked differences in the $N_{\rm HI}$ distribution. A possible interpretation is that stellar feedback is sufficiently strong to create low-density channels not only in GMCs but also on galactic scales, while high $N_{\rm HI}$ regions are kept shielded, rendering $N_{\rm HI}$ bimodal. Such a bimodal distribution is indeed what we find in our GMC simulations, although the exact locations of the two peaks may differ from those of compact galaxies. Taking one step further, if clouds in  compact galaxies share identical physical properties with those in our simulated GMCs, a higher \vsep\ for the galaxies may be achieved with little effect on $f_{900}$  if the typical column density of the optically thin sight-lines for \Lya\ ($N_{\rm HI} \la 10^{18}\,{\rm cm^{-2}}$) in the compact galaxies is generally increased or if the ISM/CGM is highly turbulent ($1<\mathcal{M}\la 4$, Equation~\ref{eq:vsep}). 


Finally, we present the correlation between \fescLyC\ and the \Lya\ equivalent width (EW) in Figure~\ref{fig:ew}. The EW is estimated by comparing the intrinsic or dust-attenuated stellar continuum flux averaged at [1200, 1230]\AA\ from the simple ray tracing (Section~\ref{sec:ray_tracing}) and the intrinsic or dust-attenuated \Lya\ luminosities obtained from the {\sc Rascas} post-processing (Section~\ref{sec:rascas}). We find that the intrinsic \Lya\ EWs are initially very large ($\sim400$--$600\, \AA$) because of the contribution from hot OB and Wolf-Rayet stars. These EWs are much larger than those of young stellar populations \citep{charlot93} or those observed in actively star-forming galaxies at high redshift \citep[e.g.,][]{malhotra02,santos20}, but quickly drop to $\sim 100\,\AA$ around 3 Myr from the onset of star formation.  In the \Lya\ bright phase ($t\la 3\, {\rm Myr}$), \fescLyC\ varies from $<10^{-4}$ to $\sim 0.1$, resulting in the flat sequence for $\fescLyC\la 0.1$. Once the GMCs are sufficiently dispersed, \Lya\ EWs decrease to about $10\, \AA$, while $\fescLyC$ keeps increasing at $\ga 0.1$; thus, there exists a negative correlation between \fescLyC\ and \Lya\ EW on GMC scales. Dust absorption tends to be more significant for the continuum photons than \Lya\ during LyC faint phases, leading to extreme \Lya\ EWs of up to $\sim 1000\,\AA$. The trend found in our GMC simulations is evidently distinct from that of the LyC leaker data at $0.2\la z\la 0.4$ \citep{gazagnes20,izotov21}, but the absorption associated with the ISM/CGM and/or the presence of an underlying stellar population should easily alleviate the difference between our GMC results and galactic measurements (Song et al. {\it in prep}).

\begin{figure*}
    \centering
    \includegraphics[width=0.48\linewidth]{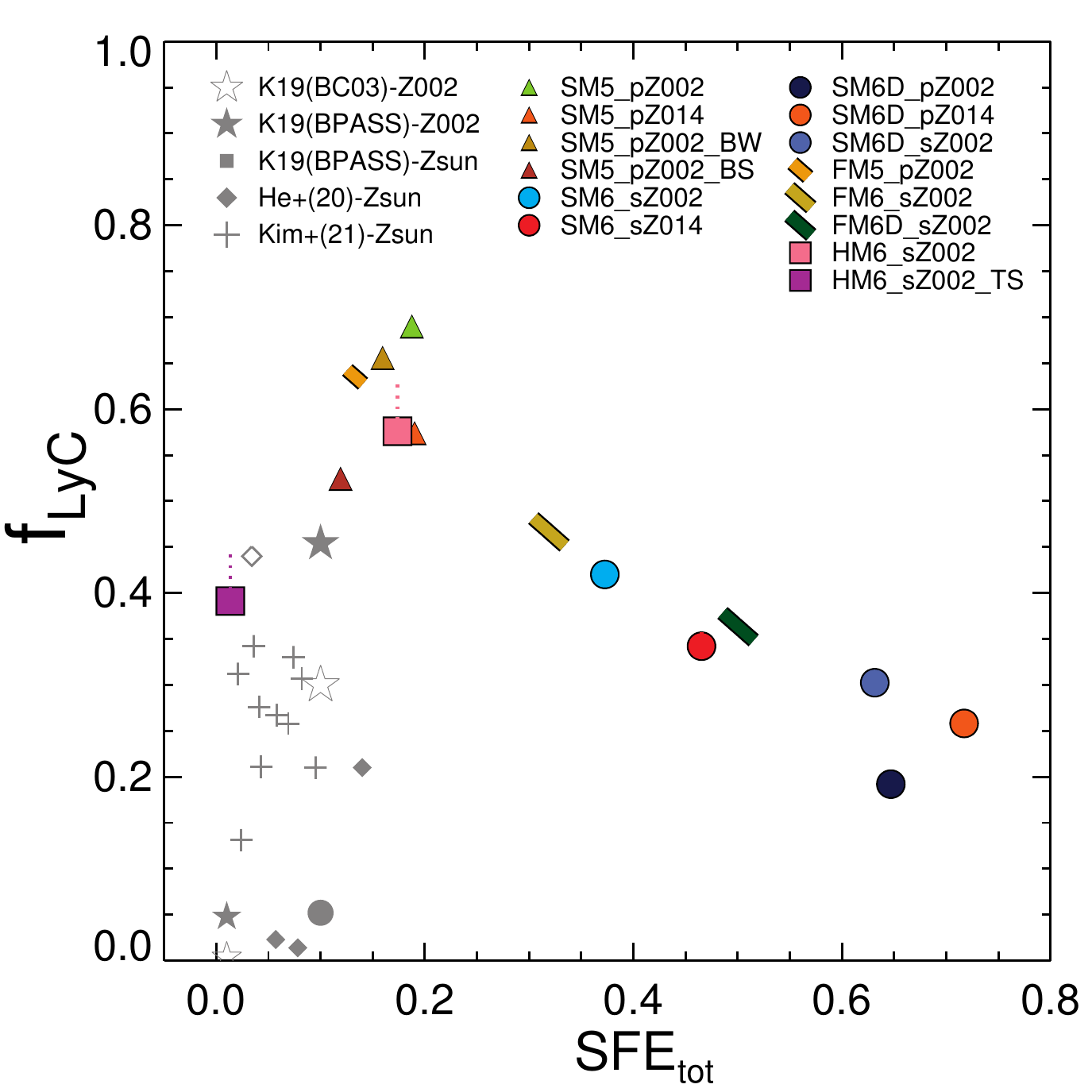}
    \includegraphics[width=0.48\linewidth]{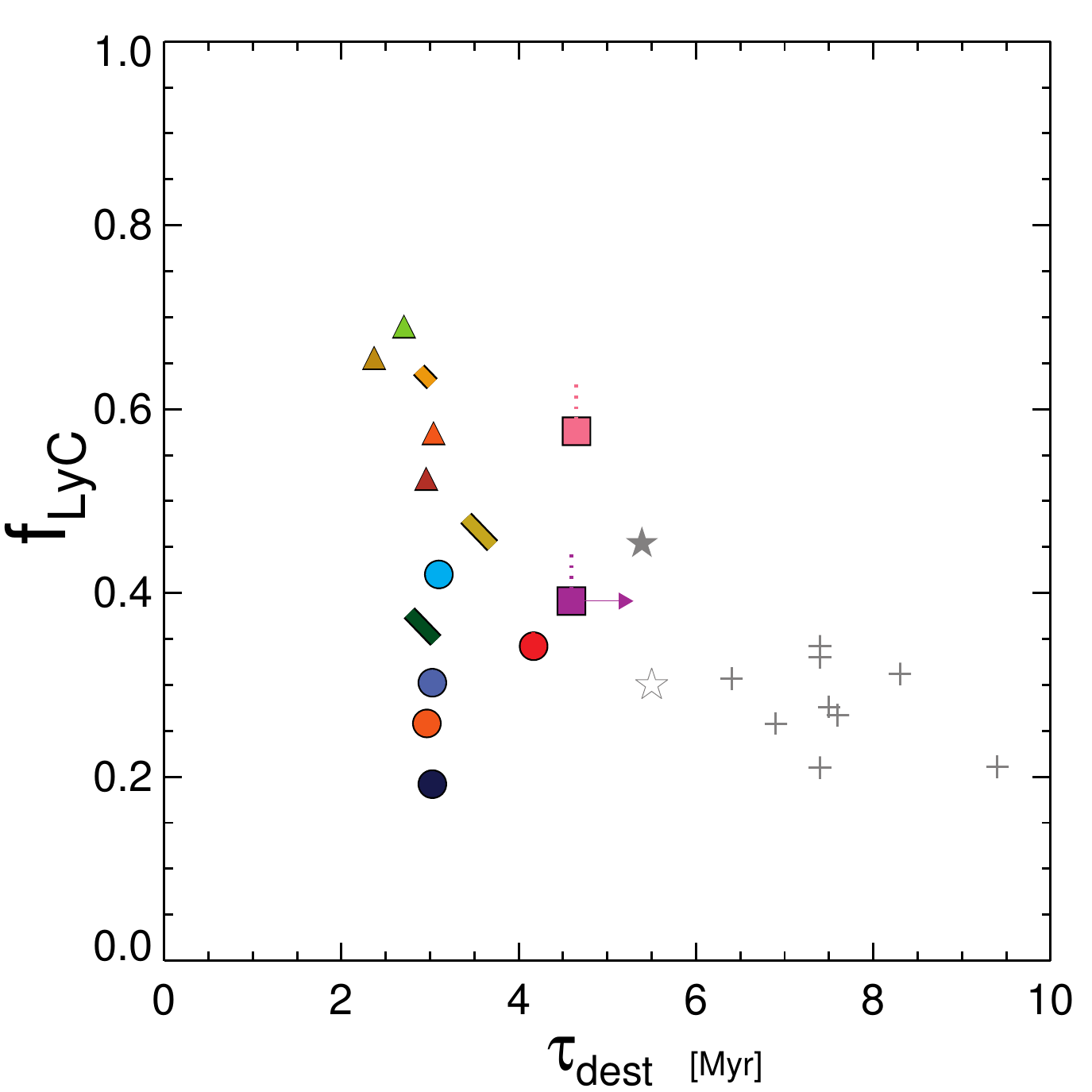}    
    \caption{ Correlation between the luminosity-weighted escape fraction and the total star formation efficiency in the simulated clouds (left). Different symbols indicate the results from the runs with different simulation parameters, as indicated in the legend. Clouds with a smaller mass ($10^5\,\msun$) are shown with smaller symbols. Dotted lines correspond to the extrapolated luminosity-weighted escape fraction measured until $t=20\,{\rm Myr}$ assuming that \fescLyC\ is kept the same as the value in the final snapshot.  Also included as grey star symbols are the escape fractions from \citet{kimm19} with the single (empty) or binary stellar evolution model (filled). Bigger star symbols indicate the results measured from $10^6\,\msun$ clouds, while the small star symbols correspond to those from  $10^5\,\msun$ clouds. The grey filled square shows the solar metallicity case from \citet{kimm19}.   Grey crosses display \fescLyCL\ measured from $10^5$ clouds by \citet{kimjg21}. The simulated data for clouds with $10^5\,\le M_{\rm cloud}\le 3\times10^5\,\msun$ from \citet{he20} are shown as filled grey diamonds. Note that there is a clear correlation between \tSFE\ and the luminosity-weighted escape fraction of LyC photons. The right panel indicates the correlation between the escape fraction and the destruction timescale of dense gas ($\tau_{\rm dest}$). We define $\tau_{\rm dest}$ as the time between the onset of star formation and the moment at which the amount of gas with $\nH\ge 100\,\cmq$ drops to 0.1 $M_{\rm cloud}$. }
    \label{fig:fesc_sfe}
\end{figure*}

\section{Discussion}
Having acquired an understanding of the detailed properties of escaping LyC and \Lya\ photons, we now discuss  the correlation between \fescLyCL\ and  or destruction timescale of the GMC ($\tau_{\rm dest}$).  We also discuss a possible way to infer \fescLyC\ from photons with different wavelengths. The limitations and caveats of this study are also presented later in this section.

\subsection{Correlation between $\left<f_{\rm esc}\right>$--\tSFE}

An interesting conclusion of \citet{kimm17} is that the escape of LyC photons is more efficient if the gas mass enshrouding young stars is smaller (see their Figure 13). Using sub-pc resolution, cosmological RHD simulations of atomic-cooling halos with a mass of $\sim 10^8\,\msun$, they measured the escape fraction for individual star formation events and demonstrated that it correlates well with the ratio between the amount of dense gas that cannot be photoionized and the photon production rate. This suggests that for a given gas mass, a more efficient star formation episode would lead to a higher \fescLyC. The results of \citet{kimm19} also support this picture in  that a larger fraction of LyC photons escapes from GMCs with an \tSFE\ of 10\% compared with those with with a \tSFE\ of 1\%.

To further investigate this idea, we plot \tSFE\ against \fescLyCL\ in Figure~\ref{fig:fesc_sfe}. GMCs with different morphologies are indicated by different symbols, and clouds with a smaller mass ($10^5\,\msun$) are shown as smaller triangles. We also re-measure  luminosity-weighted escape fractions until $t=10\,{\rm Myr}$ for the $10^6\,\msun$ clouds of \citet{kimm19}. In marked contrast to the previous results obtained from GMCs with \tSFE\  of $\le 10\%$ \citep{kimm19}, we find that in the current simulations the escape fraction is anti-correlated with \tSFE\ in actively star-forming GMCs ($20\% \la \tSFE \le 80\%$). The relation is preferentially driven by massive clouds, as \tSFE\ of the less massive clouds tend to be high overall. In particular, clouds with high surface densities convert the majority of the gas into stars, but their \fescLyCL\ is found to be the lowest. Note that the the amount of the remaining gas per stellar mass is smaller with higher \tSFE, and thus one may naively expect that the LyC escape is more efficient in clouds with high surface densities. However, the anti-correlation with \tSFE\ indicates that it is more difficult for young massive stars to photoionize their optically thick media in clouds with higher surface densities. This is because recombination rates in the ionised gas scale with the gas density squared. While the offset between the time at which the ionizing luminosity reaches 50\% of its maximum and the time at which \fescLyC\ becomes 50\% is $\approx 2\,{\rm Myr}$ for \texttt{SM6\_sZ002}, the offset is longer  ($\approx 3\,{\rm Myr}$) in a run with a higher \tSFE\ (\texttt{SM6D\_sZ002}) (see also panel (b) from Figure~\ref{fig:fescLyC_all}). The difference in the offset is modest ($\approx 1\,{\rm Myr}$), but as the majority of ionizing photons are produced in the first $t\la 5\,{\rm Myr}$, the resulting \fescLyCL\ differs by a factor of 2 ($\sim40\%$ versus $20\%$). 

In the other regime  where turbulence prevents the cloud from collapsing beyond $\approx1\%$ (\texttt{HM6\_sZ002\_TS}), \fescLyCL\ is again reduced to $\approx 40\%$, compared with  $\fescLyCL\approx 60$--$70\%$ from the less massive \texttt{M5} clouds. The decrease is largely due to the finite number of LyC photons from massive stars and thus the slower propagation of ionization fronts. \citet{kimm19} also pointed out that  $10^6\,\msun$ clouds with \tSFE\  of 1\% exhibit a small \fescLyCL\ of $\sim 1$--$5\%$ when the star particles are randomly placed at $r\la 5\, {\rm pc}$ from the cloud center. In our case, a larger \fescLyCL\ of $\approx 40\%$ is obtained in the \texttt{HM6\_sZ002\_TS} run because strong turbulence allows LyC radiation to escape through a porous medium. However, the escape fraction may be reduced if the propagation of ionization fronts is delayed because of a low \tSFE\ caused by non-thermal pressure, such as strong magnetization. Simulations of low-mass ($10^5\,\msun$),  magnetized GMCs performed by \citet{kimjg21} indeed show that \fescLyCL\ is reduced from $30\%$ in their runs with $\mu_{\rm B,0}=\infty$ where \tSFE\ is 0.08 to $13\%$ in the $\mu_{\rm B,0}=0.5$ runs where \tSFE\ is 0.02. Taken together, we are thus led to conclude that \fescLyCL\ is  likely to be maximal at intermediate \tSFE\ ($\sim 0.2$) in GMCs with $10^5$--$10^6\,\msun$\footnote{\citet{he20} performed RMHD simulations of GMCs with $3\times 10^3$--$3\times 10^5\,\msun$, and found that \fescLyCL\ decreases with increasing \tSFE. This potentially suggests that the dependence of \fescLyCL\  on \tSFE\ may be different in lower-mass clouds with $M_{\rm cloud}< 10^5\,\msun$.}. 

The relation between \tSFE\ and \fescLyCL\ shows that these quantities both result from the same evolutionary process of the clouds. As can be seen in Figure~\ref{fig:fesc_sfe}, the main drivers of this mechanism are the initial surface density and the level of turbulence (or virial parameter), which are modulated to second order by the shape, magnetic field strength, and metallicity of the gas.

\subsection{Correlation between $\left<f_{\rm esc}\right>$--$\tau_{\rm dest}$}

Because the escape of LyC radiation depends on the evolution of the GMC, it is reasonable to conjecture that \fescLyCL\ may be characterized by the destruction timescale of the GMCs. In Figure~\ref{fig:fesc_sfe} (right panel), we show $\tau_{\rm dest}$, the time between the onset of star formation and the moment at which the mass of the dense gas with $\nH\ge100\,\cmq$ falls below 10\% of its initial value\footnote{We choose this definition to facilitate a comparison with other studies. On the basis of the RHD simulations of \citet{kimm19}, we measured the timescale at which 95\% of molecular hydrogen is dissociated \citep{kimjg21} and found that this roughly matches the timescale at which 90\% of the dense gas with $\nH\ge100\,\cmq$ is destroyed.}. The figure demonstrates that in the current simulations the correlation between $\tau_{\rm dest}$ and \fescLyCL\ is rather weak.  Although the low-mass GMCs having the maximal \fescLyCL\ can be explained by the rapid disruption of the clouds,  there is non-negligible scatter in the diagram because the destruction of the dense (molecular) gas does not necessarily mean that the GMC is entirely transparent to LyC. 
For this to happen, the column density should drop below $N_{\rm HI}\sim 10^{17}\,{\rm cm^{-2}}$.

\begin{figure}
    \centering
    \includegraphics[width=7cm]{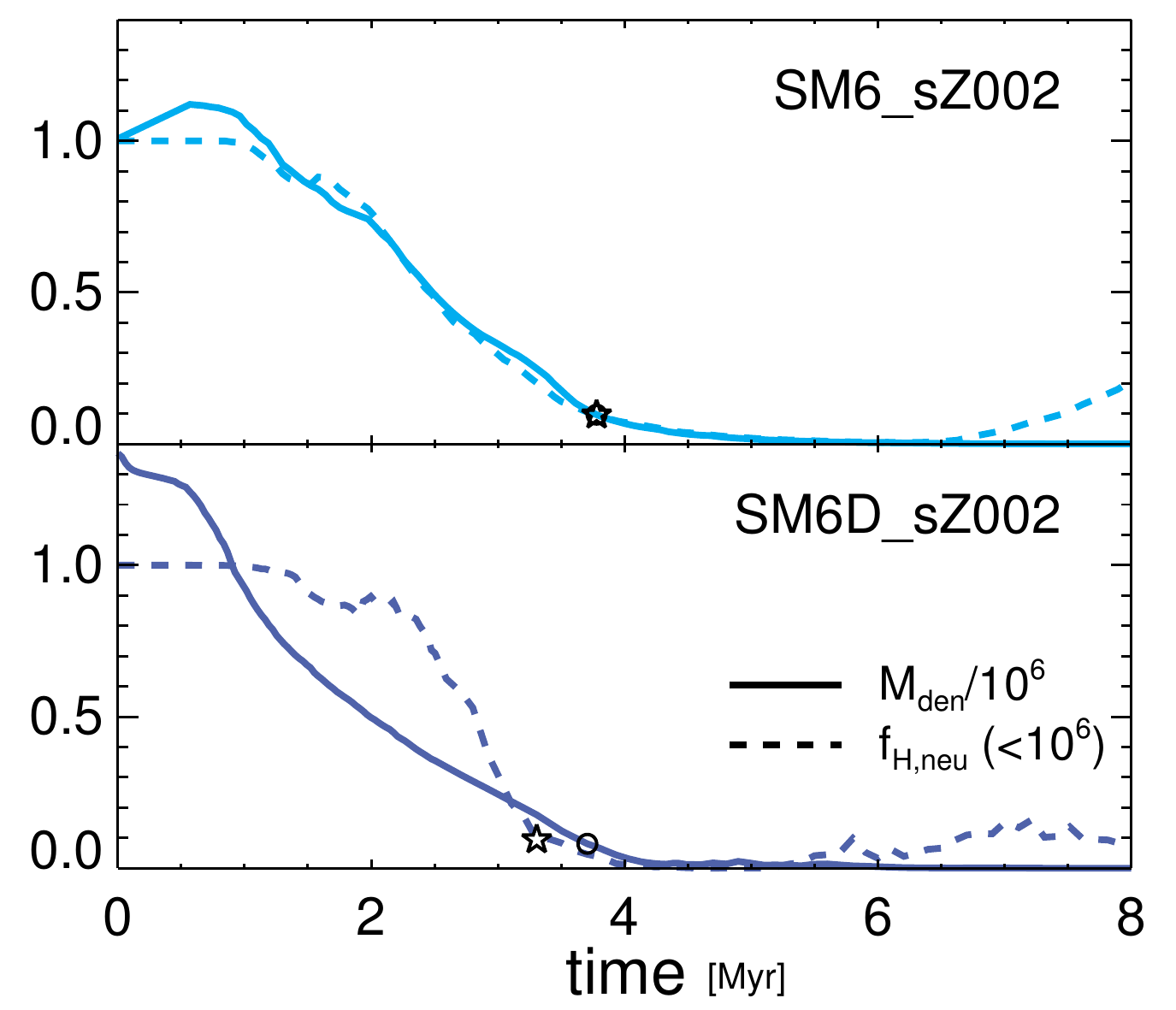}
    \caption{ Differences in the development of ionized bubbles in GMCs with normal (upper panel) and high gas surface density (lower panel). The solid lines indicate the amount of dense gas with $\nH\ge100\,\cmq$, divided by $10^6\,\msun$.  The dashed lines show the neutral fraction of hydrogen within a radius inside which the total gas mass is $10^6\,\msun$. The circle and star symbols correspond to the moment at which  the fraction of dense gas and the fraction of neutral hydrogen fall below 0.1, respectively. Note that the destruction timescales are similar for the two GMCs, but a larger portion of the gas is kept neutral for an extended period of time in the run with  higher gas surface density (compare the dashed lines at 2--3 Myr). 
    }
    \label{fig:m_ion}
\end{figure}

\begin{figure*}
    \centering
    \includegraphics[width=8.cm]{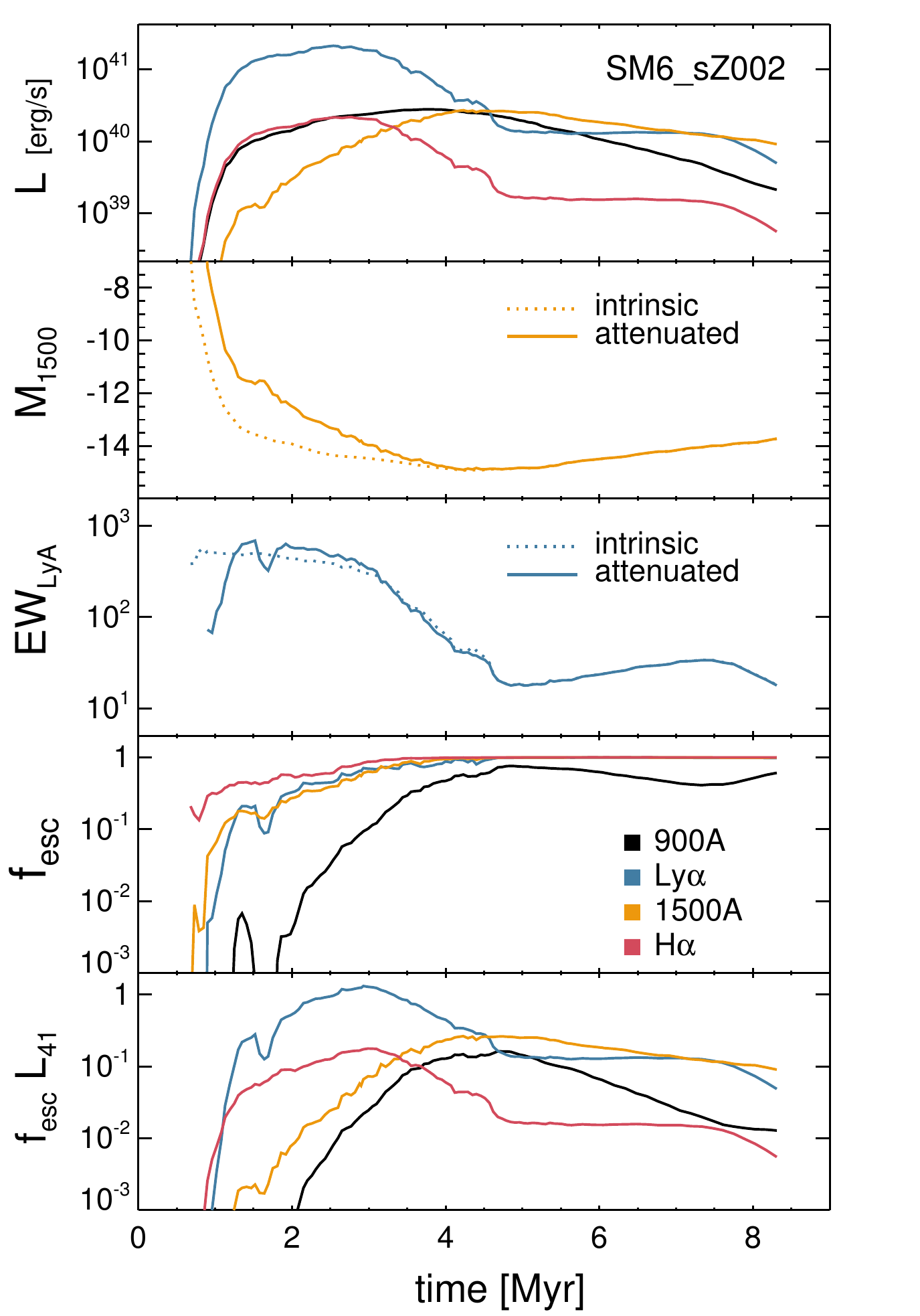}
        \includegraphics[width=8.cm]{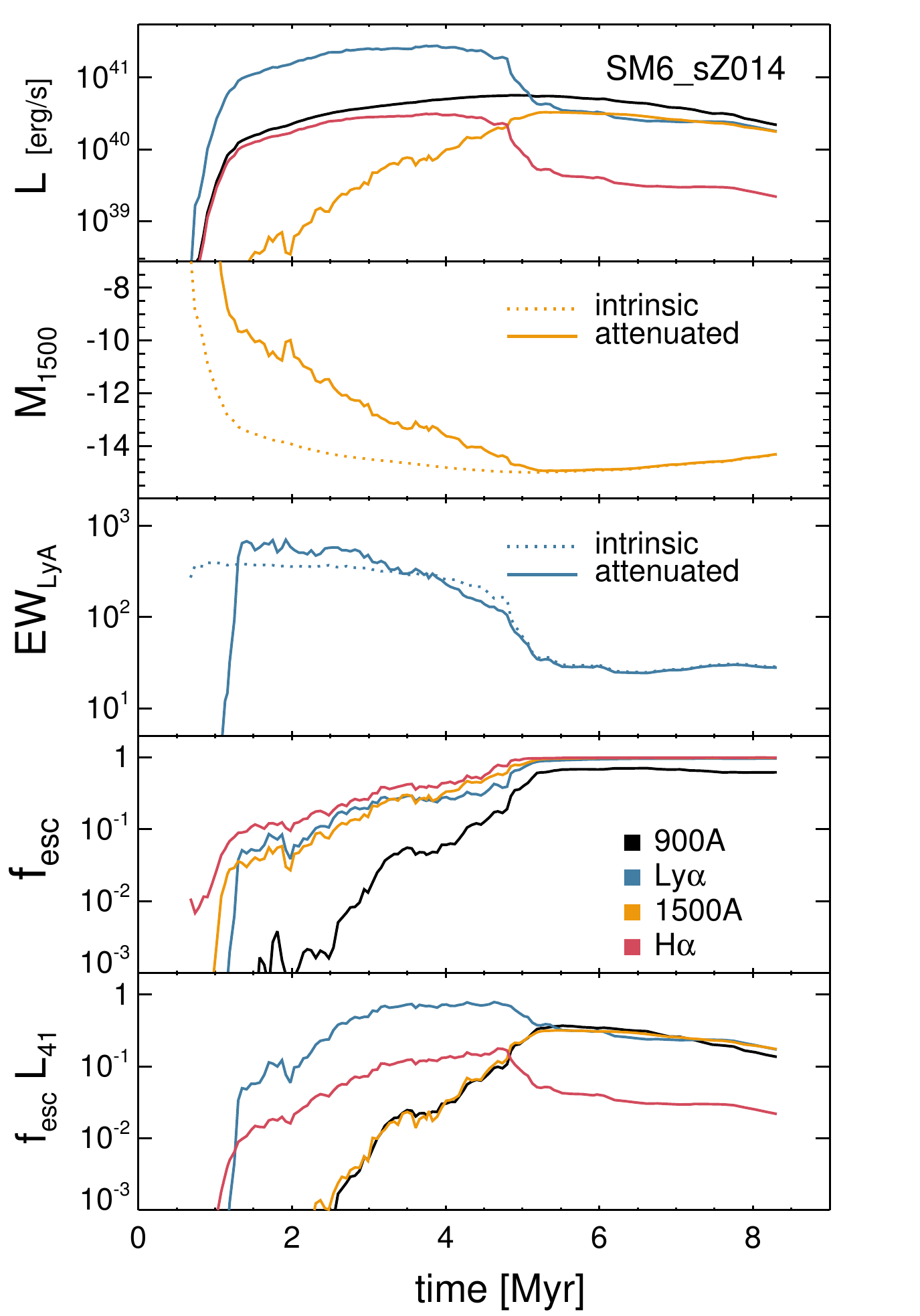}
    \caption{ The intrinsic luminosities and escape fractions of photons with various wavelengths from the metal-poor (\texttt{SM6\_sZ002}; left) and metal-rich cloud (\texttt{SM6\_sZ014}; right). From top to bottom, each row indicates the intrinsic luminosity in units of $\rm erg s^{-1}$, UV magnitude at 1500 \AA, \Lya\ equivalent width in \AA, escape fraction, and escaping luminosity, normalized to $10^{41}\,{\rm erg\,s^{-1}}$. Photons with different wavelengths are shown as different colors.
    }
    \label{fig:fesc_uv}
\end{figure*}

To single out the origin of the difference in \fescLyCL\ for clouds with similar $\tau_{\rm dest}$, we plot in Figure~\ref{fig:m_ion} the amount of the dense gas with $\nH\ge100\,\cmq$ divided by $10^6\,\msun$, and the neutral fraction of the central $10^6\,\msun$ gas in the two runs (\texttt{SM6\_sZ002} and \texttt{SM6D\_sZ002}) whose $\tau_{\rm dest}$ are similar ($3.1$ and $3.5\,{\rm Myr}$, respectively).  The initial dense gas mass in the \texttt{SM6\_sZ002} cloud is $10^6\,\msun$, but it increases slightly as the gas in the outer envelope with $\nH=50\,\cmq$ collapses. On the other hand, the outer envelope gas is considered as a dense component in the  \texttt{SM6D\_sZ002} case because of its high initial density ($\nH=176\,\cmq$), and the solid line in the bottom panel starts at $\approx1.4\times10^{6}\,\msun$. Note that stars begin to form earlier ($\Delta t\approx 0.5\,{\rm Myr}$) in the dense cloud than in the fiducial run (Figure~\ref{fig:SFH}), and therefore the neutral fraction should drop first in the dense cloud if ionization fronts propagate at a similar speed. Yet, we find that while the dense component is efficiently dispersed in the  \texttt{SM6D\_sZ002} run, the ionized bubble develops slowly, and a large fraction of the central $10^6\,\msun$ gas is maintained neutral for an extended period of time, unlike the fiducial surface density run (\texttt{SM6\_sZ002}). 
Therefore, apart from the destruction timescale for the dense gas, the ionization speed also matters for setting up \fescLyCL\ on cloud scales. 

We also note that the correlation between \fescLyCL\ and $\tau_{\rm dest}$ may be more evident if the results from the clouds with different surface densities are included in Figure~\ref{fig:fesc_sfe}.   \citet{kimjg21} found that the destruction timescales for $10^5\,\msun$  clouds with a lower $\Sigma_{\rm gas}$ are longer ($\tau_{\rm dest}\sim6-10\,{\rm Myr}$) than those of our low-mass GMCs, while their \fescLyCL\ tends to be smaller. Because the gas collapse and star formation occurs slowly, the clouds are disrupted on longer timescales, and LyC photons emitted from the stars formed early are easily absorbed. In this regard, $\tau_{\rm dest}$ may be seen as a rough indicator of \fescLyCL, but more simulations are needed to validate the connection between \fescLyCL\ and $\tau_{\rm dest}$ based on the same numerical framework.

\subsection{Inferring $f_{\rm esc}$ from Ly$\alpha$, UV, and H$\alpha$ }

In observations, constraining the escape of LyC photons from high-z galaxies is a difficult task because the ionizing flux is usually very faint owing to significant attenuation by the neutral ISM and the IGM. Alternative methods  based on [CII] 158 $\mu m$ and [OIII] 88 $\mu m$ emission lines \citep{katz20}, or UV absorption lines such as CII 1334 \AA\  \citep{mauerhofer21}, Si II 1260 \AA\ \citep{gazagnes18}, or \Lya\ \citep{verhamme15}, have thus been proposed. In a similar spirit, we compare \fescLyC\ with the escape fraction of photons at different wavelengths and examine if other often-used emission lines may be used to infer \fescLyC.

\begin{figure*}
    \centering
    \includegraphics[width=7.cm]{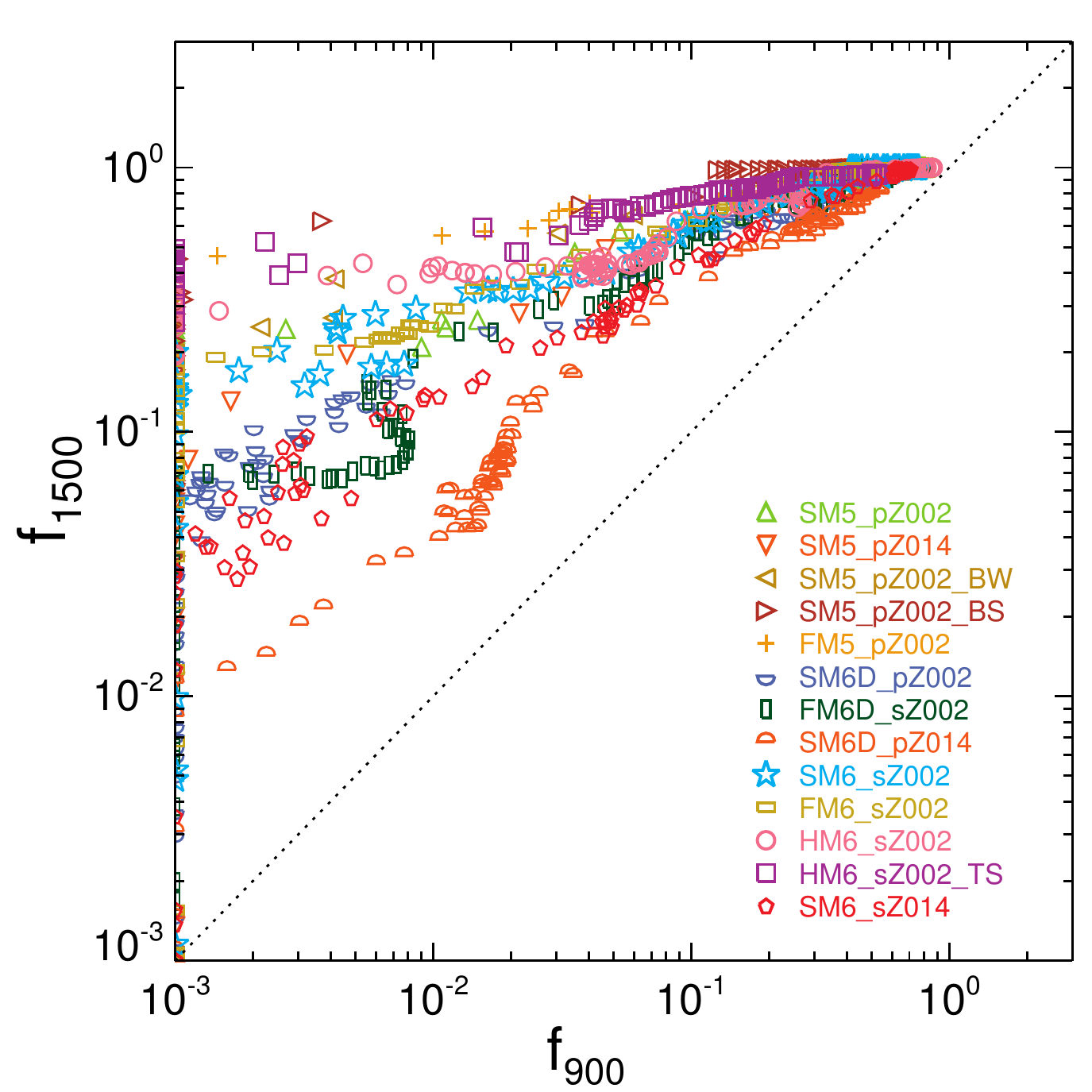}
    \includegraphics[width=7.cm]{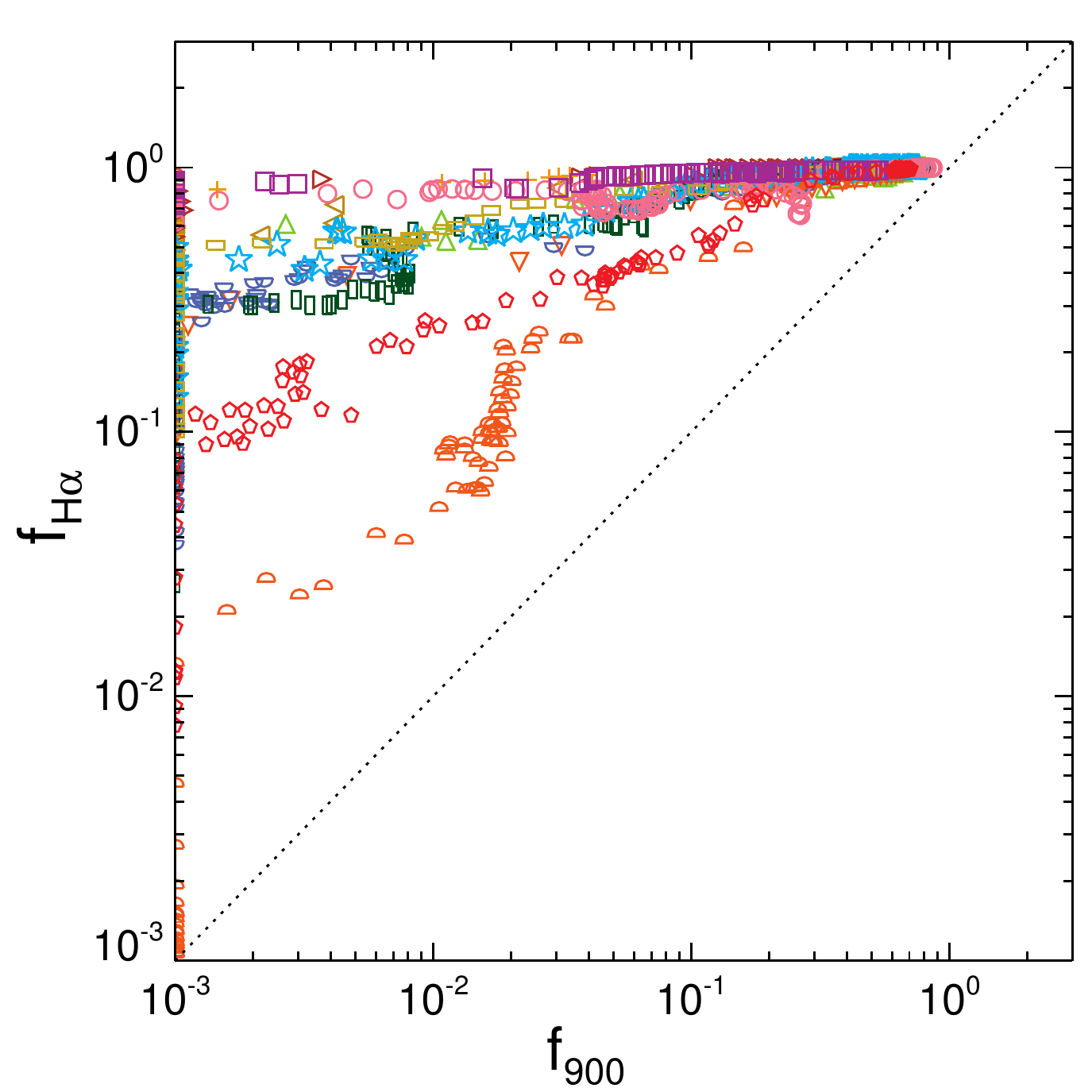}    
    \includegraphics[width=7.cm]{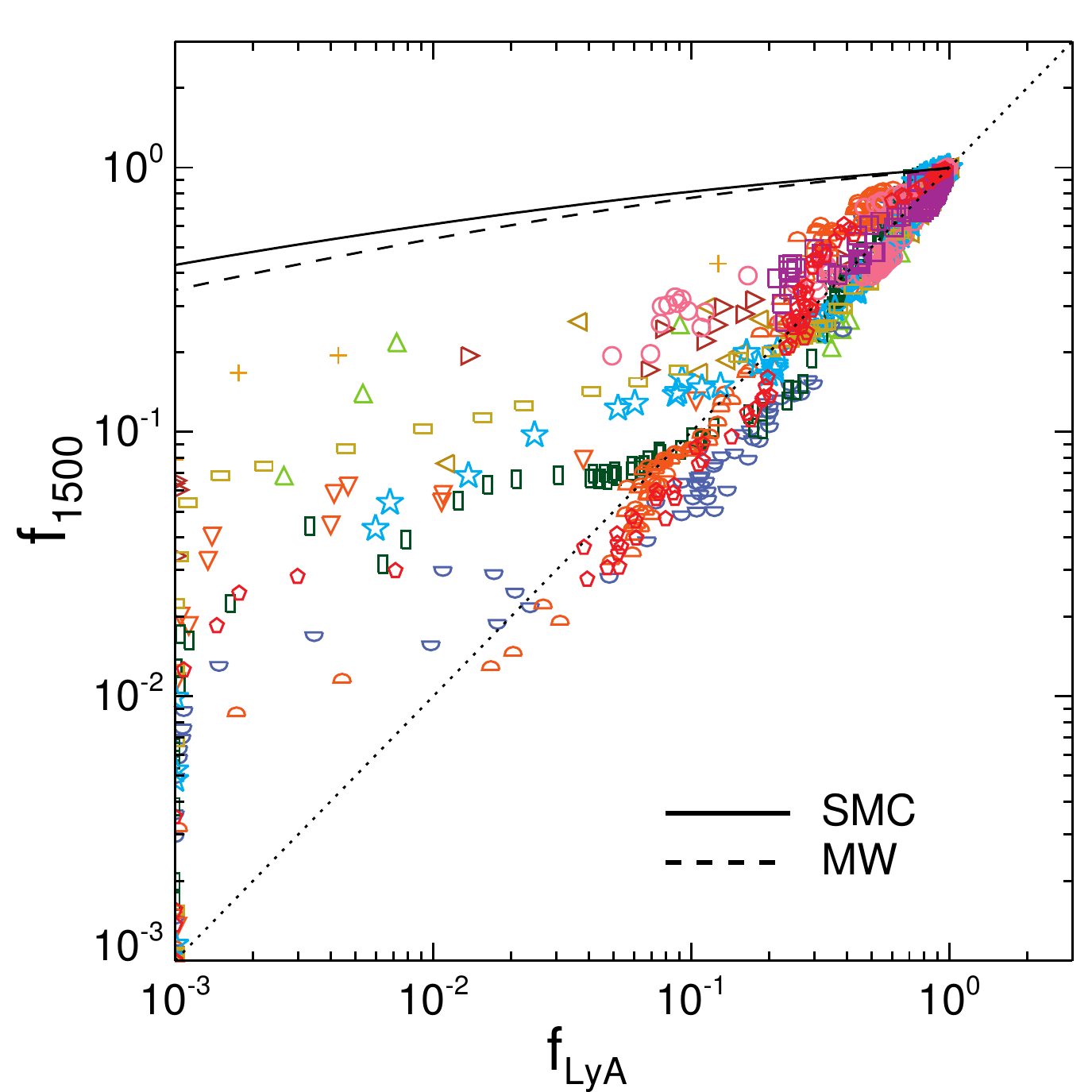}
    \includegraphics[width=7.cm]{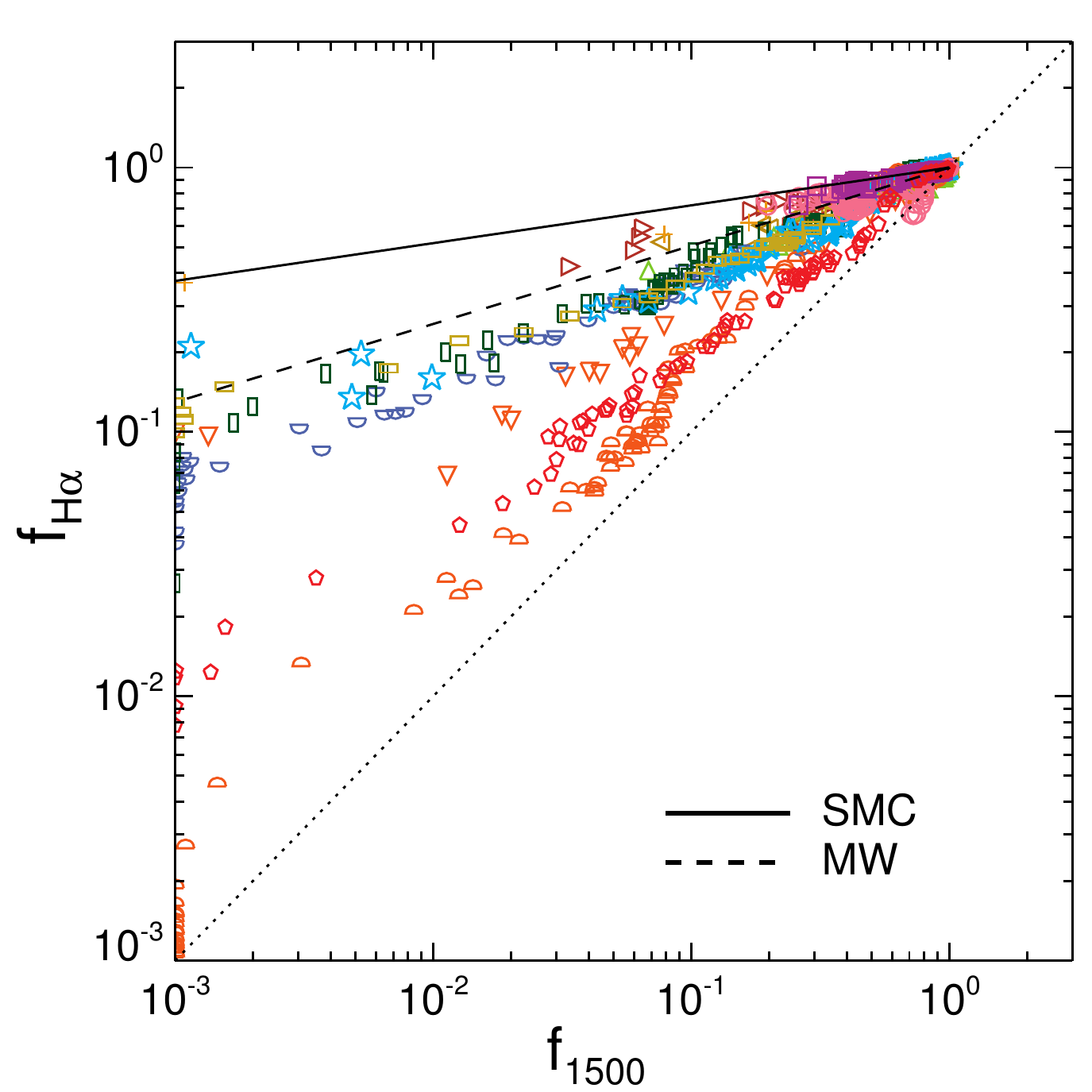}    
    \caption{ The correlation between the escape fraction of various wavelengths. The colored symbols show simulated escape fractions measured at different times from GMCs with different conditions, as indicated in the legend. The dotted line shows one-to-one correspondence. The solid and dashed lines in the bottom panels exhibit the correlation predicted from the uniform media with Small Magellanic Cloud-type or Milky Way-type dust. Note that the simulated GMCs display a steeper correlation than the uniform case, indicating that the UV photons are absorbed by clumpy gas distributions.   }
    \label{fig:fesc_comp}
\end{figure*}

In Figure~\ref{fig:fesc_uv}, we present the luminosity and  escape fraction of photons in UV and optical bands. The UV luminosity at 1500 $\AA$ is computed by integrating the flux at [1475, 1525] $\AA$, and we apply SMC-type or MW-type dust  attenuation for the clouds with $Z=0.002$ or $Z=0.014$, respectively, based on \citet{weingartner01}. The emissivity of the Balmer $\alpha$ line at 6562.8 $\AA$ is calculated as,
\begin{equation}
\varepsilon_{\rm rec,H\alpha} = n_e \, n_p \, P_{\rm B, H\alpha}(T) \, \alpha_B(T) \, e_{\rm H\alpha},
\end{equation}
where $n_e$ and $n_p$ are the number densities of electrons and protons, respectively, $P_{\rm B,H\alpha}$ is the probability for a recombination event to produce a H$\alpha$ photon \citep{storey95}, $\alpha_B(T)$ is the temperature-dependent Case B recombination coefficient \citep{hui97}, and $e_{\rm H\alpha}=1.89\,{\rm eV}$ is the energy of the photon. As is the case for \Lya\ photons, we randomly choose the initial frequency of the H$\alpha$ photons using a Gaussian distribution with the thermal Doppler broadening for individual cells. We also calculate $L_{900}$ by integrating the flux between [880, 910] $\AA$.

Figure~\ref{fig:fesc_uv} shows that \Lya\  is the most luminous among the four observables ($L_{900}$, $L_{\rm Ly\alpha}$, $L_{1500}$, $L_{\rm H\alpha}$) in the early phase ($t \la 4\,{\rm Myr}$) of the GMC evolution. $L_{\rm Ly\alpha}$ becomes comparable to the luminosities at 900 and 1500 $\AA$ later, as LyC photons escape efficiently from the clouds and the recombination process becomes inefficient. H$\alpha$ luminosities share the same trend as  $L_{\rm Ly\alpha}$, but are smaller by a factor of 8--10 because of the difference in the photon energy and the transition probability. The \Lya\ EWs are initially quite high ($\sim500\,\AA$), but decreases to a few tens of $\AA$. Recombination lines (\Lya\ and H$\alpha$), including LyC radiation, emerge very quickly as star formation proceeds, but the flux at 1500 \AA\ develops rather slowly. This can be attributed to massive stars in the zero age main sequence becoming more luminous in the early phase of hydrogen burning, as nuclear reactions are enhanced due to the temperature increase in the convective stellar core \citep[e.g.][]{iben67}. The UV light from the simulated clusters is affected by significant dust attenuation ($A_{1500}\la 2$--$4$) until the clouds are dispersed, and it then reaches the maximum luminosity of $M_{1500} \approx -15$ in AB magnitude. Such clusters are challenging to detect at the epoch of reionization even with the Hubble Space Telescope imaging \citep[e.g.,][]{livermore17,atek18}, but may be accessible with the James Webb Space Telescope or deep MUSE observations of \Lya\ emission via strong lensing \citep[e.g.,][]{vanzella20}.

\begin{figure}
    \centering
    \includegraphics[width=\linewidth]{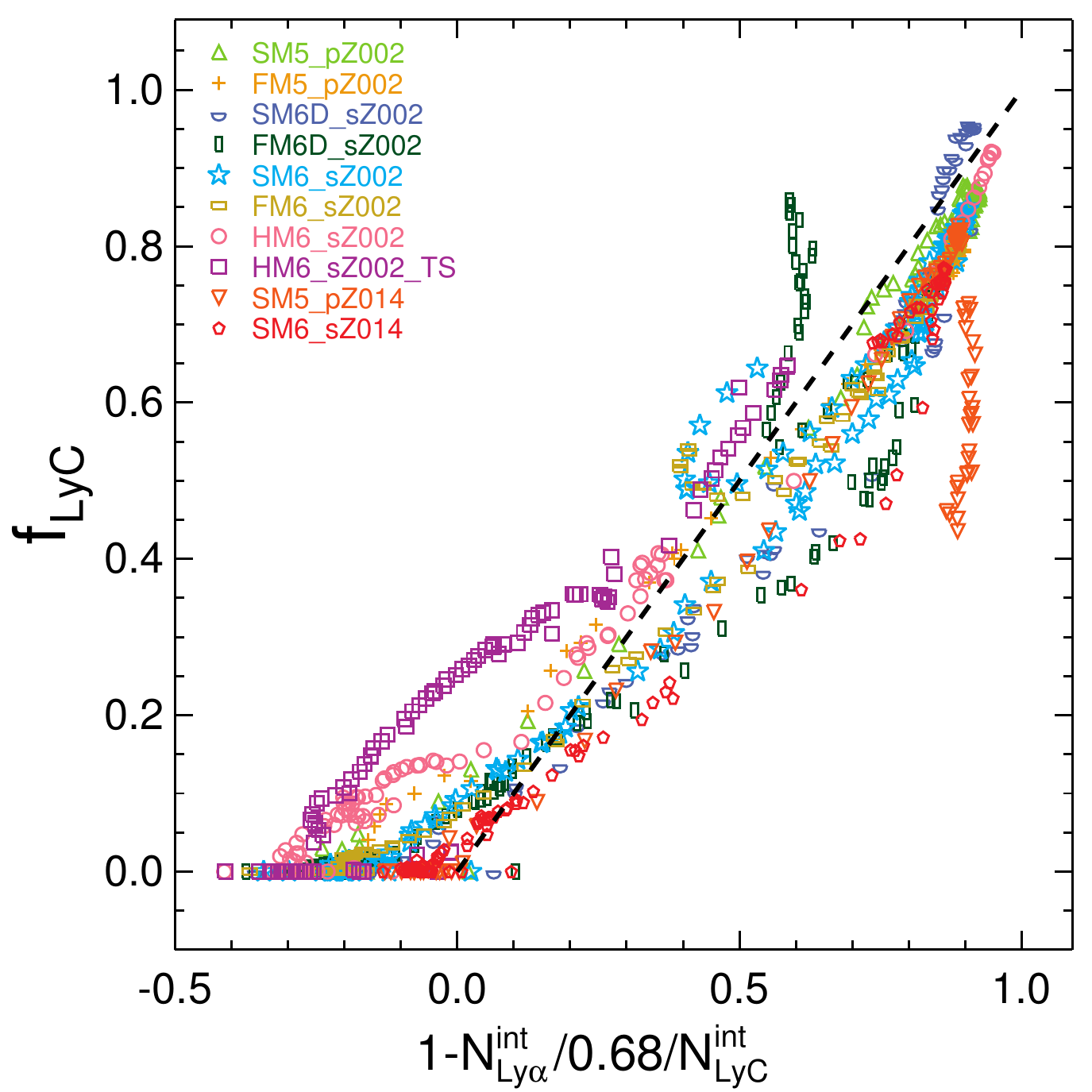}
    \caption{Escape fraction of LyC radiation deduced from the number of \Lya\ photons produced per second ($\fescLyCa\equiv1-N_{\rm Ly\alpha}^{\rm int}/0.68/N_{\rm LyC}^{\rm int}$). The runs with different input parameters are shown as different symbols and color-codings, as indicated in the legend. For comparison, we display a one-to-one relation as a dashed line. The inferred escape fractions (\fescLyCa) reasonably match the true \fescLyC\ in the range $\fescLyCa\ga 0.1$, albeit with significant scatter ($\Delta \fescLyC \sim0.2$). Note that \fescLyCa\ tends to be greater than \fescLyC\ at $\fescLyCa\ga0.4$ (see the text). }
    \label{fig:fesc_from_lya}
\end{figure}

The escape fractions of \Lya\ and the UV photons at 1500\,\AA\ appear to trace each other reasonably well (Figure~\ref{fig:fesc_uv}). This is essentially because of the clumpy gas distributions that simultaneously block both photons along some sight-lines \citep{hansen06}, not because of the similar dust absorption cross-sections. The resonant nature of \Lya\ increases the interaction probability with dust, and approximately half of the \Lya\ photons are destroyed at $N_{\rm HI}\approx 5\times 10^{19}\,{\rm cm^{-2}}$ in the case of SMC-type dust with $Z=0.002$, assuming a uniform medium of $T=10^4\,{\rm K}$ \citep[e.g.,][]{verhamme06}. By contrast, despite the proximity in wavelength, the UV photon in 1500\AA\ requires $\approx 40$ times higher $N_{\rm HI}$ to be absorbed by dust.  This is more clearly illustrated in Figure~\ref{fig:fesc_comp} (bottom left panel), where we present the escape fraction measurements of simulated GMCs along with simple analytic estimates of the escape fraction in a uniform medium. While only 0.1\% of \Lya\ photons would survive in a uniform medium regardless of the dust model assumed, a significant fraction ($\approx 40\%$) of the UV photons would be transmitted according to the simple analytic model. However, the  simulated GMCs show similar $f_{1500}$ and \fescLya\ for $\fescLya\ga 0.1$, although $f_{1500}$ becomes systemically larger for $\fescLya\la 0.1$, as $N_{\rm HI}$ distributions in the early clouds are not as clumpy as those at the late phase of the GMC evolution. 

Our results thus suggest that the escape fraction at 1500\,\AA\ can only be as useful as \Lya\ in selecting potential LyC leakers. It is indeed evident from Figure~\ref{fig:fesc_comp} (top left panel) that not only \fescLya\ but also $f_{1500}$ is systematically higher than \fescLyC, with dusty clouds being less efficient in leaking UV photons. Nevertheless, the extreme UV condition $f_{1500} \ga 0.5$ is likely to increase the probability of finding GMCs with $f_{900}\ga 0.1$, compared with a blind survey.  
The post-processing of a cosmological SPH simulation run with lower resolutions ($250\,h^{-1}$ pc and $3\times10^5\,h^{-1}\,\msun$) also showed that the UV condition may be used to find LyC leakers even on galactic scales \citep{yajima14}.

Compared with UV or \Lya\ photons,  H$\alpha$ photons are less affected by dust, and they escape more efficiently from the GMC (top right panel). Except for the metal-rich  massive cloud (\texttt{SM6\_sZ014}), the majority of  simulated data show $f_{\rm H\alpha}\ga 0.5$, making $f_{\rm H\alpha}$ the least favored indicator for selecting potential LyC leakers.  The bottom right panel in Figure~\ref{fig:fesc_comp} shows that there is a positive correlation between $f_{\rm H\alpha}$ and $f_{1500}$, but again, $f_{\rm H\alpha}$ is smaller than the simple estimate obtained from attenuation due to a uniform SMC or MW-type dust, indicating the effect of the clumpy gas distribution. The simulated clouds also show a more significant deviation from the analytic estimates (solid or dashed lines) in the $\fescLya$--$f_{1500}$ plane than in then $f_{\rm H\alpha}$--$f_{1500}$ plane. This implies that the typical column density of the clumpy gas is not significantly higher than $N_{\rm H}\sim10^{22}\,{\rm cm^{-2}}$, given that the dust absorption cross-section at 1500 \AA\ is only 7 (3.4) times that at 6563 \AA\ in the case of the SMC-type (MW-type) dust.

Although the use of escape fractions may not be practical to directly infer  \fescLyC, Ly$\alpha$ (or H$\alpha$) luminosities may be a promising alternative. If the majority of \Lya\ photons are produced by recombinative radiation, the number of \Lya\ photons produced per second ($\dot{N}_{\rm Ly\alpha, int}$) can be approximated as $\dot{N}_{\rm Ly\alpha,int} \approx P_B(T) \, \dot{N}_{\rm LyC,abs}$, where $\dot{N}_{\rm LyC,abs}$ is the number of LyC photons absorbed per second and $P_B(T)$ is the temperature-dependent transition probability. In a completely ionized medium with $T=10^4\,{\rm K}$, $P_B(T)$ is $0.68$. Collisional radiation adds to the total \Lya\ luminosity, and we find that its fractional contribution is less than $\approx 30\%$, consistent with \citet[][see their Figure 8]{kimm19}. Neglecting collisional radiation, the escape fraction can be approximately obtained as
\begin{equation}
f_{\rm LyC}' = 1 - \dot{N}_{\rm Lya}^{\rm int}/0.68/\dot{N}_{\rm LyC}^{\rm int},
\label{eq:fesc_from_lya}
\end{equation}
where $\dot{N}_{\rm LyC,int}$ is the production rate of LyC radiation. Because ground-based observations or local observations do not always have access to \Lya,  the H$\alpha$ luminosity is used instead by assuming an intrinsic ratio of $L_{\rm Ly\alpha}/L_{\rm H\alpha}\approx8.7$ after correcting for the attenuation due to dust by using the Balmer decrement ratio ($L_{\rm H\alpha}/L_{\rm H\beta}=2.86$). Together with physical information about individual stars, derived from spectroscopy or photometry, \fescLyCa\ can be inferred from HII regions or a section of a galaxy \citep[e.g.,][]{doran13,choi20}. 

Figure~\ref{fig:fesc_from_lya} examines the validity of these assumptions by comparing the true \fescLyC\ with \fescLyC\ deduced from Equation~\ref{eq:fesc_from_lya}. Several important features can be gleaned from the figure. First, although there is significant scatter ($\Delta \fescLyC \sim 0.2$), the inferred \fescLyCa\ matches \fescLyC\ reasonably well down to $\fescLyCa\approx 0.1$. Below this, the inference cannot tell whether 30\% or none of the photons are escaping. Second, there is a tendency that $\fescLyCa < \fescLyC$ at $\fescLyC\la 0.2$, whereas the opposite is true for $\fescLyC \ga 0.4$. The former essentially occurs because collisional radiation becomes important and makes a non-negligible contribution to the total budget of escaping \Lya\ photons \citep{smith19}. Note that during this phase, the typical conditions that generate \Lya\ photons are $\nH\approx10^{3.5-4.5}\,\cmq$ and $T\approx 10^{4.1}\,{\rm K}$. On the other hand, the typical density of \Lya-emitting gas drops to $\nH\sim 5$--$10\,\cmq$ in the later phases, and the relative contribution from  collision radiation to $L_{\rm Ly\alpha}$ decreases to $\sim 10\%$. More importantly, in such conditions, recombination occurs slowly and not all of the LyC photons are used to produce \Lya. As a result, Equation~\ref{eq:fesc_from_lya} tends to over-estimate the true escape fraction. The vertical trend seen near $f_{\rm esc}'=1$ in the \texttt{SM5\_pZ014} run is likely the most dramatic example\footnote{The vertical trend seen in the \texttt{FM6D\_sZ002} run occurs because a portion of SN-driven shells are shock-heated, enhancing the recombinative as well as collisional \Lya\ radiation.}. The escape fraction is no longer a simple function of the solid angle for optically thick sightlines. Strictly speaking, our Case B assumptions are unlikely to be valid in the optically thin regime, and the use of Case A coefficients would increase the number of \Lya\ photons by 25\% at $T=10^4\,{\rm K}$. The inferred \fescLyCa\ would decrease slightly, but the overall trends should remain the same. 

Finally, we remark that the escaping luminosity of the LyC radiation correlates well with the escaping luminosity at 1500\,\AA\ (Figure~\ref{fig:fesc_uv}, bottom panels). In the case of the \texttt{SM6\_sZ014} run, the escaping luminosities are virtually indistinguishable.  This is a coincidence since the escape fractions of 900\AA\ and 1500\AA\ photons are significantly different at $t\la 4 \,{\rm Myr}$. The similarity arises because the timescale for UV luminosities to reach the maximum from the zero-age main sequence roughly matches the timescale of cloud disruption ($\sim 3\,{\rm Myr}$). The \texttt{SM6\_sZ002} run shows a less tight correlation, but again because the disruption timescale is similar ($\sim 3\,{\rm Myr}$) the escaping luminosity traces \fescLyC\ reasonably well.  However, GMCs that are dispersed slowly can add complexity to this trend, and the presence of the ISM/CGM can alter the escape fraction of both photons in a different manner. It is thus not conclusive yet if the escaping UV luminosity is a promising proxy to select potential LyC leakers. Cosmological simulations that can resolve individual GMCs are required to assess this possibility.



\subsection{Caveats}

In this work, we study the propagation of LyC and \Lya\ from GMCs  using high-resolution RMHD simulations, but our work has several limitations and considerable scope for improvement. First, the parameter space we probe is quite limited due to the constraint in computational resources. For example, the only case with low \tSFE\ ($\sim1\%$) in our simulations is when strong turbulence is applied, but its physical origin is not justified. An alternative way of simulating GMCs that form stars at a slow rate may be  by considering lower gas column densities of $\Sigma\la 50\,\msun\,{\rm pc^{-2}}$ (for $10^5\,\msun$) or $\Sigma\la 100\,\msun\,{\rm pc^{-2}}$  (for $10^6\,\msun$), or by introducing strong magnetic fields  \citep[e.g.,][]{kimjg18,kimjg21}. Furthermore, we do not study the effects of different turbulent seeds on the evolution of GMCs and hence the escape of LyC photons \citep{grudic21a,kimjg21}, but choose to adopt a different geometry and a high resolution in this study. Admittedly, even the turbulent structures we employ do not necessarily represent those of high-redshift galaxies that are responsible for reionization. To improve the initial conditions of GMC simulations and to study the propagation of radiation in a realistic setting, high-resolution cosmological galactic-scale simulations are required and are likely possible in the near future.

Second, our simulations lack prescriptions for physical processes that may be dynamically important in GMC environments. Proto-stellar jets and stellar winds are known to provide extra momentum to the surroundings of massive stars, shaping the morphology of HII bubbles \citep[e.g.,][]{rogers13,dale14,grudic21b,geen21},  although winds become relatively less important at lower metallicities. Multiple scattering of \Lya\ photons \citep{dijkstra08,smith17,kimm18,tomaselli21} may accelerate the disruption of dense clumps in the early stage of star formation as well, potentially increasing  the escape fractions of LyC photons \citep{kimm19}. Inclusion of cosmic rays generated by diffusive shock acceleration can also drive gentle outflows, regulating star formation on a long-term timescale \citep[e.g.,][]{girichidis16,pfrommer17,ruszkowski17,dashyan19}. While understanding the detailed effects of these processes is certainly a challenging task, it is the natural step forward to unravel the complex evolution histories of the GMCs.

\begin{figure}
   \centering
   \includegraphics[width=\linewidth]{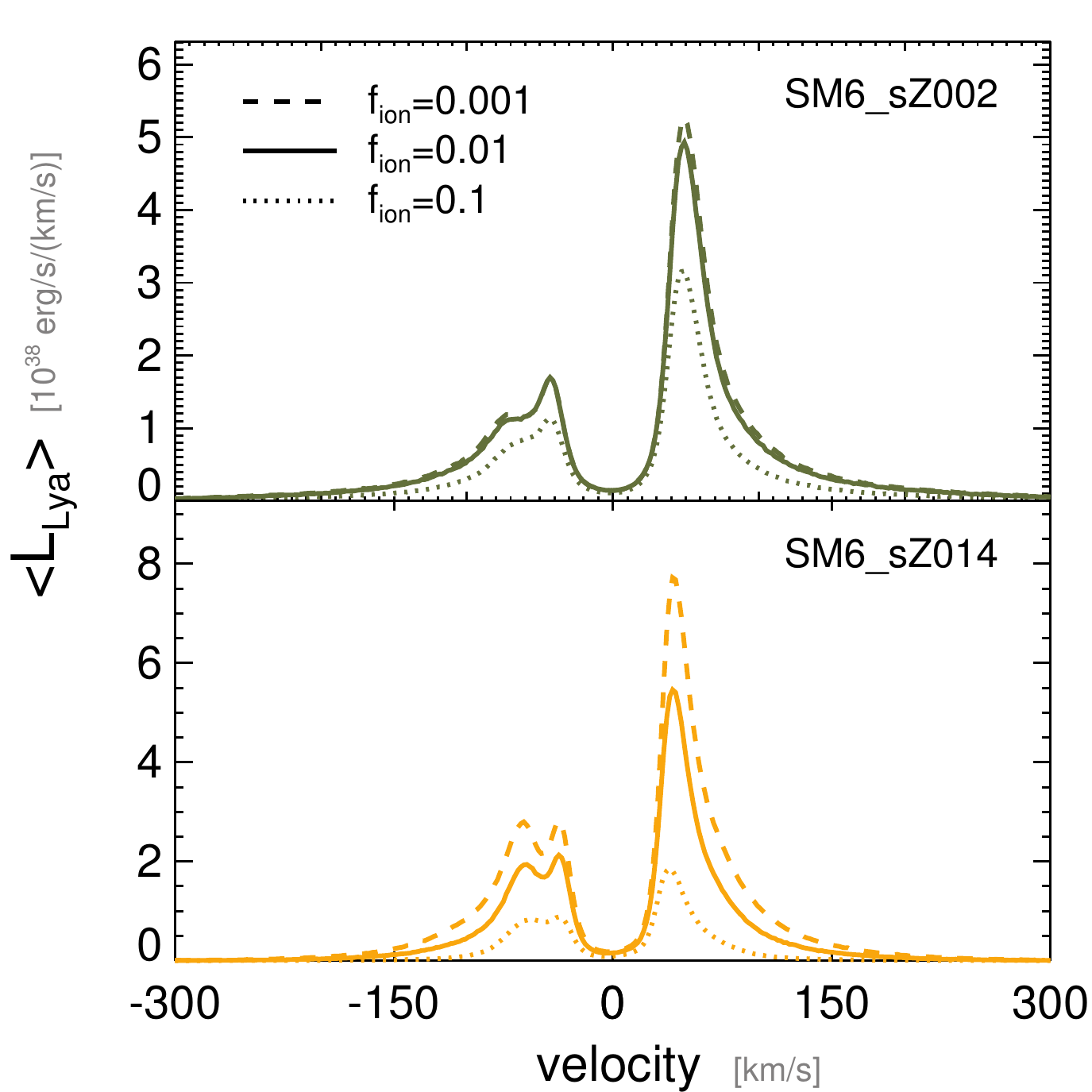} 
   \caption{Impact of the choice of \fion,  the fraction of dust that survives in ionized media, on the luminosity-weighted \Lya\ profiles in a metal-poor  (\texttt{SM6\_sZ002}, top) and metal-rich GMC (\texttt{SM6\_sZ002}, bottom). The line profiles are normalized by the bin size of the velocity. Increasing \fion\ suppresses the line fluxes, but the velocity separations of the double peaks are barely changed.  }
   \label{fig:fion}
\end{figure}

Last but not the least, our modeling of dust is likely over-simplied. We assume that 1\% of dust survives in an ionized medium (i.e. $\fion=0.01$), but observations appear to suggest a wide range of $10^{-4}\la \fion \la 10^{-1}$ \citep[see][for the discussion]{laursen09}.  Although this parameter is not very well constrained, several mechanisms that lower the dust abundance are likely to be at play in HII regions, such as radiation pressure on dust \citep{draine11b} or disruption due to radiative torque applied by UV radiation \citep{hoang19}, and thus the low \fion\ value may be justifiable. In order to gauge the uncertainties associated with \fion, we present the luminosity-weighted \Lya\ profiles calculated with $\fion=0.1$ or 0.001 in Figure~\ref{fig:fion}\footnote{Note that an increase in the amount of dust hardly affects the escape of LyC radiation as the optical depth to neutral hydrogen is already an order of magnitude larger than that of dust \citep{kimm19}, and we focus on the impact on \Lya\,  as the only agent that destroys \Lya\ is dust, in this study.}. We find that the total \Lya\ escape fraction from the \texttt{SM6\_sZ002} run (54\%) is changed to 34\% or 58\% when \fion\ is adjusted to $0.1$ or 0.001, respectively. The change in \fescLyaL\ is more pronounced in the metal-rich case (\texttt{SM6\_sZ014}), with values of 10\% ($\fion=0.1$) or 45\% ($\fion=0.001$), compared with 31\% in the fiducial case. Nevertheless, the level of uncertainties in the prediction of \fescLya\ is unlikely to have a significant impact on our conclusions, and more importantly, the velocity separation is virtually unaffected by the choice of $\fion$. 

\section{Summary and conclusions}

In this study, we investigated the escape of LyC and \Lya\ photons from simulated GMCs with diverse physical conditions, with a particular focus on dense and metal-poor clouds, which are likely to be relevant to the evolution of high-$z$ galaxies and reionization. By analyzing 18 high-resolution ($0.02\le \Delta x_{\rm min} \le 0.08 \,{\rm pc}$) RMHD simulations with a self-consistent star formation model based on the sink particle algorithm, we showed that a significant fraction  of LyC ($\approx$ 15--70\%) and \Lya\ photons ($\approx$ 15--85\%) leak from the clouds as they are dispersed rapidly due to stellar feedback. The detailed results can be summarized as follows.

\begin{itemize}
\item The escape fractions of LyC and \Lya\ photons in GMCs generally increase with time as photoionization heating disrupts the clouds very efficiently in $t_{\rm desc}$=3--5  Myr. The escaping luminosity of LyC radiation peaks at $\approx$ 3 Myr from the onset of star formation (Figure~\ref{fig:fescLyC_all}), whereas the maximum escaping luminosity of \Lya\ photons tends to occur earlier at 1--2 Myr (Figure~\ref{fig:fescLyA_all}). These timescales are larger in a homogeneous cloud where the gas surface density is lower. 

\item The time-integrated escape fraction of ionizing radiation (\fescLyCL) reveals a strong correlation with the total SFE, showing a maximum at $\tSFE\approx 20\%$ (Figure~\ref{fig:fesc_sfe}). LyC leakage is less notable in GMCs with lower \tSFE\  because of the finite number of LyC photons, while the propagation is suppressed in GMCs with high \tSFE\  because their extreme gas densities prevent the rapid development of ionized bubbles (Figure~\ref{fig:m_ion}).

\item Consistent with previous findings, less massive clouds with $10^5\,\msun$ show a larger LyC escape fraction of $\sim 60\%$, while a more massive cloud with $10^6\,\msun$ reveals a lower $\fescLyCL$ of $\sim 40\%$. The difference is partly driven by the higher gas surface density adopted in the more massive GMCs. Indeed, the smallest fraction ($\approx 15\%$) of LyC radiation escapes from the high-mass GMC with an extremely high $\Sigma_{\rm gas}$ of $1000\,\msun\,{\rm pc^{-2}}$.

\item Metal-rich clouds tend to show lower \fescLyCL\ and \fescLyaL\ than metal-poor GMCs by $15$--$31\%$ and $22$--$45\%$, respectively,  although their \tSFE\  are sometimes higher. We attribute this to additional cooling due to metals, which weakens the effect of photoionization heating.

\item The escape of LyC  photons is suppressed in a GMC with strong turbulence as \tSFE\ is dramatically reduced and the number of LyC photons become finite. In contrast, a larger fraction of \Lya\ photons  escapes from a GMC with strong turbulence. 

\item Only marginal differences  in  \fescLyCL\ and \fescLyaL\ are found between clouds with different morphologies (spherical versus filamentary), magnetic field strengths, and the presence of SNe, although filamentary structures allow for more efficient \Lya\ escape. 


\item \Lya\ photons escape most efficiently from their birth clouds in the middle of cloud disruption (Figure~\ref{fig:img_all}). The velocity separation of the double peak in the \Lya\ spectrum is initially very large ($\vsep\sim500\,\kms$), but the luminosity-weighted spectrum reveals a smaller $\vsep$ of $\approx120\,\kms$ (Figure~\ref{fig:Lratio}). The velocity separation as well as the evolutionary locus in the \vsep--$f_{900}$ plane turn out to be similar, regardless of the properties of the simulated GMCs (Figure~\ref{fig:Lratio} and \ref{fig:vsep}), suggesting that these measurements may reveal more about the properties of the ISM and CGM of galaxies rather than their molecular clouds.

\item The escape fraction of \Lya\ photons is systemically larger than that of LyC ($f_{\rm Ly\alpha}\approx f_{900}^{0.27}$) (Figure~\ref{fig:vsep}). Despite the significant difference in the  absorption probability owing to dust, the correlation with the escape of UV photons at 1500\AA\ is found to be equally strong as that of \Lya\ as clumpy gas distributions lead to an absorption of both \Lya\ and the UV photons, while some photons escape through low column density channels (Figure~\ref{fig:fesc_comp}). By contrast, the correlation between $f_{900}$ and $f_{\rm H\alpha}$ is found to be very weak, as $f_{\rm H\alpha}$  is usually very high. 

\item \fescLyC\ may be best inferred from \Lya\ luminosities by assuming recombination in the optically thick regime (Equation~\ref{eq:fesc_from_lya} and Figure~\ref{fig:fesc_from_lya}).  Although there is a significant scatter ($\Delta f \sim 0.2$), the inferred escape fraction shows a reasonable one-to-one correspondence at $\fescLyC \ga 0.1$. However, there is a fair chance that the approach may under-estimate the true \fescLyC\ at low $\fescLyC \la 0.2$ due to the neglect of collisional radiation, or may overestimate at high $\fescLyC \ga 0.4$ because the optically thick assumption is no longer valid.
\end{itemize}

We remark that the escape of ionizing radiation from our gravitationally well-bound GMCs is overall more efficient than required to fully ionize the Universe by $z=6$ \citep[e.g.,][$\fescLyCL\sim10\%$ ]{robertson15,rosdahl18}. However, this is not a concern because the absorption of LyC by the ISM is significant. As shown by \citet{yoo20}, the escape fraction in disk galaxies can easily drop from $\sim 50\%$ at the 100 pc scale to $\sim 10\%$ by the time LyC photons leave their host dark matter halo. Moreover, observations of local GMCs or star forming regions often reveal extremely large \fescLyC\ \citep{pellegrini12,doran13,mcleod19,choi20,della-bruna21}. The question remains though, whether the properties of the local GMCs can be applicable to those in high-$z$ galaxies. Another important uncertainty in understanding the UV propagation is the star formation modelling. For example, \citet{howard18} argue that the escape fraction fluctuates widely until the clouds are completely dispersed, while this study, including \citet{kimjg18}, finds values for \fescLyC\ that are rather monotonically increasing with time. Obviously, it is difficult to single out the cause of the difference, given the different numerical methods used, but we suspect that this is likely because of different star formation models. In \citet{howard18}, gas is accreted first and 20\% of this gas is converted into stars on a time scale of 0.37 Myr, while in our simulations, stars form immediately once enough gas mass is accreted onto sink particles. We believe that a correct way of modelling star formation is not yet conclusive and subject to debate. Combined efforts based on upcoming observing facilities and state-of-the-art simulations are required to further shed light on the role of LyC photons from GMCs in the evolution of galaxies and reionization, and to exploit the useful information in galactic \Lya\ spectra.

\section*{Acknowledgements}
We thank an anonymous referee for helpful comments on the manuscript. TK was supported by the National Research Foundation of Korea (NRF-2019K2A9A1A06091377 and 2020R1C1C1007079). 
SG acknowledges support from a NOVA grant for the theory of massive star formation. 
TG is supported by the ERC Starting grant 757258 ‘TRIPLE’.
This work was supported by the Programme National Cosmology et Galaxies (PNCG) of CNRS/INSU with INP and IN2P3, co-funded by CEA and CNES. We additionally acknowledge support and computational resources from the Common Computing Facility (CCF) of the LABEX Lyon Institute of Origins (ANR-10-LABX-66).  The supercomputing time for numerical simulations was kindly provided by KISTI (KSC-2019-CRE-0196),  and large data transfer was supported by KREONET, which is managed and operated by KISTI. This work also used the DiRAC Complexity system, operated by the University of Leicester IT Services, which forms part of the STFC DiRAC HPC Facility (www.dirac.ac.uk). 
Some of the simulations in this paper were performed on the Dutch National Supercomputing cluster Cartesius at SURFsara and on the draco cluster hosted by the Max Planck Computing and Data Facility (http://www.mpcdf.mpg.de/).  The authors gratefully acknowledge the data storage service SDS@hd supported by the Ministry of Science, Research and the Arts Baden-Württemberg (MWK) and the German Research Foundation (DFG) through grant INST35/1314-1FUGG.

\appendix

\section{Effects of resolution}
\label{appendix:resolution}
The evolution of GMCs is often affected by the numerical resolution because the Stromgren radius and the cooling length of SN driven shells become progressively small with increasing density. Therefore, we investigate whether our results numerically converge by comparing \tSFE\ and escape fractions. 

\begin{figure}
    \centering
    \includegraphics[width=9cm]{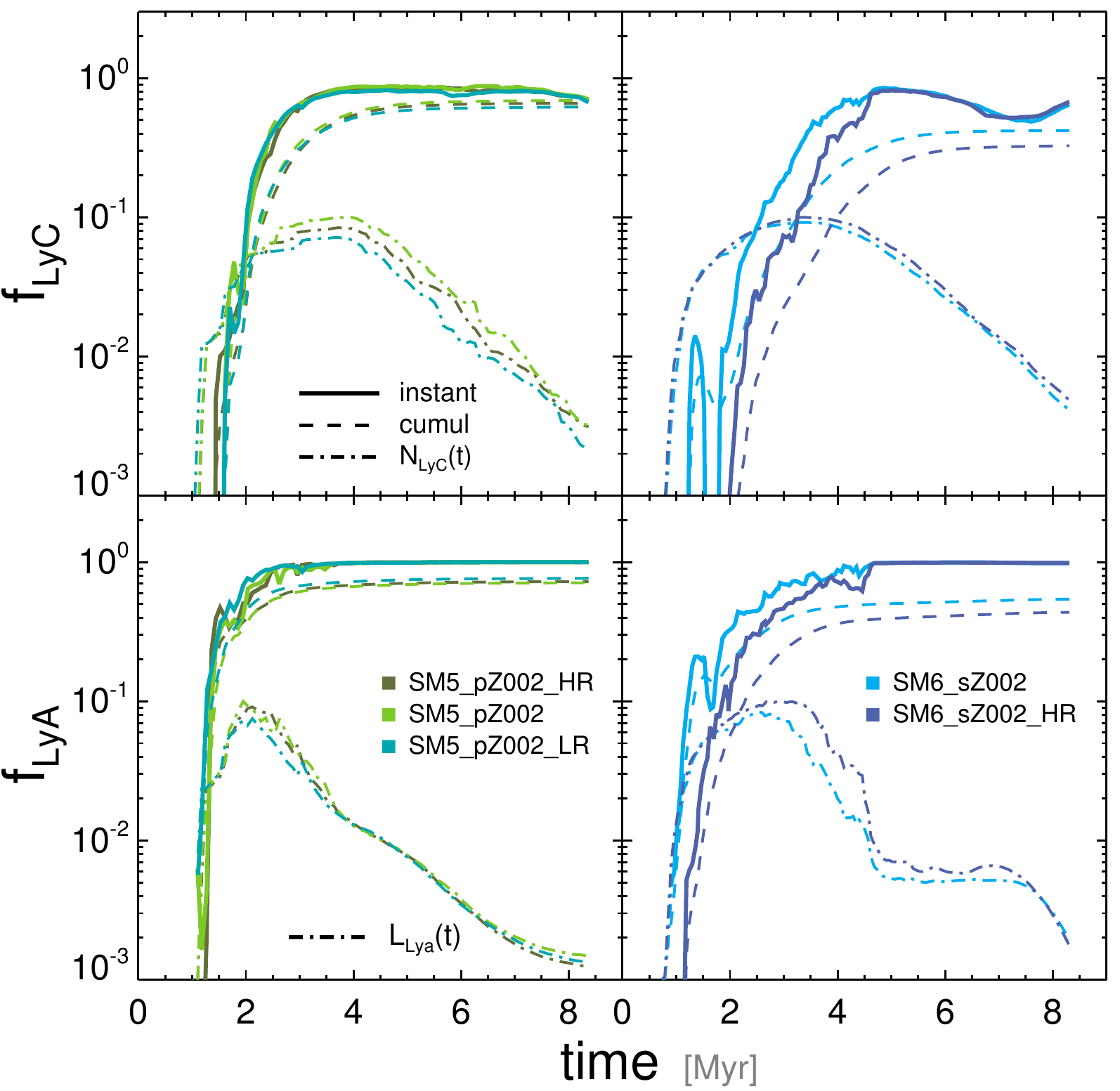}
    \caption{Effect of resolutions on the escape of LyC and \Lya\ radiation. The results from the runs with varying resolutions are shown as different color-codings. The solid and dashed lines show instantaneous and cumulative escape fractions, respectively, while the dot-dashed lines correspond to the number of LyC photons or intrinsic \Lya\ luminosity, as indicated in the legend.}
    \label{fig:resolution}
\end{figure}

The left panels in Figure~\ref{fig:resolution} show that \fescLyC\ and \fescLya\ in the small cloud of mass $10^5\,\msun$ are almost indistinguishable at  minimum cell widths in the range 0.02--0.08 pc. On the other hand, differences in \tSFE\ in different resolution runs are more perceptible. Our fiducial case with a minimum 0.04 pc resolution converts 25.7\% of gas into stars, but the fraction decreases to 19.2\,\% in a lower-resolution simulation (\texttt{SM5\_pZ002\_LR}). However, there is no systematic trend of an increase in the resolution leading to a higher \tSFE, as our highest resolution run produces 23.9\% of stars. This indicates that our simulation results are likely to converge reasonably at resolutions in the range 0.02--0.04 pc, but the results for a resolution of 0.08 pc may still be acceptable within an accuracy of about $20\%$. 

The right panels in Figure~\ref{fig:resolution} further shows the effects of the resolution in a more massive cloud with $10^6\,\msun$. Unlike small GMCs, the differences in \fescLyC\ and \fescLya\ are larger, despite the differences in \tSFE\ between the fiducial (\texttt{SM6\_sZ002}, 0.510) and higher-resolution run (\texttt{SM6\_sZ002\_HR}, 0.549) being small. The \fescLyCL\ value in the 0.08 pc resolution run is 54.2\%, but this is reduced to 43.6\%, which  again differ by $\approx 20\%$. This may not be negligible, but we argue that a relative comparison between runs with the same resolution should still be useful as it allows us to study the physical origin of the potential differences in \fescLyC\ and \fescLya\ between different settings. 

\bibliography{refs}
\bibliographystyle{aasjournal}



\end{document}